\documentclass[1p,preprint]{elsarticle}
\usepackage{graphicx}
\usepackage{color}
\usepackage{multirow}

\usepackage[linesnumbered,noend,vlined, scleft,nofillcomment,ruled]{algorithm2e}

\usepackage{subfigure}
\usepackage{amsmath} 
\newtheorem{definition}{Definition}[section]

\begin{document}

\begin{frontmatter}

\title{GraCT: A Grammar-based Compressed Index\\ for Trajectory Data \tnoteref{t1}}

\tnotetext[t1]{ A shorter, preliminary version of this paper appeared
in {\em Proc.\ SPIRE 2016} \citep{RodriguezBrisaboa16}.
}

 \author[cor]{Nieves R. Brisaboa}
 \ead{brisaboa@udc.es}
 \author[cor]{Adri\'an G\'omez-Brand\'on\corref{cor1}}
 \ead{adrian.gbrandon@udc.es}
 \author[chi]{Gonzalo Navarro}
  \ead{gnavarro@dcc.uchile.cl}
 \author[cor]{Jos\'e R. Param\'a}
  \ead{jose.parama@udc.es}

\cortext[cor1]{Corresponding author. Tel. +34981167000 Fax.
+34981167160.}

 \address[cor]{Universidade da Coru\~na,  Facultade de Inform\'atica, CITIC, Campus de Elvi\~na s/n, 15071 A Coru\~na, Spain}
 
 \address[chi]{Millennium Institute for Foundational Research on Data (IMFD), Department of Computer Science, University of Chile,
 Beauchef 851, Santiago, Chile}

\begin{abstract}
We introduce a compressed data structure for the storage of free trajectories of moving objects (such as ships and planes) that efficiently supports various spatio-temporal queries. Our structure, dubbed GraCT, stores the absolute positions of all the objects at regular time intervals (snapshots) using a $k^2$-tree, which is a space- and time-efficient version of a region quadtree. Positions between snapshots are represented as logs of relative movements and compressed using Re-Pair, a grammar-based compressor. The nonterminals of this grammar are enhanced with MBR information to enable fast queries.
 
The GraCT structure of a dataset occupies less than the raw data compressed with a powerful traditional compressor such as \textit{p7zip}. Further, instead of requiring full decompression to access the data like a traditional compressor, GraCT supports direct access to object trajectories or to their position at specific time instants, as well as spatial range and nearest-neighbor queries on time instants and/or time intervals.
 
Compared to traditional methods for storing and indexing spatio-temporal data, GraCT requires two orders of magnitude less space, and is competitive in query times. In particular, thanks to its compressed representation, the GraCT structure may reside in main memory in situations where any classical uncompressed index must resort to disk, thereby being one or two orders of magnitude faster.

\end{abstract}


\end{frontmatter}

\section{Introduction}

More than two decades after it emerged, the field of moving object databases is still an active area of research. The data collected from the GPS positions of large sets of cars, ships, planes, smartphones and wearable devices, has lead to a growing interest in applications that exploit trajectory information, for example with data mining purposes. As these datasets grow in size and the applications managing moving objects become more sophisticated,  more space/time-efficient storage techniques are required.

A trajectory is the path followed by a moving object through space as a function of time. Due to storage requirements and the limitations of the devices used to acquire the object positions, the continuous movement of an object is usually approximated with discrete samples of spatio-temporal location points. The more samples taken, the more accurate the trajectory. High sampling rates, however, result in large amounts of data, which increase storage, transmission and processing needs. Even when storage, network and processing capacity grows rapidly, the collected data grows even faster, and thus it is necessary to aim for reduced trajectory representations \cite{zheng11}. 


Traditional methods for compressing trajectories include line generalization (or simplification) techniques, which keep only some of the trajectory points and discard the rest. 
This approach results in some loss of information on the real trajectory. A lossless strategy to obtain compression is the use of {\em delta compression}, where each new position is stored as the difference with the previous one. This idea exploits the fact that consecutive positions are expected to be closer to each other, and that smaller numbers can be stored using fewer bits. Extracting a whole trajectory with this arrangement is easy. Efficiently accessing the position of an object at a given time, instead, requires sampling some absolute positions at regular time intervals, which introduces a space/time tradeoff.

Another issue is how to index the trajectory data in order to answer queries other than just retrieving a whole trajectory or finding the position of an object at a given time instant. A number of indexes have been proposed since the 90's to handle a rich set of queries on trajectory data. Most indexes were modifications of the R-tree \cite{Guttman:1984}  augmented with another dimension to deal with the time. None of those works, however, compress the data. Rather, they are designed to work on disk, which is much slower than the main memory. 

With the increasing gap in the access time of main memory versus disk, compressing the trajectories in order to query them in main memory is an attractive option. Some recent proposals following this trend \cite{TrajStore, Wang:2014} build on delta compression, coupled with an encoding that favors small numbers. The optimal codes for delta compression can be obtained with a statistical encoder that exploits frequency bias (typically, smaller numbers are more frequent).

In this article we introduce a new way to compress trajectories and demonstrate that it is much more effective than statistical compression on real-world data. Instead of exploiting the higher frequency of smaller differences, we exploit the fact that, in many applications, trajectories tend to be similar to others, wholly or piecewise. We resort to {\em grammar compression} \cite{KY00}, which is a method from the family of dictionary-based compression methods. Grammar compression generates a context-free grammar that generates (only) a given input sequence. Our input sequence is the concatenation of all the trajectories in differential form. Grammar compression exploits the similarities between trajectories: if many trajectories are similar, then a small grammar can be found that generates the whole dataset.


Our index, called {\em Grammar-based Compressed Representation of Trajectories (GraCT)}, is an in-memory representation that combines grammar compression of trajectories with additional data structures to support efficient queries of various kinds on the dataset, without the need of decompressing it. On one hand, {\em snapshots} of the positions of all the objects at regular time intervals are created, and those positions are indexed in a compressed quadtree representation called $k^2$-tree \cite{ktree}. Such spatial index allows us to restrict the set of candidate objects when answering queries. The movements of the objects between two consecutive snapshots are represented in a sequence called the {\em log of movements}, which is represented in differential form and compressed with Re-Pair, an effective grammar compression method \cite{larsson2000off}. To avoid decompressing and sequentially traversing the logs to answer queries, the nonterminals of the grammar are enhanced with additional information, most notably the minimum bounding rectangle (MBR) of all the (relative) positions represented by the nonterminal. MBRs allow us to skip whole nonterminal symbols without the need to decompress them, thereby speeding up the traversal of logs. This arrangement enables GraCT to efficiently answer \textit{time-slice}, \textit{time-interval} and \textit{k-nearest neighbor} queries.

GraCT requires discretizing the data in order to obtain compression. When the data must be used with full precision, GraCT can be used as an {\em in-memory cached index} using a sufficient degree of discretization. The candidate set of answers delivered by GraCT are then read from disk and verified.


 

We evaluate our techniques on three real-life datasets: ships at sea, commercial aircrafts, and city taxis.
We show that the grammar compression of GraCT is highly effective, reducing the raw data to 4\%--7\% of its original size. This is even less than the space achieved by \textit{p7zip}, a powerful dictionary-based compressor that does not support direct access nor querying. It is also less than half the space used by a baseline index we implement that replaces grammar compression by statistical symbolwise compression. Compared to a traditional method for storing and indexing spatio-temporal data (concretely, the MVR-tree \cite{PapadiasT01}), the compressed representation of GraCT occupies two orders of magnitude less space. In relation to query times, GraCT is up to 4 times faster than the statistically compressed index, except for retrieving whole trajectories, where it is slightly slower. GraCT is also faster than the MVR-tree running in main memory (which, as explained, is feasible for small datasets only) for all queries except on nearest-neighbor queries, on which it can be clearly slower. GraCT is one or two orders of magnitude faster, however, when the MVR-tree cannot be maintained in main memory. Finally, a simple combination with a relational storage of the full-precision trajectory points shows that an in-memory cache based on GraCT is an order of magnitude faster than the MVR-tree operating on disk for the basic time-slice queries.

The paper is structured as follows. Section \ref{sect:related} covers related work. Section \ref{sect:gract} introduces the GraCT structure and Section \ref{sect:comp} explains in detail the  compression of the logs. Sections \ref{sect:quep} and \ref{query} describe how the various supported queries are processed in GraCT. Section \ref{sect:exp} presents the experimental results. Finally, Section \ref{sect:con} presents our conclusions and directions for future research.

\section{Related Work}\label{sect:related}



 

This section describes the previous work on indexing and compressing trajectories. Section~\ref{sec:indexing} describes the related work on indexing trajectories, Section~\ref{sec:compression} on compressing trajectories, and Section~\ref{sec:compression+indexing} on combining both. Readers willing to go directly to the description of our new index may jump directly to Section~\ref{sec:cds}, where the basic concepts of compact data structures are covered.

\subsection{Indexing trajectories} \label{sec:indexing}

Indexes that handle free trajectories of points in the space are also called spatio-temporal indexes.
One type spatio-temporal index builds on a classic multidimensional spatial index, in the form of a temporally augmented R-tree \cite{Guttman:1984}. While the R-tree uses two-dimensional Minimum Bounding Rectangles (MBRs) to enclose the spatial objects stored in the database, the 3DR-tree \cite{Vazirgiannis1998} uses Minimum Bounding Boxes (MBBs), where the third dimension is time, to enclose the segments of the trajectories. The problem is that the three-dimensional space covered by an MBB can be large, even in small segments, resulting in high levels of overlap and limited power of discrimination. The STR-Tree \cite{PfoserJT00} is an extension of the 3DR-tree designed to overcome this problem, by modifying the insertion/split strategy that produces the MBBs. The same work also proposes an index called TB-tree, where several segments of a trajectory are bundled into partial trajectories that are inserted as MBBs of an R-tree.

A second kind of index is a versioned R-tree, which creates an R-tree for each timestamp and a B-tree to select the relevant R-trees. Creating an R-tree for each timestamp requires large amounts of space. To overcome this, instead of storing the complete R-tree for each timestamp, these techniques store only the part of the R-tree that is different from the R-tree in the previous timestamp. Examples of indexes of this type are MR-tree \cite{Xu90}, HR-tree \cite{NascimentoS98}, HR+-tree \cite{TaoP01} and MV3R-tree \cite{PapadiasT01}.

A third family of indexes, called grid-based indexes, partition the space and build a temporal index for each of the spatial partitions. The Scalable and Efficient Trajectory Index (SETI) \cite{ChakkaEP03} divides the spatial dimension into static, non-overlapping cells. Within each cell, the trajectory segments are indexed by time using an $R^*$-tree. If a segment crosses the boundary between two cells, it is clipped and stored twice in the $R^*$-trees of both partitions. Other examples from this family include Multi Time Split B-Tree \cite{Zhou:2005}, Compressed Start-End tree \cite{WangZXM08}, and PIST \cite{Botea2008}. More recently, GCOTraj \cite{YANG20181} presented a work along the same line, which also divides the space into cells, and all the subtrajectories in a cell are stored in the same disk page. Pages are ordered on the disk following a space filling curve to optimize I/O times. 

PA-trees \cite{4138203} use a completely different approach to avoid MBBs. The index is built with linear and order-\textit{k} polynomial approximations of the trajectories. Of course, the approximation deviates from the real trajectory, thus  the index also stores  the maximum deviation with respect to the actual trajectory to avoid false negatives.

A recent active research line is the management of trajectories in the distributed computing framework. 
 PRADASE \cite{Ma:2009:QPM:1651263.1651266}  and CloST \cite{Tan:2012:CHS:2396761.2398589} are trajectory storage systems  built on top of Hadoop and extended with a spatio-temporal index. MD-HBase \cite{Nishimura2013}, R-HBase \cite{7022647}, and GeoMesa \cite{GeoMesa} use distributed key/value stores. 
TrajSpark \cite{Zhang} is an in-memory system based on Spark and equipped  with a two-level spatio-temporal index. While we do not address distributed computing in this article, we remark that, in general, a compressed storage enables the distribution of spatiotemporal data across fewer computers, thereby reducing the communication costs.

The closest predecessor of our GraCT index is SEST-Index \cite{GGORN05,Worboys05}. This was the first structure considering snapshots and logs of events, but it did not consider compression and was oriented to storing ``events'' (objects that appear in or disappear from an area).

\subsection{Compressing trajectories}
\label{sec:compression}

A simple way to reduce the size of a trajectory is to produce a new trajectory that approximates the original one by selectively removing some of its points. This approach, called trajectory simplification, is based on so-called line generalization techniques. This strategy  is typically used as a preprocessing step that is applied before the trajectories are compressed or indexed \cite{zheng11}. Therefore, these methods are compatible with other compression and indexing techniques.

The simplest trajectory simplification method is uniform sampling \cite{1644324}, which selects points at regular intervals (e.g., 5th, 10th, 15th, etc.), and discards the rest. Though simple to apply, the resulting trajectory may lose too much precision.

More evolved algorithms analyze the points in order to identify those containing more information, while discarding more redundant surrounding points. One of the best-known algorithms of this type is Douglas-Peucker \cite{Douglas:1973:ARN}. Other methods in this family include top-down time ratio \cite{MeratniaB04}, sliding window \cite{989531}, SQUISH \cite{Muckell2014}, and OPERB \cite{Lin:2017}.

Trajectory compression techniques based on speed and direction include points in the approximated trajectory provided they represent a significant change in the course of the trajectory. Examples of methods from this family include dead reckoning \cite{TrajcevskiCSWV06}, threshold-guided sampling, and STTrace \cite{1644324}.

Finally, some methods use the knowledge of an underlying network to decide which points are retained (e.g., those at bus stops or at street intersections) \cite{SchmidRL09,TaLCF16}.

Independently of whether and how trajectory simplification is applied, the resulting trajectories can be compressed. The typical alternative is {\em delta compression}, which stores the first point in the trajectory in absolute form and all the others as the difference with the previous point. This applies to all three components (x, y, and timestamp). 

Trajic \cite{Trajic} uses a strategy related to delta compression. It uses a predictor of the next point of a trajectory and an encoder of the residuals, that is, the difference between the prediction and the actual point. If the residual is small, the encoder uses few bits. 

Finally, a somewhat related research line faces the compression of traffic data over networks, mainly flow at intersections and average speed, using low-dimensional models such as principal component analysis or low-rank approximation \cite{PC3,PC2}. Basically, these techniques approximately represent a matrix as the product of smaller matrices. Similar techniques are used in 3D animated video to compress the trajectories formed by the positions of the vertices of animated synthetic figures \cite{PC4}, and could probably be applied to compress GPS trajectories. 

GraCT does not discard points, but discretizes the positions according to an underlying grid, whose size is parameterized by the needs of the applications (e.g., it makes no sense to store ship positions with a precision of centimeters). It also uses a discrete temporal scale, interpolating the measures so that each trajectory has a point at each time instant. GraCT is unique in using grammar-based compression on the trajectories.

\subsection{Trajectory compression and indexing}
\label{sec:compression+indexing}

There are just a few systems that combine trajectory compression and indexing.

TrajStore \cite{TrajStore} is similar to SETI \cite{ChakkaEP03}, as it also slices the trajectories into subtrajectories that fit in the cells into which the space was divided. The difference is that the data of each cell is compressed and that the cells are adapted to the distribution of the data. TrajStore uses delta compression and a cluster-based compression, which groups several similar trajectories and stores only one of them; thus it is a lossy method. 
In addition, the  cells are spatially indexed by a global quadtree, and each cell has a temporal index of the disk pages that make up the cell.

SharkDB \cite{Wang:2014} is an in-memory trajectory storage system that combines a column-oriented data structure and compression. SharkDB divides the time into frames of a given length (e.g., one minute) and stores one point for each trajectory within that frame. All of the points within the frame are stored as a column of a column-oriented database management system. To reduce data size, it also uses delta compression. SharkDB uses multithreading to speed up query processing.

PRESS \cite{Song:2014} is a system that stores compressed trajectories and supports queries, but the trajectories must be paths in an underlying network (e.g., train stations, bus stops). We only consider free trajectories in this article.

\subsection{Compact data structures and compression methods} \label{sec:cds}

Compact data structures \cite{Nav16} are a new family of data structures that combine query support with compressed data representation. They often use space close to that of a plain compressed representation, while supporting queries on the data in time competitive with classical data structures. Our index makes use of  various compact data structures and compression methods.

\subsubsection{Rank and Select}

The following two operations over bit sequences (or bitmaps) are basic components of most compact data structures: $rank_b(B, p)$ counts the number of occurrences of bit $b$ in bitmap $B$ up to position $p$, whereas $select_b(B, n)$ returns the position in bitmap $B$ where the $n$-th occurrence of bit $b$ is located.

Some data structures (see \cite{Munro96}, for example) allow these operations to be solved in constant time, using $n + o(n)$ bits of total space (in practice, approximately 5\% extra space over the original
bitmap).

\subsubsection{ $k^2$-tree}\label{k2}

The $k^2$-tree is a compact data structure originally designed to represent Web graphs within reduced space, allowing them to be navigated directly in compressed form \cite{ktree}. In general, a $k^2$-tree can be used to represent the adjacency matrix of any graph, as well as binary matrices.

\begin{figure}[t]
\begin{center}

\includegraphics[scale=0.15]{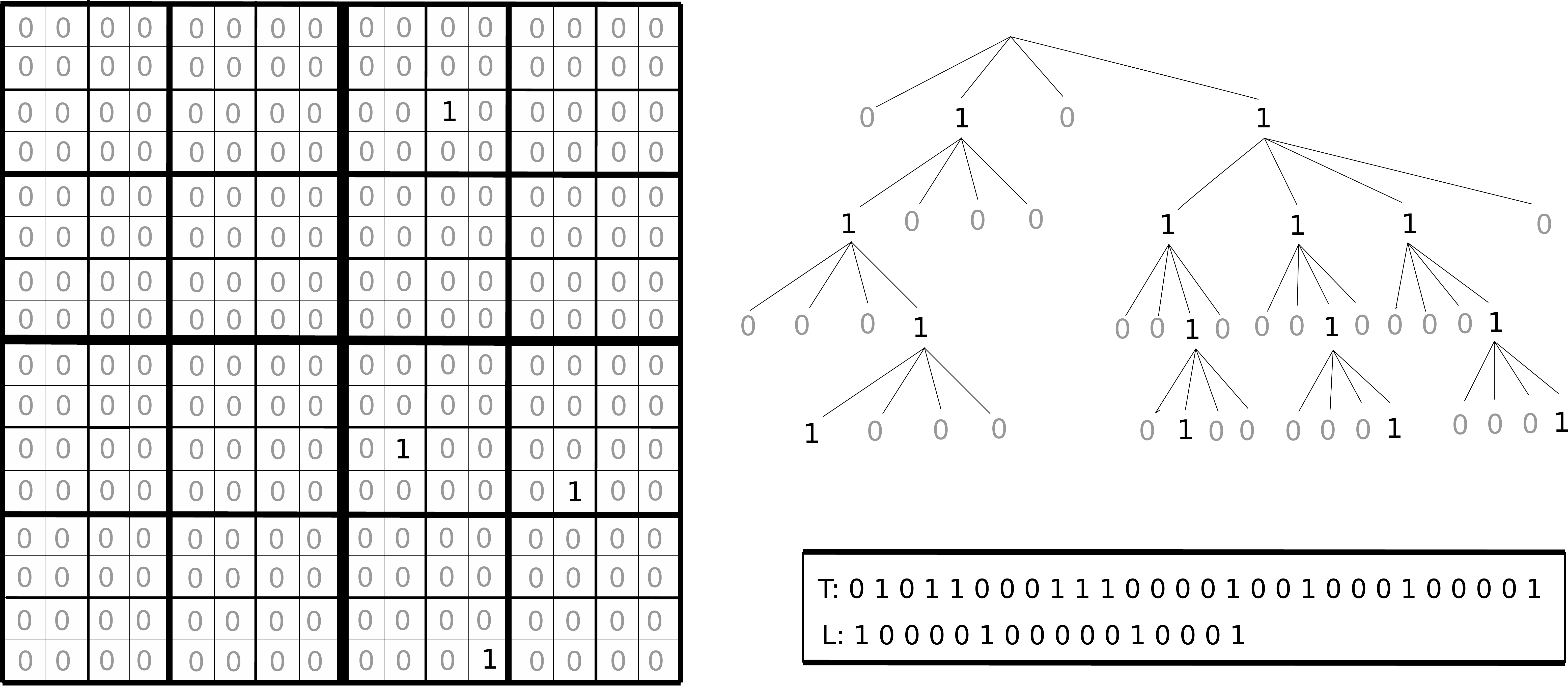} 

\end{center}
\caption{Example of a binary matrix (left), the associated $k^2$-tree conceptual representation (top right), and its compact representation (bottom right), for $k=2$.}
\label{fig:example}
\end{figure}

Conceptually, the $k^2$-tree is a $k^2$-ary tree built from a binary matrix by recursively subdividing the matrix into $k^2$ submatrices of the same size. It starts by subdividing the original $n \times n$ matrix into $k^2$ submatrices of size $n^2/k^2$. The submatrices are ordered from left to right and from top to bottom. Each submatrix generates a child of the root node whose value is $1$ if there is at least one $1$ in the cells of that submatrix, and $0$ otherwise. The subdivision proceeds recursively for each child with value $1$ until it reaches a submatrix full of 0s, or until it reaches the cells of the original matrix (i.e., submatrices of size $1\times 1$). Figure~\ref{fig:example} shows an example.

Instead of using a pointer-based representation, the $k^2$-tree is compactly stored using bitmaps $T$ and $L$ (see Figure \ref{fig:example}). \textit{T} stores all the bits of the $k^2$-tree, except those in the last level. The bits are placed according to a level-wise traversal: first the $k^2$ binary values of the children of the root node, then the values of the second level, and so on. $L$ stores the last level of the tree, consisting of cell values of the original binary matrix.

Any cell, row, column, or region of the matrix can be obtained efficiently via $rank$ and $select$ operations over bitmaps $T$ and $L$. For example, assuming a value of 1 at position $p$ in $T$, its $k^2$ children start at position $p_{children}=rank_1(T, p) \cdot k^2$ of $T$. If the children of a node return a position $p_{children} > |T|$, the actual values of the cells are retrieved by accessing $L[p_{children}-|T|]$. Similarly, the parent of position $p$ in $T:L$ is $q-(q \!\!\mod k^2)$, where $q = select_1(T,\lfloor p/k^2\rfloor)$, and $q \!\!\mod k^2$ indicates the submatrix of $p$ within that of its parent.

\subsubsection{Re-Pair}

\begin{figure}[t]
\begin{center}
\includegraphics[scale=0.33]{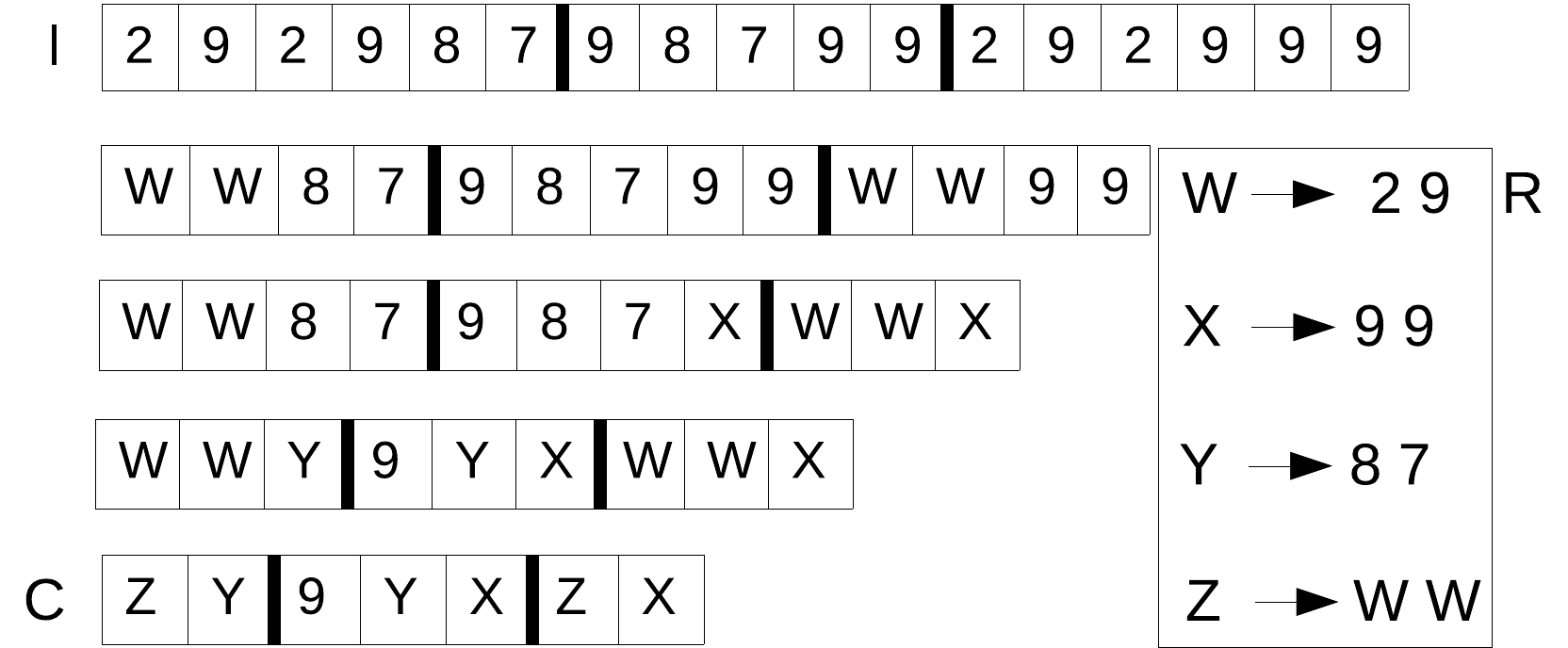}
\caption{Example of Re-Pair compression.}\label{re-pair}
\end{center}
\end{figure}

Re-Pair~\cite{larsson2000off} is a grammar-based compression method. Given a sequence of integers $I$ (called {\em terminals}), the method proceeds as follows: (1) it obtains the most frequent pair of integers $ab$ in $I$; (2) it adds rule $s \rightarrow ab$ to a dictionary $R$, where $s$ is a new symbol not present in $I$ (called a {\em non-terminal}); (3) every occurrence of $ab$ in $I$ is replaced by $s$, and (4) steps 1-3 are repeated until all pairs in $I$ appear only once (see Figure \ref{re-pair}). The resulting sequence after compressing $I$ is called $C$. Every symbol in $C$ represents a phrase (a sequence of one or more of the integers in $I$). If the length of the represented phrase is 1, then the phrase consists of an original (terminal) symbol; otherwise, the phrase is represented by a new (non-terminal) symbol. Re-Pair can be implemented in linear time and a phrase may be recursively expanded in optimal time (i.e., proportional to its length).

\subsubsection{DACs}
Directly Addressable Codes (DACs) \cite{BLN13} are a variable-length encoding scheme for sequences of integers, which support fast direct access to any given position in a sequence. In other words, they allow the $i^{th}$ integer to be decoded without the need to decompress the preceding integers. If the sequence of integers has many small numbers and few large ones, then DACs obtain a very compact representation.

Given a sequence of integers $X$ = $x_1, x_2, \dots, x_n$, DACs take the binary representation of that sequence and rearrange it into a level-shaped structure as follows: the first level, $B_1$, contains the first (least significant) $n_1$ bits of the binary representation of each integer. A bitmap $C_1$ is added to indicate whether the binary representation of each integer requires more than $n_1$ bits (1) or not (0). In the second level, $B_2$ stores the next $n_2$ bits of the integers with a value of 1 in $B_1$. A bitmap $C_2$ marks the integers that need more than $n_1+n_2$ bits, and so on. This process is repeated for as many levels as needed. The number of levels $ \ell$ and the number $n_l$ of bits at each level $l$, with $1\le l \le \ell$, is calculated in order to maximize compression. Each value $x_i$ is then retrieved using less than $\ell$ $rank$ operations on the bitmaps $C_l$ and extracting chunks from the arrays $B_l$.

\subsubsection{Permutations} \label{Sectperm}

A permutation of the integers $[1..n]=\{1, \ldots, n\}$ is a reordering of the values $\{1, \ldots, n\}$. The two main operations over the reordered sequence are $\pi(i)$ and $\pi^{-1}(j)$. The first yields the number at position $i \in [1..n]$ in the sequence, while the second identifies the position in the sequence where $j \in [1..n]$ appears.

The simplest way to store a permutation is an array $P[1..n]$, where $P[i] = \pi(i)$, so that $\pi(i)$ is answered in constant time. In order to answer $\pi^{-1}(j)$ efficiently, we can double the space by storing a second array. We instead use an intermediate option that uses $(1+\epsilon)n$ instead of $2n$ cells and answers $\pi^{-1}(j)$ queries in time $O(1/\epsilon)$ \cite{munro2012succinct}.
 
\section{The GraCT Index}\label{sect:gract}
 
\subsection{Discretization}
 
GraCT represents moving objects that follow free trajectories in space. The method assumes that the positions of all the objects are synchronized and stored at regular time instants (e.g., every minute). Since GPS devices may report the positions of objects at times that do not match the required time instants, the positions in the desired time instants are obtained from the raw timestamped positions using interpolation.

A raster model is used to represent the space, which is divided into cells of a fixed size, and objects are assumed to fit in one cell. The size of the cells and the length of the period between represented time instants are parameters that can be adapted to the specific domain. The shorter the length of the period is and the smaller the size of the cell is, the more accurate the trajectory data will be, though achieving less compression. This is a tradeoff between space usage and precision in the storage. 

Alternatively, we may regard this discretization as a way to obtain a more compact in-memory index, while the full data is maintained on disk and accessed only to filter out false-positives from a usually small set of candidate answers. We will also experiment with this variant of the index, showing that this is better than directly operating on the full data on disk.

\subsection{Snapshots}

Every $d$ time instants, GraCT uses a data structure based on $k^2$-trees to represent the absolute positions of all the objects. These data structures are called \textit{snapshots}. The distance $d$ between snapshots is another parameter of the system, which trades space for query time. Between two consecutive snapshots, the trajectory of each moving object is represented as a \textit{log}, which is an array of movements represented in differential form. 

Let ${\cal S}_{h}$ denote the snapshot representing the position of all the objects at time instant $k$. The components of ${\cal S}_{h}$ are:

\begin{itemize}
\item The \textit{time instant} represented by the snapshot. 
\item A $k^2$-tree storing the positions in the space (i.e., the cells of the raster) where there are objects. 
\item An array of integers {\em perm} storing the identifiers of the objects at each cell in the space where there are objects.
\item A bitmap \textit{Q}, which is an auxiliary data structure of \textit{perm} that is used to determine the correspondence between positions in the space and the positions of \textit{perm}.

\end{itemize}

The $k^2$-tree represents a binary matrix where a cell set to 1 indicates that the cell contains one or more objects. However, we need to know what objects are in the cell set to 1. Each 1 in the binary matrix corresponds to a bit set to 1 in bitmap $L$ of the $k^2$-tree. The list of object identifiers corresponding to each of those bits set to 1 is stored in {\em perm}, where the object identifiers are sorted according to their order of appearance in $L$. This array turns out to be a permutation of all the object identifiers. Bitmap \textit{Q} is aligned with \textit{perm}; a $0$ at a given position of $Q$ means that the object identifier aligned in \textit{perm} is the last object in the cell, while a 1 signals that the next object identifier in {\em perm} is in the same cell.

\begin{figure}[t]
\begin{center}

\includegraphics[scale=0.17]{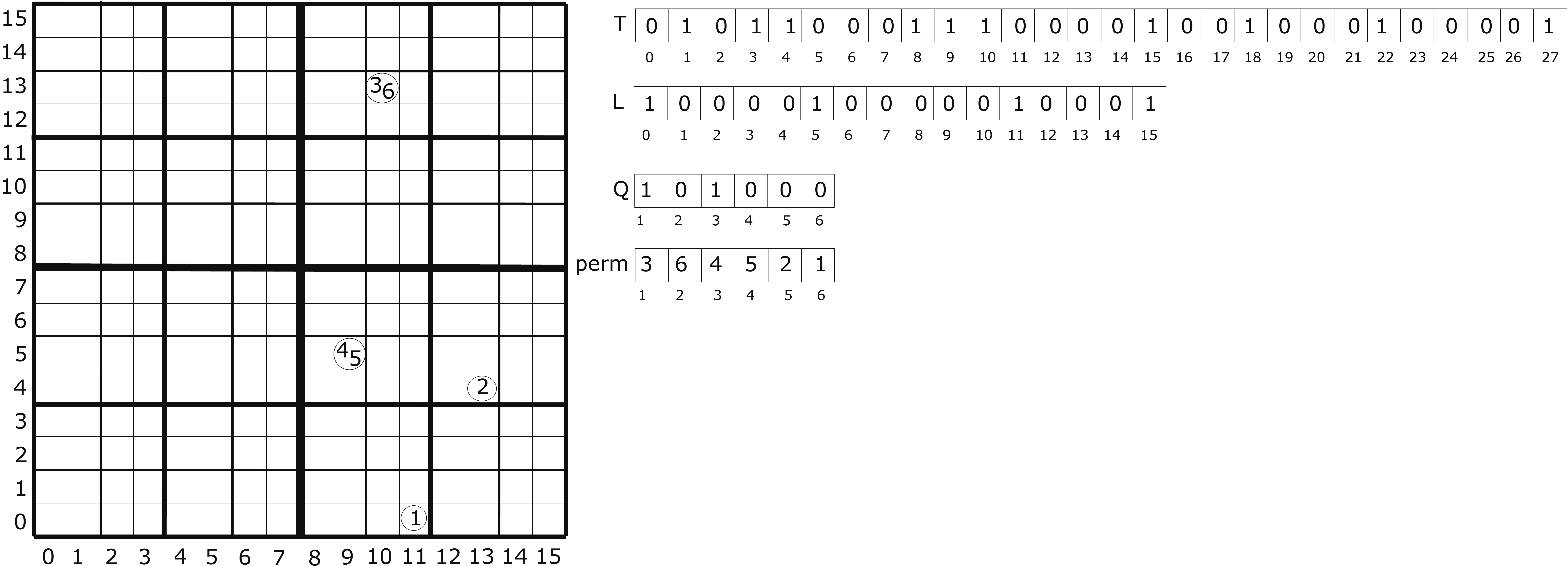}
\end{center}

\caption{Position of objects in the space (left) and the arrays (right) that make up a snapshot representing that situation.}\label{snapshot}
\end{figure}

Figure \ref{snapshot} shows an example. The left side shows the matrix representing the space and the objects placed in the cells (these cells are the same containing a 1 in  Figure \ref{fig:example}). The four arrays on the right are the actual data structures that represent the snapshot information. Arrays $T$ and $L$ are the same as those in Figure \ref{fig:example} and mark the positions having objects. The objects in those positions are stored in arrays $Q$ and {\em perm}. For instance, the object identifiers corresponding to the second 1 in $L$ (which is at position 5 in $L$ and corresponds to the cell at position (9,5)) are stored starting at position 3 in $perm$, since the first 0 (signaling the end of the entries corresponding to the first 1 in $L$) is at position 2. In order to find out how many objects there are in the corresponding cell, we access $Q$ starting at position 3 and search for the first 0 after that position, which is at position 4; this shows that there are two objects in the inspected cell. By accessing positions 3 and 4 in {\em perm}, we obtain the object identifiers 4 and 5. The object identifiers corresponding to the third 1 in $L$ start at position 5 in {\em perm}, and so on.

With these structures used to represent the absolute positions of all moving objects at the snapshots, we can answer two types of queries in time $O(\log_k s + output)$, where $s \times s$ is the grid size and $output$ is the size of the query output:

\begin{itemize}
\item \textit{Find the objects in a given cell}: First, we traverse the tree from the root until we reach position $n$ in $L$ corresponding to that cell. Next, we count the number of $1$s in the array of leaves $L$ until position $n$; this gives us the number of leaves with objects up to the $n$-th leaf, $x = rank_1(L, n)$. Then we calculate the position of the $(x-1)$-th $0$ in $Q$, which indicates the last bit of the previous leaf with objects, and we add $1$ to obtain the first position in our leaf. That is, $p = select_0(Q, x-1)+1$ is the position in $perm$ of the first object identifier corresponding to the searched position. From $p$, we read all the object identifiers aligned with 1s in $Q$ until we reach a 0, which signals the last object identifier in that leaf.

\item \textit{Find the position of a given object in the space}. First, by using the $\pi^{-1}$ operation (see Section \ref{Sectperm}), we obtain the position $k$ in \textit{perm} of the searched object. Next, we have to find the leaf in $L$ corresponding to the $k$-th position of $perm$. For this, we calculate the number of leaves with objects before the object in position $k$ of {\em perm}; that is, we count the number of $0$s until the position before $k$, $y = rank_0(Q, k-1)$. We then find in $L$ the position of the $(y+1)$-th $1$; that is, $select_1(L, y+1)$. With that position in $L$, we can traverse the $k^2$-tree upwards in order to obtain the position in the space of that cell, and thus the position of the object.
\end{itemize}

\subsection{Log of relative movements}

The movements of objects between two consecutive snapshots are recorded in sequences of consecutive object positions called {\em logs}, with one separate sequence per object. Logs actually represent the positions in relative form with respect to the previous one, where the first absolute position can be obtained from the preceding snapshot. Therefore, logs can be thought of as the sequences of consecutive {\em movements} of the objects along time. As we will see soon, this model must be extended to consider cases where objects appear and disappear. 

\begin{figure}[t]
\begin{center}

\includegraphics[scale=0.5]{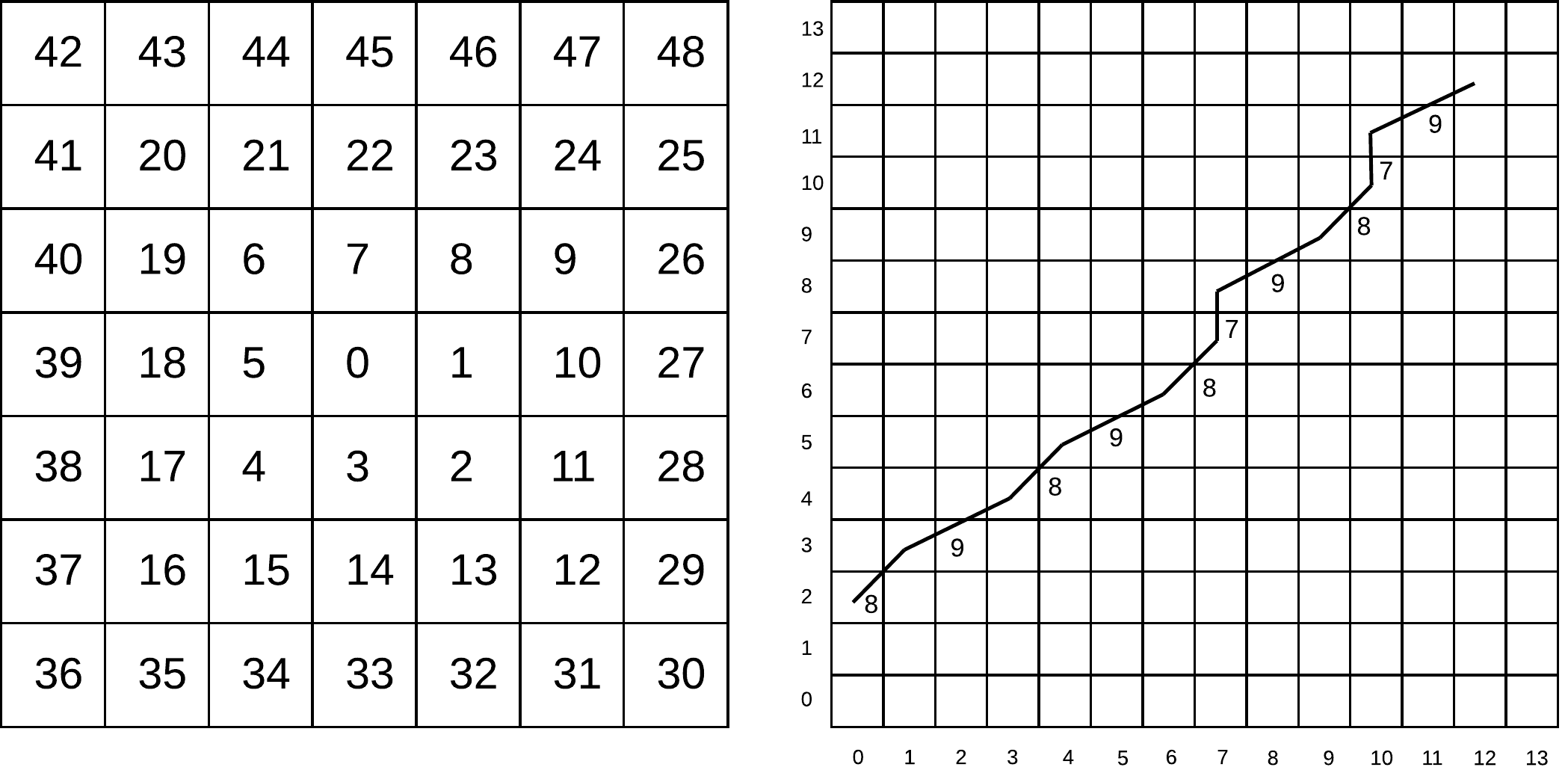}
\end{center}

\caption{Encoding movements.}\label{espiral}
\end{figure}

Objects change their positions along the two Cartesian axes, so every movement in the log can be described with two integers. In order to save space, our method encodes the two values using a single positive integer. The cells around the actual position of an object are enumerated following a spiral in which the origin is the initial position of the object, as shown in Figure \ref{espiral} (left). As an example, assume that an object moves one cell to the East and one cell to the North with respect to the previous known position. Instead of encoding the movement as the pair (1,1), it is encoded as an 8. Figure \ref{espiral} (right) shows the trajectory of an object starting at cell (0,2). Each number indicates a movement between two consecutive time instants. Since most relative movements involve short distances, this technique usually produces a sequence of small numbers.

Between two consecutive snapshots ${\cal S}_{h}$ and ${\cal S}_{h+d}$, each object is represented by a log, ${\cal L}_{h,h+d}(id)$, where $id$ is the identifier of the object.

\begin{figure}[t]
\begin{center}

\includegraphics[scale=0.16]{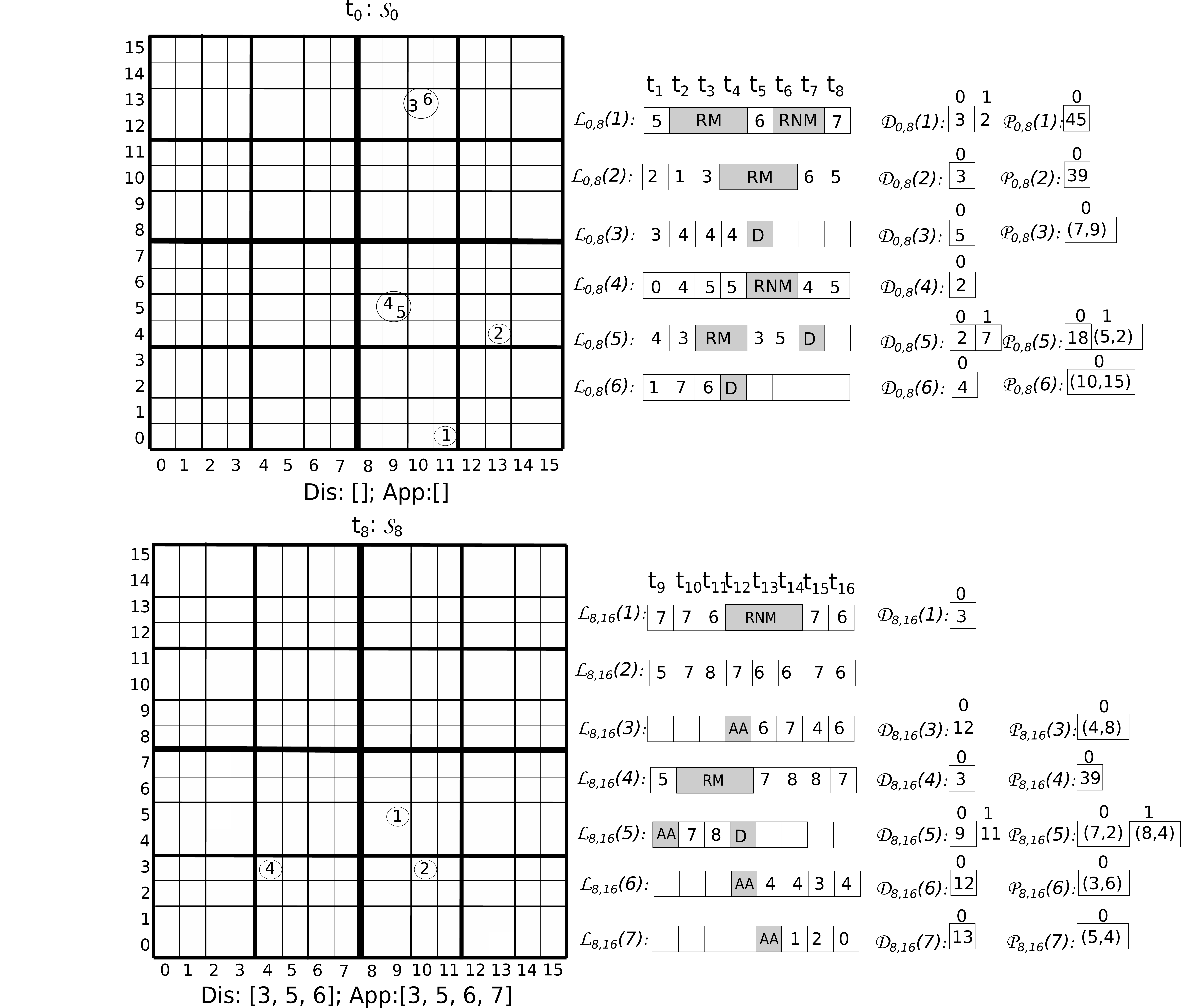}
\end{center}
\caption{Snapshots and logs.}\label{components}
\end{figure}

Figure \ref{components} presents a situation in which there is a snapshot every eight time instants. Time instants $t_0$ and $t_8$ are represented with snapshots, while the time instants in the intervals $[t_1,t_8]$ and $[t_9,t_{16}]$ are represented with logs. Note that, although $t_8$ is represented with a snapshot, it is also included in the preceding logs. This duplicity allows us to traverse the logs in both directions, which accelerates certain queries, as seen later. The meaning of the other components of the figure is also explained in the sequel.

\subsubsection{Managing appearances and disappearances} \label{disapp}

Because the model does not explicitly store the time instant of each recorded position, a mechanism is required to signal which objects are active at each time instant and to determine to which time instant each log entry corresponds.

For this purpose, the log manages three events: when an object starts emitting its positions, when it stops, and when it stops emitting signals for periods of time. In fact, the three cases are homogeneously managed by the following events, each identified in the log with a special codeword:

\begin{itemize}
\item {\em Disappearances} ($D$). The occurrence of codeword $D$ in ${\cal L}_{h,h+d}(id)$ means that object $id$ stopped emitting its position at the corresponding time instant of that portion of the log, and that $id$ did not restart emitting (at least) until the time instant of the next snapshot (${\cal S}_{h+d}$).

To enable backward traversal of the logs, each disappearance must store the time instant and the absolute coordinates of the object at the moment where it disappears.

\item {\em Absolute appearance} ($AA$). The occurrence of codeword $AA$ in ${\cal L}_{h,h+d}(id)$ means that $id$ started to emit its position in the corresponding  time instant of that portion of the log, and that $id$ had not emitted its positions  (at least) since the previous snapshot (${\cal S}_h$). \textit{AA} may signal the first appearance of an object or its reappearance after having disappeared in a preceding log.

To enable forward traversal of the logs, an absolute appearance stores the time instant and the absolute coordinates where the object appears. Note that, even if the object is reappearing, it might do so far away from the position where it had disappeared in a previous log.

\item {\em Relative disappearance}. The occurrence of a codeword indicating a relative disappearance in ${\cal L}_{h,h+d}(id)$ in an entry corresponding to time instant $t_i$ means that $id$ stopped emitting positions at that time instant, and that $id$ restarted emitting its positions in a time instant between $t_i$ and the time instant of the next snapshot (${\cal S}_{h+d}$). In other words, a relative disappearance occurs when the object disappears and reappears within the same log.

A relative disappearance must store the number of time instants in which $id$ was not emitting its positions and, depending of the type of relative disappearance, also a movement:

\begin{itemize}
\item {\em Relative disappearance with movement} ($RM$) means that $id$ appears in a different position with respect to its last known position. This requires the log to store the relative movement (using the spiral encoding) with respect to that last known position.

\item {\em Relative disappearance without movement} ($RNM$) means that $id$ appears in the same position where the signal was lost, therefore there is no need to store a movement.

\end{itemize}
\end{itemize}

The extra information is stored in two auxiliary arrays associated with each log ${\cal L}_{h,h+d}(id)$:

\begin{itemize}
\item Array ${\cal D}_{h,h+d}(id)$ stores:
\begin{itemize}
\item in the case of relative disappearances, the time the disappearance lasted;
\item in the case of an absolute appearance or a disappearance, the absolute time instant of that event.
\end{itemize}
\item Array ${\cal P}_{h,h+d}(id)$ stores:
\begin{itemize}
\item in the case of relative disappearances with movement, the relative movement with respect to the last known position;
\item in the case of an absolute appearance or a disappearance, the absolute position of that event.
\end{itemize}
\end{itemize}

Figure \ref{components} shows two snapshots; ${\cal S}_0$ is the snapshot of Figure \ref{snapshot}. The right-hand side shows the logs representing the movements of objects in the time instants after those snapshots, thus representing the positions of objects between time instants $t_0$ and $t_{16}$. Object 1 disappears at $t_2$ and then reappears at $t_5$ (see ${\cal L}_{0,8}(1)$). Since the time instants occur between the same pair of snapshots, this is a relative disappearance. The number of time instants that the disappearance lasts is stored in array ${\cal D}_{0,8}(1)$. In the first position of ${\cal D}_{0,8}(1)$ there is a 3, which means that the first disappearance of ${\cal L}_{0,8}(1)$ lasted three time instants. In the same portion of the log, there is another relative disappearance between time instants $t_6$ and $t_7$, therefore in the second entry of array ${\cal D}_{0,8}(1)$ there is a 2, indicating that the disappearance lasted two time instants. The relative disappearance that starts at $t_2$ is of type $RM$, so the log must store the position where object 1 reappears. Array ${\cal P}_{0,8}(1)$ stores this position using spiral encoding with respect to the last known position. In the example, a 45 means that object 1 reappeared three cells to the North.

In the entry in ${\cal L}_{0,8}(3)$ corresponding to time instant $t_5$, codeword $D$ means that object 3 disappears at that time instant and does not reappear in ${\cal L}_{0,8}(3)$. Therefore, ${\cal D}_{0,8}(3)$ stores a 5 in its first position (there are no prior relative reappearances), indicating that the object disappeared at $t_5$, and in ${\cal P}_{0,8}(3)$, the only entry stores the coordinates (7,9), which denote that the object was in that position when it disappeared. Object 3 reappears in ${\cal L}_{9,16}(3)$, in the entry corresponding to $t_{12}$. This reappearance is signaled with codeword $AA$ aligned with time instant $t_{12}$, but this is only to improve the readability of the figure. In fact, codeword $AA$ is the first entry in ${\cal L}_{9,16}(3)$, so, in order to find out to which time instant codeword $AA$ corresponds, ${\cal D}_{9,16}(3)$ stores a 12. Additionally, in order to find out the position of object 3 at $t_{12}$, ${\cal P}_{9,16}(3)$ stores the absolute coordinates of the object at the time instant of reappearance, which in our example are $(4,8)$.

Finally, to help manage disappearances and appearances during queries, two additional arrays are stored with each snapshot ${\cal S}_h$:

\begin{itemize}
\item $Dis$, a list of the objects that were active in the previous snapshot and stopped emitting before the time instant represented by ${\cal S}_h$. 
\item $App$, a list of objects that are not present in ${\cal S}_h$ and appear (or reappear) before the next snapshot.
\end{itemize}

Figure \ref{components} shows that ${\cal S}_8.Dis$ contains objects 3, 5 and 6, meaning that they were active in their portions of the log immediately before ${\cal S}_8$, but are missing at time instant $t_8$. Similarly, ${\cal S}_8.App$ contains objects 3, 5, 6 and 7, indicating that they are missing at $t_8$, but appear in the portion of the log immediately following $t_8$.

\section{Compressing the Log} \label{sect:comp}

\begin{figure}[t]
\begin{center}

\includegraphics[scale=0.55]{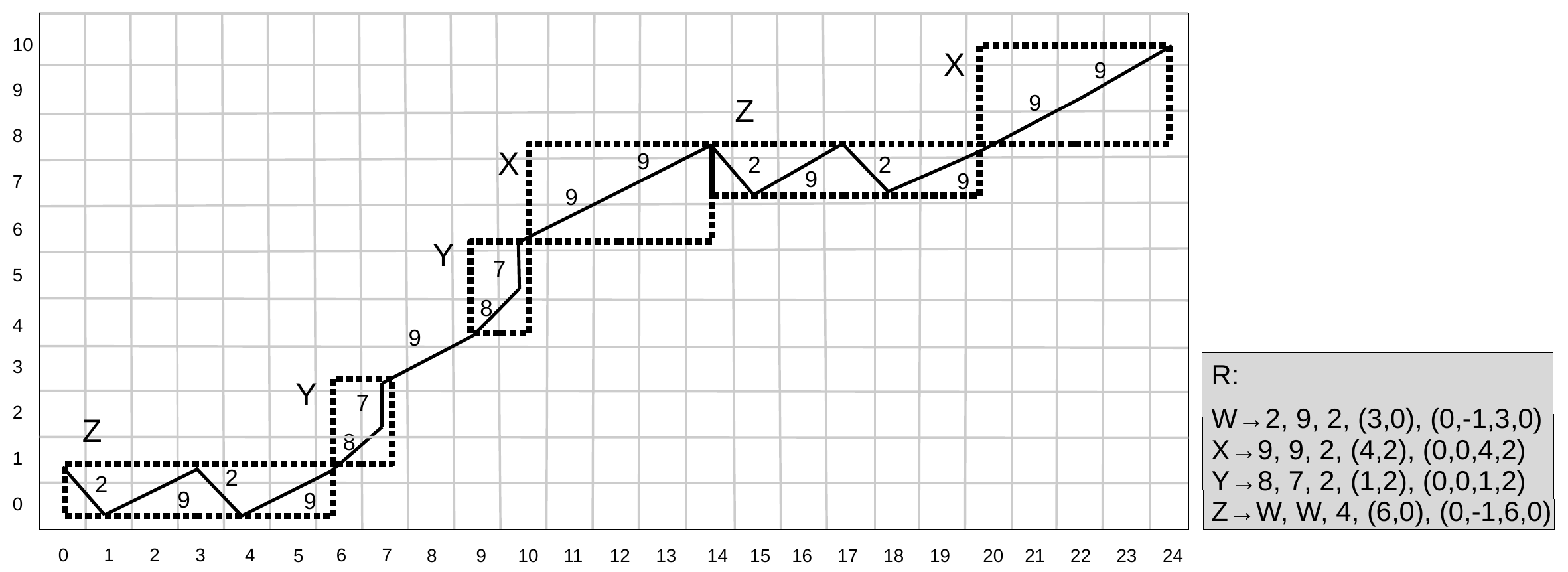}
\end{center}

\caption{Movements of $C$ array in Figure \ref{re-pair}, MBRs of the rules, and enriched rules. }\label{ruta}
\end{figure}

In many applications, objects spend most of their time either stopped or moving along a specific course at a fixed speed. This generates long sections of the log with numbers representing the same or contiguous values of the spiral. For example, the moving object in Figure \ref{ruta} follows a NE trajectory, moving one or two cells in the time elapsed between two consecutive time instants. Its log represents the series of relative movements 2,9,2,9,8,7,9,8,7,9,9,2,9,2,9,9,9 (see array $I$ in Figure \ref{re-pair}). These series of similar movements are compressed very efficiently using a grammar compressor.

In order to query trajectories without completely decompressing them, the grammar rules in GraCT not only include the symbols to be replaced, but are also enriched with additional information. Specifically, each rule in $R$ will have the following information: $s \rightarrow a, b, \#t, x, y, MBR$, where:

\begin{itemize}
\item $s \rightarrow ab$ is the normal rule of Re-Pair,
\item $\#t$ is the number of time instants covered by the rule,
\item $(x,y)$ are the relative coordinates  of the final position of the object after the application of the rule (that is, the total displacement in both coordinates), and
\item $MBR$ is the minimum bounding rectangle enclosing the movements of the rule. MBR is represented as $(x1, y1, x2, y2)$, where $(x1, y1)$ corresponds to the bottom-left corner and $(x2, y2)$ to the top-right corner.
\end{itemize}

For example, the rules in Figure \ref{re-pair} are enriched as follows. The first rule of $R$ is $W \rightarrow 2,~9, ~2,~ (3,0), ~(0,-1,3,0)$: $2$ and $9$ are the substituted symbols; the next $2$ indicates that the rule represents a sequence of two movements; $(3,0)$ indicates the position of the object after the application of the rule if we start at $(0,0)$, and the last four values are two corners (bottom-left and top-right) defining a rectangle that encloses all the movements encoded by the rule.

With this additional information, most of the non-terminal symbols of a compressed log file $\cal L$ do not need to be decompressed in order to obtain the position of an object at a given time instant between two snapshots. Consider, for example, the compressed log ${\cal L} = Z,Y,9,Y,X,Z,X$ of Figure \ref{ruta} (this is the array $C$ of Figure \ref{re-pair}).
Assume that we want to find the position of the object at the 5th time instant, that is, $(7,2)$. From the preceding snapshot we obtain that the absolute position of the object at the beginning of the log is (0,1). Next, we inspect the log $\cal L$ from the beginning. The first value is a $Z$. The enriched rule indicates that this symbol represents four time instants, after which the object is displaced six columns to the East while remaining in the same row. Since it started at $(0,1)$ the object will be at position $(6,1)$ at $t_4$. Since our target time instant is $t_5 > t_4$, we can skip the whole nonterminal $Z$. The next symbol is $Y$, which lasts two time instants. This would take us to $t_6$, which surpasses our target time instant. Therefore, and only in the last step of the search, we have to decompress the rule and process its components: $Y \rightarrow 8~7$. The $8$ is a terminal symbol that lasts one time instant, and thus it is enough to reach our target time instant. An 8 moves one column to the East and one to the North, which applied to the current position $(6,1)$, takes us to the correct position $(7,2)$.

The $MBR$ component also helps speed up other queries, as seen later.

\begin{figure}[t]
 \begin{center}   
 \includegraphics[scale=0.16]{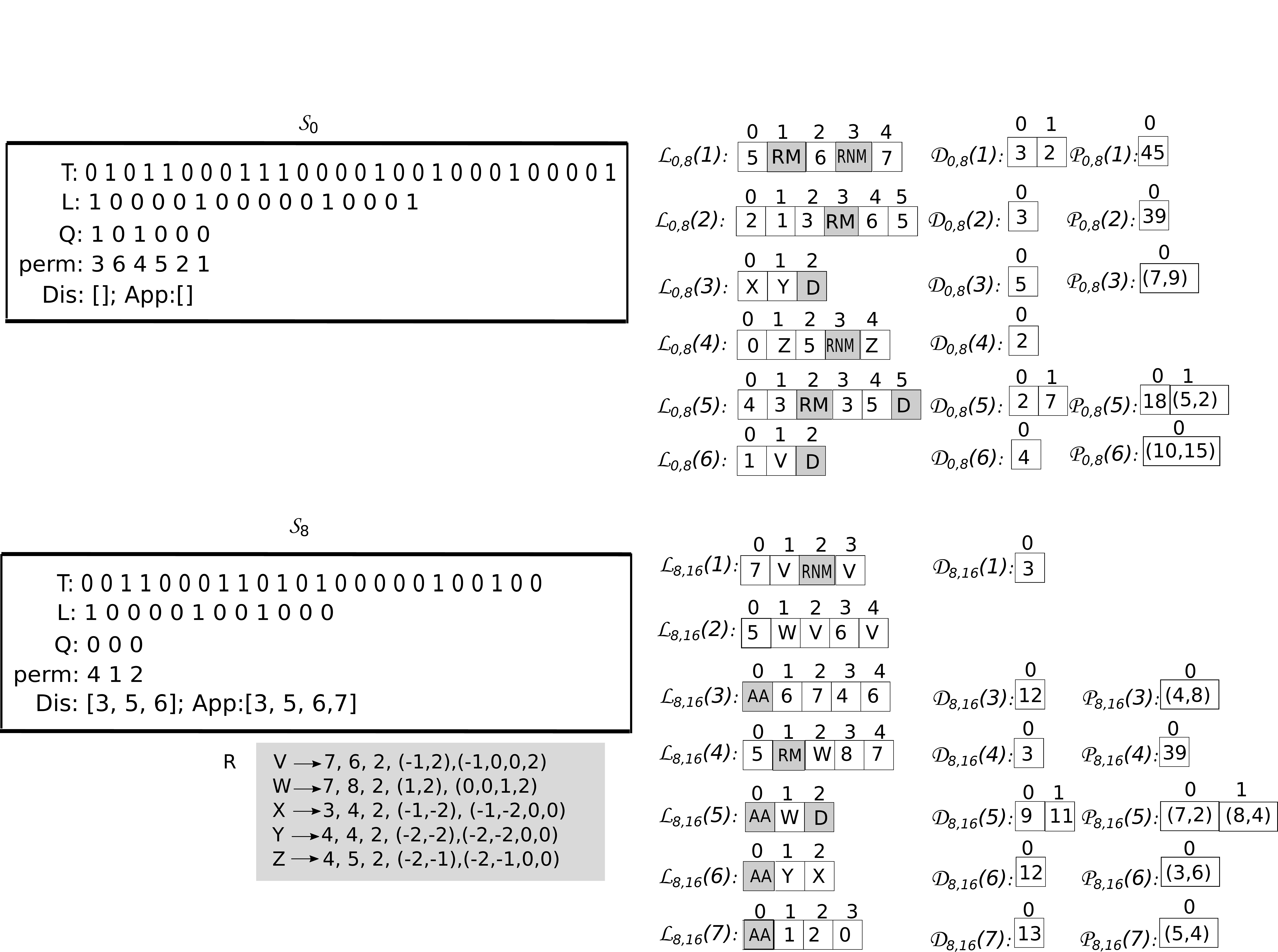}
 \end{center}

 \caption{GraCT structures.}\label{graCT}
 \end{figure}

These additional elements that enrich the rules are compressed with DACs. To obtain better compression, the times of all of the rules are compressed using one DAC, separately from the three pairs of coordinates of all of the rules, which are compressed using a separate DAC.
Figure \ref{graCT} shows the resulting data structures that make up the GraCT index for our running example in Figure \ref{components}.

\begin{figure}[t]
 \begin{center}   
 \includegraphics[scale=0.45]{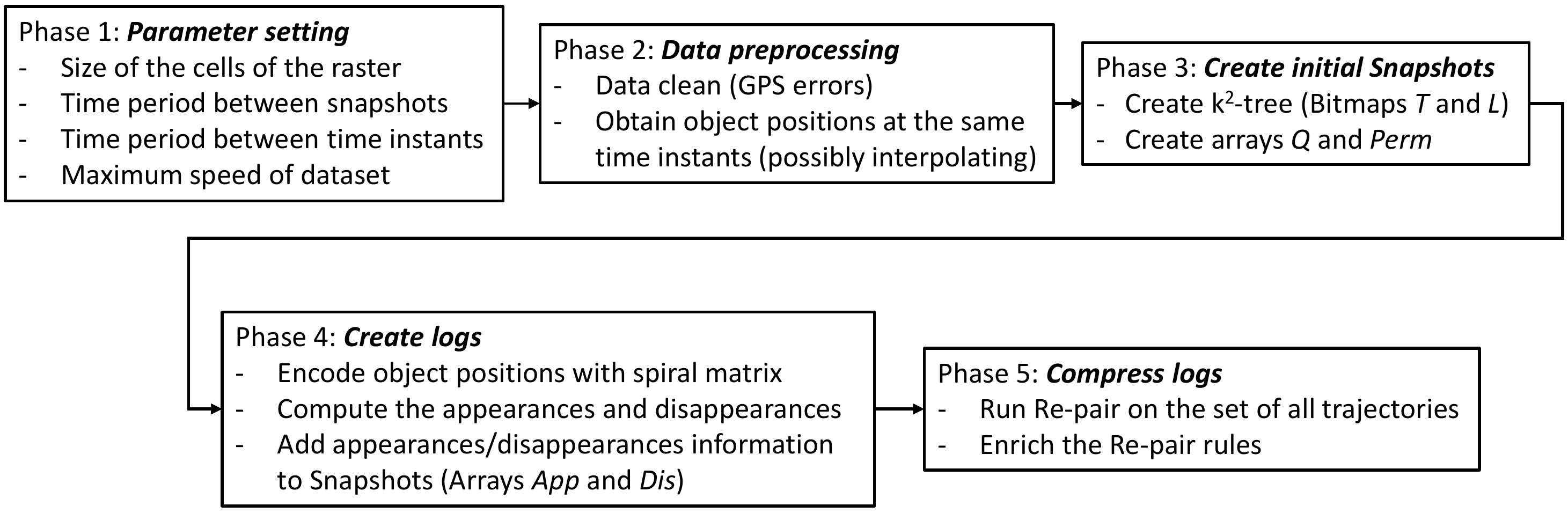}
 \end{center}

 \caption{Flow chart of the process to create a GraCT index.}\label{flowchart}
 \end{figure}

Figure \ref{flowchart} shows a flow chart that summarizes the whole process to create a GraCT data structure.

\section{Query Preliminaries} \label{sect:quep}

This section presents a series of definitions for later use.

\begin{definition}
Let $t_b$ and $t_e$ be two time instants where $t_b \leq t_e$. Then $CT_O(t_b, t_e) = (m_{1}, m_{2}, ..., m_{j})$ represents the compressed trajectory (that is, the movements) of object $O$ between time instants $t_b$ and $t_e$.
Each $m_i$ can be:
\begin{itemize}

\item ${\cal S}_{(x,y)}$, an absolute position $(x,y)$ obtained from a snapshot.

\item  NT/T, a grammar symbol encoding the movements of $O$, where NT denotes a non-terminal and T denotes a terminal.

\item $D_{t,(x,y)}$, a disappearance code, where $t$ is the time instant of the disappearance and $(x,y)$ is the object position at that time instant.
\item $RNM_{\#}$, a reappearance without movement code, where $\#$ is the duration of the disappearance.
\item $RM_{\#,value}$, a reappearance with movement code, where $\#$ is the duration of the disappearance and {\em value} is the spiral code representing the movement.
\item $AA_{t,(x,y)}$, an absolute appearance code, where $t$ is the time instant of the appearance and $(x,y)$ is the object position at that time instant.
\end{itemize}
\end{definition}

If object $O$ is missing at $t_b$, then $m_{1}$ will refer to the first time instant after $t_b$ containing information about the position of $O$. In the same way, if the object is missing at $t_e$, then $m_{j}$ will be the movement of $O$ corresponding to the closest time instant with information about the position of $O$ prior to $t_e$.

The sequence
$CT_O(t_b, t_e)$ is obtained by extracting the portions of the logs and the positions at the snapshots of $O$ covering the time interval $[t_b,t_e]$.

For example, in our running example in Figures \ref{components} and \ref{graCT}, for object $O_6$, the compressed trajectory between 0 and 13 is $CT_{O_6}(0, 13) =$ (${\cal S}_{(10, 13)}, 1, V, D_{4,(10,15)}, AA_{12,(3,6)}, Y)$. Note that the last code ($Y$) covers time instants 13 and 14, since it is a non-terminal symbol. In contrast, in $CT_{O_5}(11,15)=(AA_{11,(7,2)},W,D_{13,(8,4)})$. The last code ($D_{13,(8,4)}$) corresponds to time instant 13, since there are no positions after that time instant in the requested time interval.

\begin{definition}
If $m_i$ is an element of $CT_O(t_b, t_e)$,  $p_c$ is the absolute position of object $O$ at time instant $t_c$, and $t_b \leq t_c \leq t_e$, then operation $moveJ(p_c,t_c, t_e,m_i)$ returns either:

\begin{itemize}

\item $\langle t_{nc}, p_{nc}\rangle$, where $p_{nc}$ is the position resulting from the application of the movements encoded in $m_i$ to $p_c$, and $t_{nc} \le t_e$ is the time instant of that position, or

\item $\langle t_e, p_e\rangle$, where $p_e$ is the position at $t_e$, if $m_i$ encodes movements surpassing $t_e$.

\end{itemize}

\end{definition}

The name $moveJ$ is the abbreviation of ``\textit{move jump}". If $m_i$ is a non-terminal code, it uses $m_i.(x,y)$, the relative coordinates of the enriched information included in the definition of the non-terminal symbol $m_i$, which, when applied to the previous position, gives the new position of the object after the application of  all the movements encoded by $m_i$. For example, $moveJ((9,5),t_1,t_{3},Z)$ uses the relative $(x,y)$ coordinates of the definition of $Z\rightarrow 4,5,2,(-2,-1),(-2,-1,0,0)$, that is $(-2,-1)$, which, when applied to $(9,5)$, obtain the new position $(7,4)$. Note that the process never obtains the position at $t_2$ but jumps from $\langle t_1,(9,5)\rangle$ to $ \langle t_3, (7,4)\rangle$.

Note also that if we issue $moveJ((9,5),t_1,t_{2},Z)$, where the application of $Z$ would produce the position at $t_3 > t_2 = t_e$, the operation has to decompress $Z$ into its components (4 and 5) and apply them one by one until it reaches $t_e$. In this case, therefore, only code $4$ is applied, to move the object to $(8,4)$. The decompression of non-terminal codes in \textit{moveJ} only occurs in this situation. 

In this example we fully decompress the nonterminal that exceeds $t_e$, but in general we continue advancing recursively on the terminals and nonterminals of the right-hand side of its rule. That is, if the application of the next nonterminal $Z$ would exceed $t_e$, and the rule associated with $Z$ is $Z \rightarrow Z_1 Z_2$, then we replace $Z$ by $Z_1 Z_2$ and try again. This means that, if applying $Z_1$ also exceeds $t_e$ we will continue recursively with $Z_1$, and otherwise we will process the whole $Z_1$ in constant time and continue recursively with $Z_2$.
Therefore, if we are indexing $t$ time instants in total and use a balanced grammar (i.e., of height $O(\log t)$), then \textit{moveJ} can be implemented in time $O(1)$ if $m_i$ does not exceed $t_e$ and $O(\log t)$ if it does.

\begin{definition}
If $m_i$ is an element of $CT_O(t_b, t_e)$, $p_c$ is the absolute position of object $O$ at time instant $t_c$, and $t_b \le t_c \le t_e$, then $RmoveJ(p_c,t_b, t_c,m_i)$ returns either:
	
\begin{itemize}
		
\item $\langle t_{nc}, p_{nc}\rangle$, where $p_{nc}$ is the position resulting from undoing the movements encoded in $m_i$ to $p_{c}$, and $t_{nc}$ is the time instant of that position, or
		
\item $\langle t_b, p_b\rangle$, where $p_b$ is the position at $t_b$, if $m_i$ encodes movements that obtain a time instant earlier than $t_b$.

\end{itemize}

\end{definition}

This is a reverse version of $moveJ$, and can be implemented within the same time complexity.

  \begin{table}[t]
\setlength{\tabcolsep}{1pt}
 \begin{center}
 \scriptsize
  \begin{tabular}{l p{9.5cm}}

   \hline

  $CT_O(t_b, t_e) $  &  compressed trajectory of $O$ between time instants $t_b$ and $t_e$\\
   $moveJ(p_c,t_c,t_e, m_i)$& returns the pair $\langle t_{nc}, p_{nc}\rangle$ resulting from applying the movements in $m_i$ to position $p_c$, but if that application produces a position surpassing $t_e$, then the operation returns $\langle t_{e}, p_{e}\rangle$.\\
   $RmoveJ(p_c,t_b,t_c, m_i)$& returns the pair $\langle t_{nc}, p_{nc}\rangle$ resulting from undoing the movements in $m_i$ to position $p_c$, but if that application produces a position previous to $t_b$, then the operation returns $\langle t_{b}, p_{b}\rangle$.\\ 
   
   $moveS(p_c,t_c,t_e, m_i)$& returns $( \langle(t_{c+1},p_{c+1}\rangle,\ldots, \langle t_h,p_h\rangle)$, the list of time instants/positions resulting from applying $m_i$ to $p_c$, except if $m_i$ encodes movements that surpass $t_e$, in which case $h=e$.\\

    $ER_r(t_b, t_e)$ &  expansion of region $r$ from $t_b$ to $t_e$ using the maximum speed\\
 $ER_{O}(t_b, t_e)$  &  expanded region of object $O$ from $t_b$ to $t_e$ using the maximum speed\\
     $\overline{p_i p_j}$ &  distance in cells of the shortest trajectory from point $p_i$ to point $p_j$\\
   $\overline{p r}$ &  distance in cells of the shortest trajectory from point $p$ to region $r$\\
   \hline
  \end{tabular}
 \end{center}
 \vspace{-5mm}
 \caption{Summary of definitions.}
 \label{table:annotations}
 \end{table}

\begin{definition}
If $m_i$ is an element of $CT_O(t_b, t_e)$,  $p_c$ is the absolute position of object $O$ at time instant $t_c$, and $t_b \leq t_c \leq t_e$, then $moveS(p_c,t_c, t_e,m_i)$ returns the list of pairs $(\langle t_{c+1},p_{c+1}\rangle,\ldots, \langle t_h,p_h\rangle)$ where $t_h \leq t_e$, which is the list of time instants/positions resulting from applying $m_i$ to $p_c$. If $m_i$ encodes movements that surpass $t_e$, however, the decompression ends at $t_e$.

\end{definition}

The name $moveS$ is the abbreviation of ``\textit{move step by step}". Unlike $moveJ$, if $m_i$ is a non-terminal code, $moveS$ decompresses the symbol until it obtains terminal codes, then applies those terminal symbols to $p_c$ and returns the positions of each symbol.
 
For example, $moveS((9,5),t_1,t_{3},Z)$ decompresses $Z$ to obtain the terminal codes $4$ and $5$, and thus returns $\langle t_{2},(8,4)\rangle, \langle  t_{3},(7,4)\rangle$. In contrast to $moveJ$, $moveS$ does obtain the position at $t_2$. This operation can be implemented in time $O(h-c+\log t)$ on a balanced grammar.

\begin{definition}\label{def:er}
Let $r = [x_1,y_1] \times [x_2,y_2]$ be a rectangular region in the two-dimensional space. Let $t_b$ and $t_e$ be two time instants, where $t_b < t_e$, and let $M_{sp}$ be the maximum speed of the dataset. Then $ER_r( t_b, t_e)$ denotes the expanded region of $r$ from $t_b$ to $t_e$ using $M_{sp}$:  $ER_r(t_b, t_e) = [x_1-\delta, y_1-\delta] \times [x_2+\delta,y_2+\delta]$,
where $\delta = M_{sp}\cdot(t_e-t_b)$. 
\end{definition}

The maximum speed of the dataset is the maximum speed of any object during the period of time covered by the dataset.
$ER_{r}(t_b, t_e)$ is the rectangle resulting from expanding $r$ in all directions (North, South, West and East) so that any object within $r$ at $t_b$ is bound to be within $ER_{r}(t_b, t_e)$ at $t_e$. In other words, to obtain $ER_{r}(t_b, t_e)$, $r$ is enlarged by $M_{sp} \cdot (t_e-t_b)$ cells in all four directions.

\begin{definition}
Let $t_b < t_e$ be two time instants, and let $M_{sp}$ be the maximum speed of the dataset. Then $ER_{O}(t_b, t_e)$ denotes the expanded region of object $O$ from $t_b$ to $t_e$ using the maximum speed $M_{sp}$. That is, if $O$ is at cell $(x,y)$, then $ER_{O}(t_b, t_e)=ER_{r_{(x,y)}}(t_b, t_e)$, where $r_{(x,y)}= [x,y] \times [x,y]$.
\end{definition}

That is, $ER_{O}(t_b, t_e)$ delimits the area where object $O$ is bounded to belong at $t_e$ even if running at maximum speed $M_{sp}$ in any direction, starting at time instant $t_b$.

\begin{definition}
Given a point $p_i$ and a point $p_j$, the {\em distance} between $p_i$ and $p_j$ ($\overline{p_i p_j}$) is the length of the shortest trajectory (in number of cells) from $p_i$ to $p_j$, using Euclidean distance.
\end{definition}

\begin{definition}
Given a point $p$ and a region $r$, the {\em distance} between $r$ and $p$ ($\overline{p r}$) is the length of the shortest trajectory (in number of cells) between $r = [x_1,x_2] \times [y_1,y_2]$ and $p=(x,y)$. This is zero if $x_1 \le x \le x_2$ and $y_1 \le y \le y_2$. Otherwise, it is $\min(|x-x_1|,|x-x_2|)$ if $y_1 \le y \le y_2$ and $\min(|y-y_1|,|y-y_2|)$ if $x_1 \le x \le x_2$. Otherwise, it is the minimum Euclidean distance from $p$ to the four corners of $r$.
\end{definition}

Table \ref{table:annotations} summarizes these definitions.

\section{Supporting Queries}\label{query}

We now describe how each of the supported queries are answered using the structures of GraCT.

\subsection{Obtain the position of an object in a given time instant}

Algorithm \ref{alg:searchObject} shows the pseudocode to answer this query. The algorithm first identifies a time instant close to $t_q$ where the position of the object is represented in absolute form and, from that time instant, applies the relative movements in order to obtain the absolute position of the object at time instant $t_q$.

The first option to find that time instant is the snapshot ${\cal S}_h$ that is nearest to the queried time instant (line 1 of the algorithm). If $t_q$ coincides with the time instant of ${\cal S}_{h}$, then the position of the object in the snapshot is returned (line 28; the position could be $null$ if the object is not present in ${\cal S}_h$).
The other possible cases are handled in lines 4-15 (when $t_q$ follows the snapshot) and 16-27 (when $t_q$ precedes the snapshot).

\begin{algorithm}[t!]
	\footnotesize	
	\caption{{\bf SearchObject}($O_{id}$, $t_{q}$)}
	\label{alg:searchObject}
	${\cal S}_{h} \gets nearestSnapshot(t_{q})$;\\
	$p_c \gets {\cal S}_{h}.findObject(O_{id})$; \tcp*[h]{Finds position $O_{id}$ in the closest snapshot}\\
	$t_c \gets {\cal S}_{h}.time$;\\
	\If(\tcp*[h]{Move forward from preceding snapshot}){$t_c < t_q$}{ 
		
		\If(\tcp*[h]{$O_{id}$ is not in the snapshot}){$p_c = null$ }
		{ 
			\If(\tcp*[h]{Appears in ${\cal L}_{h,h+d}(O_{id})$}){  $O_{id} \in {\cal S}_{h}.App$ }{
				$t_c \gets {\cal D}_{h,h+d}(O_{id})[0]$; \tcp*[h]{Obtain the time of appearance}\\
				\If  {$t_c \leq t_q$}{
					$p_c \gets {\cal P}_{h,h+d}(O_{id})[0]$; \tcp*[h]{Obtain the location of appearance}
					
				}
				\lElse{\Return \textit{null}}
				
			}
			\lElse{\Return \textit{null}}
		} 
		$T \gets CT_{O_{id}}(t_{c+1},t_q)$; \tcp*[h]{The trajectory, as a stack (first on top)}\\
		\While{$T$ is not empty}{
			$m \gets T.pop$;\\ 
			$\langle t_c,p_c \rangle \gets  moveJ(p_c,t_c, t_q,m)$;\\
		}
	}
	\ElseIf(\tcp*[h]{Move backward from following snapshot}){$t_c > t_q$} 
	{
		\If(\tcp*[h]{$O_{id}$ is not in the snapshot}){$p_c = null$ }
		{
			
			\If(\tcp*[h]{Disappears in ${\cal L}_{h-d,h}(O_{id})$}){$O_{id} \in {\cal S}_{h}.Dis$  }
			{
				$t_c \gets {\cal D}_{h-d,h}(O_{id})[last]$; \tcp*[h]{Obtain the time of disappearance}\\
				\If  {$t_c \geq t_q$}{
					$p_c \gets {\cal P}_{h-d,h}(O_{id})[last]$; \tcp*[h]{Obtain the location of disappearance}
					
				}
				\lElse{\Return \textit{null}}
			}
			\lElse{\Return \textit{null}}
		} 
		$T \gets CT_{O_{id}}(t_{q},t_{c-1})$; \tcp*[h]{The trajectory, as a stack (last on top)}\\
		\While{$T$ is not empty}{
			$m \gets T.pop$;\\ 
			$\langle t_c,p_c \rangle \gets  RmoveJ(p_c,t_q,t_c,m)$;\\
		}
	}	
\lIf{$t_c=t_q$} {\Return $p_c$}
\lElse{\Return $null$}
	

\end{algorithm}

In the first case, $t_c$ stores the time instant and $p_c$ the absolute position at any step of the log traversal. In line 4, $t_c$ and $p_c$ store the information of the object at ${\cal S}_h$. If the object is missing at ${\cal S}_h$ (i.e., $p_c=null$), line 6 checks if $O_{id}$ appears in the immediate next portion of the log (${\cal L}_{h,h+d}(O_{id})$), as signaled by the presence of $O_{id}$ in ${\cal S}_{h}.App$. In line 7, the algorithm obtains the time instant of the appearance of $O_{id}$ from the first entry of ${\cal D}_{h,h+d}(O_{id})$. If the appearance is prior to or occurs exactly at $t_q$, then the algorithm obtains the absolute position from ${\cal P}_{h,h+d}(O_{id})$. Otherwise, the object does not have a position at $t_q$ and the algorithm returns $null$.

Once we have the absolute position of the object, lines 12-15 traverse the compressed trajectory using \textit{moveJ} operations from time instant $t_c$ (either the closest snapshot or an absolute appearance) to $t_q$. If the algorithm is able to reach $t_q$, the position is returned.

Lines 16-27 perform the same process as lines 4-15 but in reverse, for the case where $t_q$ occurs before the snapshot.


If there are $t$ time instants, $n$ objects, the grid is of size $s \times s$, $k$ is the arity of our $k^2$-tree, $d$ is the number of time instants between two snapshots, and our grammar is balanced, then this query takes time $O(d + \log t + \log_k s + \log n)$.

\subsection{Obtain the trajectory of an object between two time instants}


Algorithm \ref{alg:searchTrajectory} shows the pseudocode to answer this query. It first obtains a time instant preceding $t_b$ with an absolute position of the object, and then it simply follows the trajectory until it reaches $t_e$. Lines 1-12 are responsible for finding a close preceding time instant with an absolute value, lines 13-17 then advance until reaching $t_b$, and finally lines 18-26 collect the trajectory from $t_b$ to $t_e$.

Line 1 obtains the preceding snapshot closest to $t_b$ (${\cal S}_h$). If the snapshot has information about the position of the object, $p_c$ stores that position and $t_c$ is the time instant of the snapshot. In lines 4-12, the algorithm deals with the case in which the object is not in ${\cal S}_h$. If the object appears in the next portion of the log ($O_{id} \in {\cal S}_h.App$), then the algorithm obtains from ${\cal D}_{h,h+d}(O_{id})[0]$ the time instant of that appearance ($t_c$); if that time instant does not exceed $t_e$, it also obtains the absolute position, which is stored in $p_c$. Otherwise, the algorithm moves to the next snapshot, until finding an absolute position or exceeding the time instant $t_e$.

If the flow reaches line 13, the algorithm has found a suitable timestamp $t_c \le t_b$ for which the absolute position of the object ($p_c$) is known. If $t_c < t_b$, lines 13-17 skip entries of the log(s) until $t_c = t_b$ and $p_c$ is the position of the object at $t_b$. Finally, lines 18-26 traverse from instants $t_b$ to $t_e$, this time collecting all the positions in a sequence $Answer$, which is finally returned.

Compared to the complexity of Algorithm~\ref{alg:searchObject}, we have an additional cost of $O((t_e-t_b+1)(1+(\log n + \log_k s)/d))$, which can be made the optimal $O(t_e-t_b+1)$ by choosing a sufficiently large $d = \Omega(\log n + \log_k s)$. This is, however, a worst-case complexity that ignores the fact that the log is compressed: on average, the term $O((\log n + \log_k s)(t_e-t_b+1)/d))$ must be multiplied by the compression ratio.


 \begin{algorithm}[t!]
     	\footnotesize
 		\caption{{\bf  Search Trajectory}($O_{id}$, $t_{b}, t_{e}$)}
         \label{alg:searchTrajectory}
			${\cal S}_{h} \gets nearestPreviousSnapshot(t_{b})$; \tcp*[h]{The closest snapshot preceding $t_b$}\\
			$p_c \gets {\cal S}_{h}.findObject(O_{id})$; \tcp*[h]{The position in the preceding snapshot}\\
			$t_c \gets {\cal S}_h.time$;

                \While{$p_c = null$ and $t_c \leq t_e$}
                {
                    \If(\tcp*[h]{Appears before next snapshot}){$O_{id} \in {\cal S}_{h}.App$}
                  {$t_c \gets {\cal D}_{h,h+d}(O_{id})[0]$; \\

                   \lIf{$t_c \leq t_e$} {$p_c \gets {\cal P}_{h,h+d}(O_{id})[0]$ }
                  }
                  \Else(\tcp*[h]{Move to next snapshot})
                       { 
                    ${\cal S}_h \gets {\cal S}_h.nextSnapshot()$;\\
                    $p_c \gets {\cal S}_{h}.findObject(O_{id})$;\\
                    $t_c \gets {\cal S}_{h}.time$;\\
                       }
                   }
              
              \lIf(\tcp*[h]{Does not appear before $t_e$}){$t_c > t_e$} { \Return $null$}
              
               \If(\tcp*[h]{Skip until reaching $t_b$}){$t_c<t_b$}
               	  {
               	   
               	   	$T\gets CT_{O_{id}}(t_c,t_{b-1})$; \tcp*[h]{The trajectory, as a stack (first on top)}\\               	   		
               	   	\While{$|T|>1$}
               	   	   {$m \gets T.pop$;\\
                        $\langle t_c, p_c\rangle \gets moveJ(p_c,t_c,t_{b-1},m)$;\\                    	   
                     }
                   
                   }
                  $Answer \gets \langle \rangle$; \\
                  $T\gets CT_{O_{id}}(t_{c+1},t_e)$; \tcp*[h]{The trajectory, as a stack (first on top)}\\
                  \While{$T$ is not empty}
                  {$m \gets T.pop$;\\
                  	\ForEach{$\langle t_j,p_j\rangle$ in $moveS(p_c,t_c,t_e,m)$}                                              { \lIf{$t_j \ge t_b$}{$Answer.add(\langle t_j,p_j\rangle)$}
                           }
                  	$t_c \gets t_j$;\\
                  	$p_c\gets p_j$;\\                  	
                  }
              \Return \textit{Answer}
	\end{algorithm}


\subsection{Time slice query}

Given a time instant $t_q$ and a window rectangle $r = [x_1,x_2] \times [y_1,y_2]$ of the space, this query returns the objects lying within $r$ at time $t_q$, and their positions. Algorithm \ref{alg.timesile} shows the pseudocode.

The process starts by obtaining the snapshot nearest to $t_q$, ${\cal S}_h$. The algorithm considers three cases. Firstly, if $t_q$ is the time instant of ${\cal S}_h$, the algorithm just needs to access the snapshot in order to obtain the objects within $r$. The second case occurs when $t_q$ is between the snapshots ${\cal S}_h$ and ${\cal S}_{h+d}$. In this case, the algorithm has to follow the trajectory of the objects from ${\cal S}_h$ until time $t_q$, to see if they end up within $r$. Instead of following all of the objects in the snapshot, the algorithm tracks only those within $ER_r({\cal S}_h.time,t_q)$, ignoring objects that have no chance of being in the answer. The algorithm must also include in the tracked objects those that are not present in ${\cal S}_h$ but appear at a later time instant, not exceeding $t_q$, and are capable of reaching $r$ at $t_{q}$.

Once this list of candidates has been obtained, the algorithm follows the log to obtain the trajectory of the tracked objects. However, the algorithm prunes the tracking further as it processes the log: a candidate object may follow a direction that takes it away from region $r$, so the algorithm rechecks the condition after applying each code of the compressed trajectory,  discarding the objects as soon as they become incapable of reaching $r$ at time $t_q$.

The third case is analogous, and occurs when $t_q$ is between ${\cal S}_{h-d}$ and ${\cal S}_h$.

\begin{figure}[t!]
\begin{center}

\includegraphics[scale=0.155]{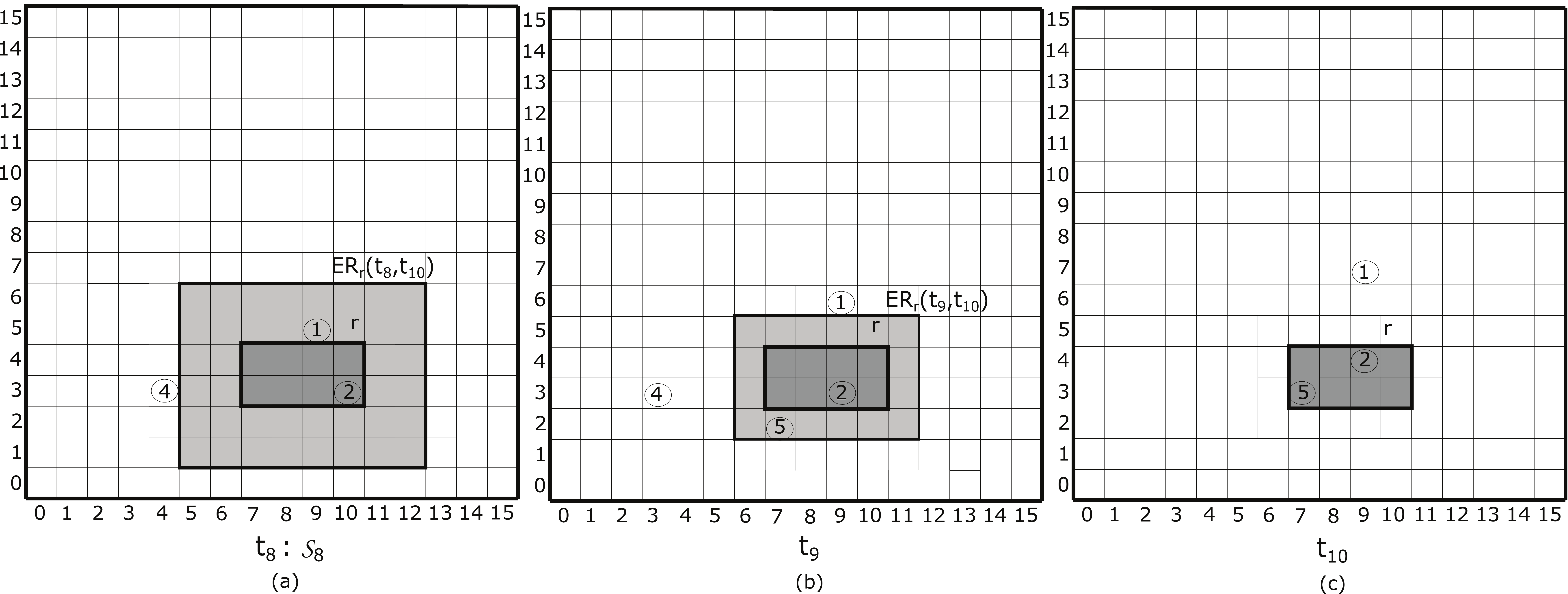}
\end{center}
\vspace*{-5mm}
\caption{Representation of a time-slice query ($t_9$ and $t_{10}$ are not represented with snapshots).} \label{timeSlice}
\end{figure}

To illustrate the idea using the running example shown in Figures \ref{components} and \ref{graCT}, Figure \ref{timeSlice} outlines the process of a time slice query where $t_q=t_{10}$ and $r=[7,10]\times[3,4]$. Since the target time instant is $t_{10}$, the closest snapshot is ${\cal S}_8$. Figure \ref{timeSlice}(a) shows the target region $r$ and the extended region $ER_r(t_8,t_{10})$ superimposed. In this example, $M_{sp}=1$ (that is, an object can only move one position per time instant), so $ER_r(t_8,t_{10})$ is $r$ enlarged by two cells in all directions.

Line 1 obtains the snapshot closest to $t_{10}$ (${\cal S}_8$). Since $t_q$ is not the time instant of the snapshot, the algorithm retrieves in line 4 all the objects in the extended region $ER_r({\cal S}_h.time,t_q) = [5,12] \times [1,6]$, which are stored in the variable \textit{Candidates}; in our example, these are the objects $O_2$ and $O_1$. Next, in lines 7-11, for each object in ${\cal S}_h.App$, the algorithm obtains the time and position of its appearance ($t_c$ and $p_c$); if that time is $t_q$ or earlier, and that position is within $ER_r(t_c,t_{10})$, then the object is added to the candidates to be tracked. In our example, objects $O_3$, $O_6$ and $O_7$ appear after $t_{10}$, therefore they are discarded. Instead, $O_5$ appears at $t_9$, so it is added to the list of candidates, because it is also within $ER_r(t_9,t_{10})$ (see Figure \ref{timeSlice}(b)). The resulting list of candidates in our example comprises $O_1,~O_2$ and $O_5$.

In lines 12-20, the algorithm follows the movements of each candidate object using the log, until it reaches $t_q$ or until the object can be discarded. There are two ways to preempt the log traversal of an object. One is that we reach a nonterminal that already covers $t_q$ while its MBR is disjoint with $r$; this is discarded in lines 16-17. If this does not happen, the log entry is processed in line 18 and we consider the second condition: at this point, we can ensure that the object will not reach the region $r$. This is verified in line 19. Finally, line 20 includes the object in the result if we reach $t_q$ and the $p_c$ is in $r$. Lines 22-35 process the logs backwards from the following snapshot.

Assuming we first process $O_1$, Step 1 of Table \ref{tab:timeSlice} shows the initial state of the algorithm. The algorithm traverses the trajectory of $O_1$ and checks if it can reach $r$ at $t_q$ after applying each $m_i$ of $CT_{O_1}(t_9,t_{10})=\langle 7,V\rangle$. The state after applying the first code ($7$) is shown in Step 2. The algorithm checks if the new position, $(9,6)$, lies inside the extended region corresponding to $t_c=t_9$. As shown in Figure \ref{timeSlice}(b), $O_1$ is outside the region, so the object is discarded.


\begin{algorithm}[t!]
	\footnotesize
	\caption{{\bf Time Slice}($r$, $t_{q}$)}\label{alg.timesile}
	${\cal S}_h \gets nearestSnapshot(t_{q})$;\\
	$t_c \gets {\cal S}_h.time$;\\
	\lIf{$t_c = t_{q}$}{
		\Return  ${\cal S}_h.findObjectsInRegion(r)$
	}
    $Candidates \gets {\cal S}_h.findObjectsInRegion(ER_{r}(t_c, t_q))$;\\
    $Answer \gets \langle \rangle$; \\
	\If{$t_c < t_q$}{
		\ForEach{$O_j \in {\cal S}_h.App$}
		{
			$t_c \gets {\cal D}_{h,h+d}(O_j)[0]$;\\
			$p_c \gets {\cal P}_{h,h+d}(O_j)[0]$;\\
			\If{$t_c \leq t_q$ and $p_c \in ER_{r}(t_c, t_q)$}{$Candidates.add(\langle O_j, t_c, p_c \rangle)$}
		}
		\ForEach{$\langle O_j, t_c, p_c \rangle \in Candidates$}
		{
			$T \gets CT_{O_j}(t_{c+1},t_q)$; \tcp*[h]{The trajectory, as a stack (first on top)}\\
			\While{$T$ is not empty}
			{$m \gets T.pop$; \\
				\If {$m = NT_{value}$ and $t_c+value.\#t \ge t_q$ and $(p_c+value.MBR) \cap r = \emptyset$ }
					{ \textbf{break} \tcp*[h]{The nonterminal is disjoint with $r$, skip to next object}}                                             
				$\langle t_c,p_c\rangle \gets moveJ(p_c,t_c,t_q,m)$;	\\
				\lIf {$p_c \not\in ER_{r}(t_c, t_q)$}
				{ \textbf{break};  \tcp*[h]{$O_j$ cannot reach $r$, skip to next object}}
				\lIf{$p_c \in r$ and $t_c=t_q$}{$Answer.add(\langle O_j, p_c\rangle)$}            	
			}
		}
    }
	\Else(\tcp*[h]{$t_c > t_q$}){
		\ForEach{$O_j \in {\cal S}_h.Dis$}
		{
			$t_c \gets {\cal D}_{h-d,h}(O_j)[last]$;\\
			$p_c \gets {\cal P}_{h-d,h}(O_j)[last]$;\\
			\If{$t_c \geq t_q$ and $p_c \in ER_{r}(t_q, t_c)$}{$Candidates.add(\langle O_j, t_c, p_c \rangle)$}
		}
		\ForEach{$\langle O_j, t_c, p_c \rangle \in Candidates$}
		{
			$T \gets CT_{O_j}(t_q,t_{c-1})$; \tcp*[h]{The reversed trajectory, as a stack (last on top)}\\
			\While{$T$ is not empty}
			{$m \gets T.pop$; \\
				\If {$m = NT_{value}$ and $t_c-value.\#t \le t_q$ and $(p_c-value.MBR) \cap r = \emptyset$ }
					{ \textbf{break} \tcp*[h]{The nonterminal is disjoint with $r$, skip to next object}}                                             
				$\langle t_c,p_c\rangle \gets RmoveJ(p_c,t_c,t_q,m)$;	\\
				\lIf {$p_c \not\in ER_{r}(t_q, t_c)$}
				{ \textbf{break};  \tcp*[h]{$O_j$ cannot reach $r$, skip to next object}}
				\lIf{$p_c \in r$ and $t_c=t_q$}{$Answer.add(\langle O_j, p_c\rangle)$}            	
			}
		}
     }
	\Return $Answer$; \\
\end{algorithm}

The algorithm now processes $O_2$. Its initial position, $p_c=(10,3)$, and its time instant, $t_c=t_8$, are obtained from the candidate list; this state is shown in Step 3. The state of Step 4 is obtained by applying the first code of $CT_{O_2}(t_9,t_{10})=\langle 5,W\rangle$. Since $(9,3)$ is within $ER_r(t_9,t_{10})$, as shown by Figure \ref{timeSlice}(b), the algorithm applies the second code of the trajectory: $m_i=W$ (a non-terminal symbol). The algorithm checks that the application of the non-terminal symbol reaches or exceeds $t_{q}$ ($t_c+value.\#t \ge t_q$). In our example, $t_c=t_9$ and $W.\#t=2$, so the rule would lead us to $t_{11}$. The MBR of $W$, however, is $[0,1] \times [1,2]$, which added to the current position $(9,3)$ is $[9,10] \times [4,5]$. This is not disjoint with $r$, so $O_2$ cannot be discarded. The algorithm then has to decompress $W$, to determine if the object is within $r$ after applying the part of $W$ that spans exactly up to $t_q$. Since $moveJ((9,3),t_9,t_{10},W)$ yields that the position of $O_2$ at $t_{10}$ is $(9,4)$, as shown in Step 5, the object $O_2$ is reported.

Finally, Step 6 shows that we start processing $O_5$ at $t_9$, where it appears. The case of this object is similar to $O_2$, and it is also reported in Step 7.

\begin{table}[t!]
\centering
\footnotesize
\begin{tabular}{|l|l|l|l|c|c|l|}
\hline
&$O_j$&$t_c$&$i$&$m_i$&$p_c$&Answer\\
\hline
\hline
Step 1&$O_1$&$t_8$&&&(9,5)&\\
Step 2&$O_1$&$t_9$&0&7&(9,6)&\\
Step 3&$O_2$&$t_8$&&&(10,3)&\\
Step 4&$O_2$&$t_9$&0&5&(9,3)&\\
Step 5&$O_2$&$t_{10}$&1&W&(9,4)&$O_2$\\
Step 6&$O_5$&$t_9$&&&(7,2)&$O_2$\\
Step 7&$O_5$&$t_{10}$&0&W&(7,3)&$O_2, O_5$\\
\hline
\end{tabular}
\caption{Trace of Algorithm \ref{alg.timesile} with the example of Figure \ref{timeSlice}.}\label{tab:timeSlice}
\end{table}

An average-case analysis of the time complexity of this query can be made assuming that the objects distribute uniformly in the space and without taking advantage of the compression of the log. Let the query region $r$ be of dimensions $\ell_x \times \ell_y$, and the query instant be at distance $t$ from its closest snapshot. The spatial query on the $k^2$-tree, for a range of size $(\ell_x + 2t) \times (\ell_y + 2t)$ and retrieving $c$ candidates, takes $O(\ell_x + \ell_y + t + (c+1) k \log_k s)$ time \cite[Sec.~10.2.1]{Nav16}. On average, $c = \frac{(\ell_x+2t)(\ell_y+2t)}{\ell_x \ell_y}\cdot o$, where $o$ is the actual output size. Further, we traverse all the objects that are within ranges of size $(\ell_x + 2t) \times (\ell_y + 2t)$ (at the snapshot time) decreasing by one until size $\ell_x \times \ell_y$ ($t$ instants later or earlier). The sum of all those costs is, on average, $o \cdot \sum_{i=0}^t \frac{(\ell_x+2i)(\ell_y+2i)}{\ell_x \ell_y} = O(o \cdot (t + t^2(\ell_x+\ell_y+t)/(\ell_x \ell_y)))$. In total, the query has a fixed cost of $O(\ell_x+\ell_y+t+k\log_k s)$ plus an average cost of $O(k\log_k s + t + (t(\ell_x+\ell_y)\log_k s + t^2(\ell_x+\ell_y+t+k\log_k s))/(\ell_x \ell_y))$ per object reported.

The analysis shows that, for this query, it is especially worthwhile to start from the closest snapshot, be it preceding or following $t_q$, so that $t$ is on average $d/4$. In addition to having to process fewer log entries, the distance $t$ between ${\cal S}_h.time$ and $t_q$ directly impacts the size of the snapshot area from where we have to collect the candidate objects, and therefore the average number of candidates. The analysis also shows that the effect of $t$ is reduced when the query area is larger.

\subsection{Time interval query}

Time-interval queries return the objects that were within $r$ at any time instant during time interval $[t_b, t_e]$. An easy solution to this query is to use a time-slice algorithm with two small modifications: $(i)$ use the expanded region with respect to $t_e$, $(ii)$ when the processed time instant ($t_c$) is within $[t_b, t_e]$ and the object is inside $r$, report and stop tracking it. Our actual solution speeds up this basic procedure by skipping whole nonterminals whenever possible.

We split the query interval $[t_b,t_e]$ into the log portions it spans, and treat each portion separately, accumulating the answers. Algorithm \ref{alg.time.int} calls \textit{ProcessPortion} (Algorithm \ref{alg.portion}) for each portion of the log overlapped by $[t_b,t_e]$.

\begin{algorithm}[t]
	\footnotesize
	\caption{{\bf Time interval}($r$, $t_{b}, t_{e}$)}\label{alg.time.int}
	
	${\cal S}_h \gets nearestPreviousSnapshot(t_{b})$;\\
	
	$Answer \gets \langle \rangle$;\\ 
	\While{${\cal S}_{h+d}.time < t_e$}
	{
		$ProcessPortion(r, t_{b}, t_{e}, {\cal S}_h, Answer)$;\\
		${\cal S}_h \gets {\cal S}_h.nextSnapshot()$;\\
		
	}

	\Return \textit{Answer}

\end{algorithm}

{\em ProcessPortion} is a modification of the time-slice algorithm. In lines 3-10, it collects the candidates to verify from the snapshot and the appearances that follow, avoiding the objects already reported. Lines 11-26 treat the trajectory of each candidate. A novelty with respect to time-slice queries is in lines 18-19: since we only report the objects and not their positions, we can directly report an object if the MBR of its next nonterminal is fully contained in the query region $r$. Otherwise, we must expand the nonterminal (lines 20-21) unless its MBR is disjoint with $r$, in which case we can process it as a whole, even if it exceeds $t_{last}$. Note that, since we manage the expansion of the nonterminals in line 21, line 23 always treats a complete symbol of the trajectory, knowing that it cannot exceed $t_{last}$ (or that, if it exceeds $t_{last}$, it is because its MBR is disjoint with $r$). Finally, lines 24-26 process a single step, adding the objects found inside $r$ and skipping them when they are shown to be unable to reach $r$ within the time left until $t_{last}$.

 \begin{algorithm}[t!]
	\footnotesize
	\caption{{\bf ProcessPortion}($r$, $t_{b}, t_{e}, {\cal S}_h, Answer$)}\label{alg.portion}
 
	$t_c \gets {\cal S}_h.time$;\\
	$t_{last} \gets  \min(t_e,{\cal S}_{h+d}.time)$;\\
    $Candidates \gets \langle\rangle$; \tcp*[h]{The candidates to be considered, as a list}\\

		\ForEach{$\langle O_j, t_c, p_c \rangle \in {\cal S}_h.findObjectsInRegion(ER_{r}(t_c, t_{last}))$}
		{
			\lIf{$O_j \not\in Answer$}{$Candidates.add(\langle O_j, t_c, p_c \rangle)$}
		 }
		\ForEach{$ O_j \in {\cal S}_h.App$}
		{
			$t_c \gets {\cal D}_{h,h+d}(O_j)[0]$;\\
			$p_c \gets {\cal P}_{h,h+d}(O_j)[0]$;\\
			\If{$t_c \leq t_{last}$ and $p_c \in ER_{r}(t_c, t_{last})$ and $O_j \not\in Answer$}{$Candidates.add(\langle O_j, t_c, p_c \rangle)$;}
		}
		
		\ForEach{$\langle O_j, t_c, p_c \rangle  \in Candidates$}
		{			
			\lIf{$p_c \in r$ and $t_c \in [t_b,t_e]$}{$Answer.add(O_j)$}
			\Else{
			$T \gets CT_{O_j}(t_{c+1},t_{last})$; \tcp*[h]{The trajectory, as a stack (first on top)}\\
			\While{$t_c < t_{last}$ and $T$ is not empty}
			{

				$m \gets T.pop$; \\
                \If {$m = NT_{value}$} {
				\If{$(p_c+value.MBR)  \subseteq r $ }
				 {$Answer.add(  O_j )$; 
				  \textbf{break}; \tcp*[h]{Include object and skip}}
				\ElseIf{$(pc+value.MBR) \cap r \not= \emptyset$ and $value \rightarrow v_1\,v_2$}
				  { $T.push(v_2)$; $T.push(v_1)$; \tcp*[h]{Replace nonterminal}
		         }}
				\Else                       
				{
				$\langle t_c,p_c\rangle \gets moveJ(p_c,t_c,{\cal S}_{h+d}.time,m)$; \\
                \If{$p_c \in r$ and $t_c \in [t_b,t_e]$}{$Answer.add(  O_j )$;
					\textbf{break}; \tcp*[h]{Include object and skip}\\
			}
			    \lElseIf {$t_c < t_{last}$ and $p_c \not\in ER_{r}(t_c, t_{last})$}
				{ \textbf{break}; \tcp*[h]{Skip candidate}					
				}
				} 
			}
		}
		}	 
\end{algorithm}

We have omitted other possible speedups for simplicity. For example, let $r = [x_1,x_2] \times [y_1,y_2]$ and $pc+value.MBR = [x'_1,x'_2] \times [y'_1,y'_2]$. Then, in  line 18, we can also directly add $O_j$ to the answers if $t_c + value.\#t \le t_e$ and, either $x_1 \le x'_1 \le x'_2 \le x_2$ and $[y_1,y_2] \cap [y'_1,y'_2] \not= \emptyset$, or if $y_1 \le y'_1 \le y'_2 \le y_2$ and $[x_1,x_2] \cap [x'_1,x'_2] \not= \emptyset$. In both cases, we know that $O_j$ must be inside $r$ at some moment not exceeding $t_e$.

The analysis for these queries is similar to that for time-slice queries, now taking $t$ as the distance between $t_e$ and the closest snapshot preceding $t_b$ (and ignoring the help of intermediate snapshots). While omitted for simplicity, we can also process this query from the snapshot following $t_e$ with the aim of reducing $t$.

\subsection{Nearest neighbor queries}


Nearest neighbor ({\em knn}) queries return the $K$ objects nearest to a given query point $p_q$ at a given time instant $t_q$. To solve this query, we obtain the candidate objects from the snapshot (${\cal S}_h$) that is nearest to $t_q$, follow their trajectory up to $t_q$, and retain the $K$ candidates that are closest to $p_q$ at time $t_q$. The problem is that, unlike in time-slice queries, we cannot easily determine which objects cannot be part of the solution. This condition changes dynamically as we find better and better candidates for the answer set: given the $K$ closest objects to $p_q$ found up to now, the distance from $p_q$ to the $K$-th closest object in $t_q$ can be used to compute an upper bound to the distance between a candidate and $p_q$ in ${\cal S}_h$, so that if the candidate is farther away, it has no chance of getting closer than the current $K$-th closest object at $t_q$. 

Since this upper bound becomes more restrictive as we improve the answer with closer objects, it is important to obtain objects close to $p_q$ as soon as possible, as this allows us to discard other objects earlier. For this purpose, we will use the hierarchical space partitioning induced by the $k^2$-tree of ${\cal S}_h$, and will traverse the regions prioritized by their potential relevance to the query (i.e., those closer to $t_q$ will be processed first).

The traversal is carried out with a priority queue, which stores $k^2$-tree regions $r$ prioritized by increasing values of $d(r) = \overline{p_q r}$. Initially the whole grid, corresponding to the $k^2$-tree root, is inserted in the priority queue, and then we iteratively extract the most promising node $v$ (i.e., covering the region $r$ with minimum $d(r)$ value). Let $\{r_0, r_1, ..., r_{k^2-1}\}$ be the subregions of $r$ induced by the children of $v$. The nonempty subregions $r_i$ corresponding to children $v_i$ of $v$ are then reinserted in the queue with priority $d(r_i)= \overline{p_q r_i}$. When the children of the region $r$ that we extract are leaves of the $k^2$-tree, the corresponding objects $O_j$ are processed in a different way. 

Those candidate objects $O_j$ we find are inserted in a second priority queue, using a similar ordering on $\overline{p_q p_c}$, where $p_c$ is the position of $O_j$; we also record the time instant $t_c$ of the snapshot ${\cal S}_h$ associated with the objects. In this second queue we initially accept every object that does not disappear in its log by time instant $t_q$, until we have $K$ elements. From this point, objects $O_j$ are inserted only if at time $t_q$ they can get closer to $p_q$ than the $K$-th object closest to $p_q$ in the second queue. More precisely, let $O$, which was known to be at position $p$ in time instant $t$, be the $K$-th object closest to $p_q$ in the priority queue (this is easily maintained by keeping the $K$ closest objects in a separate queue prioritized by decreasing distance). Then the maximum distance to $p_q$ it can reach at time $t_q$ is $d_{max}=\overline{p_q p} + M_{sp} \cdot (t_q-t)$. Let $O_j$ be at position $p_c$ in time instant $t_c$. Then, at instant $t_q$, $O_j$ cannot be closer to $p_q$ than $d_{min}(O_j)=\overline{p_q p_c} - M_{sp} \cdot (t_q-t_c)$. Therefore, we insert $O_j$ as a candidate only if $d_{min}(O_j) < d_{max}$. When we populate the queue of candidates, we must also consider the objects $O_j$ in ${\cal S}_h.App$ appearing at times $t_c \le t_q$ (and not disappearing before $t_q$). We prioritize the candidates $O_j$ using $d_{min}(O_j)$.


This condition to insert an object as a candidate gives also a criterion to stop the processing of the priority queue of regions. If we extract region $r$ from the queue, no object from $r$ can be, at time instant $t_q$, at a distance to $p_q$ below $d_{min}(r) = \overline{p_q r} - M_{sp} \cdot (t_q-t_c)$. Therefore, if $d_{min}(r) \ge d_{max}$, there is no need to consider any point from region $r$, nor from any region yet to be extracted from the first priority queue. At that point we can discard the first queue and start processing our queue of candidates. Algorithm \ref{alg.cand} gives the code (for simplicity, we consider only the nearest snapshot preceding $t_q$).

\begin{algorithm}[t!]
	\footnotesize
	\caption{{\bf ObtainCandidates}(${\cal S}_h$, $K$, $t_q$, $p_q$)}\label{alg.cand}	
	
	$v \gets {\cal S}_h.root$; \tcp*[h]{The root of the $k^2$-tree of the grid, $v.r$ is its region}\\
    $Q_R \gets \emptyset$; \tcp*[h]{Priority queue of pairs (region,dist), sorted by increasing dist} \\
    $Q_C \gets \emptyset$; \tcp*[h]{Priority queue of tuples (obj,time,pos,dist), sorted by increasing dist} \\
    $Q_R.add(\langle v,0 \rangle)$; \tcp*[h]{Insert the whole region as the first pair}\\
    $d_{max} \gets +\infty$; \tcp*[h]{Infinite until obtaining $K$ candidates}\\
	$t_c \gets {\cal S}_h.time$;\\
	\While{$Q_R$ is not empty}{
		$\langle v,dr\rangle \gets Q_R.extractMin$; \tcp*[h]{Extract minimum distance from queue}\\
		$d_{min} \gets dr - M_{sp}\cdot(t_q-t_c)$; \\
        \lIf(\tcp*[h]{Best region cannot beat $K$-th candidate}){$d_{min} \ge d_{max}$}{\textbf{break}}
		
		\If{children of $v$ are not leaves}{
			\ForEach{$v'$ nonempty child of $v$}{
				$Q_R.add(\langle v',\overline{p_q v'.r} \rangle)$;\\ 	}
						
		}
		\Else{
			\ForEach{$O_j$ inside $v.r$}{
                \If{$O_j$ is present in the log ${\cal L}_{h,h+d}$ by time $t_q$}{
                
				$p_c \gets {\cal S}_h.findObject(O_j)$; \tcp*[h]{Fast to compute from $v$}\\
                $d_{min} \gets \overline{p_q p_c} - M_{sp}\cdot(t_q-t_c)$; \\
				\If{$d_{min} < d_{max}$}
				{ $Q_C.add(\langle O_j, t_c, p_c,d_{min}\rangle)$;\\
                  \If{$|Q_C| \ge K$}
                     { Let $\langle O,t,p,d\rangle $ be the $K$-th smallest distance in $Q_C$;\\
                       $d_{max} \gets \overline{p_q p} + M_{sp}\cdot(t_q-t)$
                     }
                }
                   }
                   }
                   }
		}
    \ForEach(\tcp*[h]{Finally, add candidates that appear in the log}){$O_j \in {\cal S}_h.App$}{
                \If{$O_j$ is present in the log ${\cal L}_{h,h+d}$ by time $t_q$}{
                
				$p_c \gets {\cal P}_{h,h+d}(O_j)[0]$; 
				$t_c \gets {\cal D}_{h,h+d}(O_j)[0]$; \\
                $d_{min} \gets \overline{p_q p_c} - M_{sp}\cdot(t_q-t_c)$; \\
				\If{$d_{min} < d_{max}$}
				{ $Q_C.add(\langle O_j,t_c,p_c,d_{min}\rangle)$;\\
                  \If{$|Q_C| \ge K$}
                     { Let $\langle O,t,p,d\rangle $ be the $K$-th smallest distance in $Q_C$;\\
                       $d_{max} \gets \overline{p_q p} + M_{sp}\cdot(t_q-t)$
                     }
                }
                   }
                   }
	\Return $Q_C$
	
\end{algorithm}

\begin{figure}[t!]
	\begin{center}

		\includegraphics[scale=0.2]{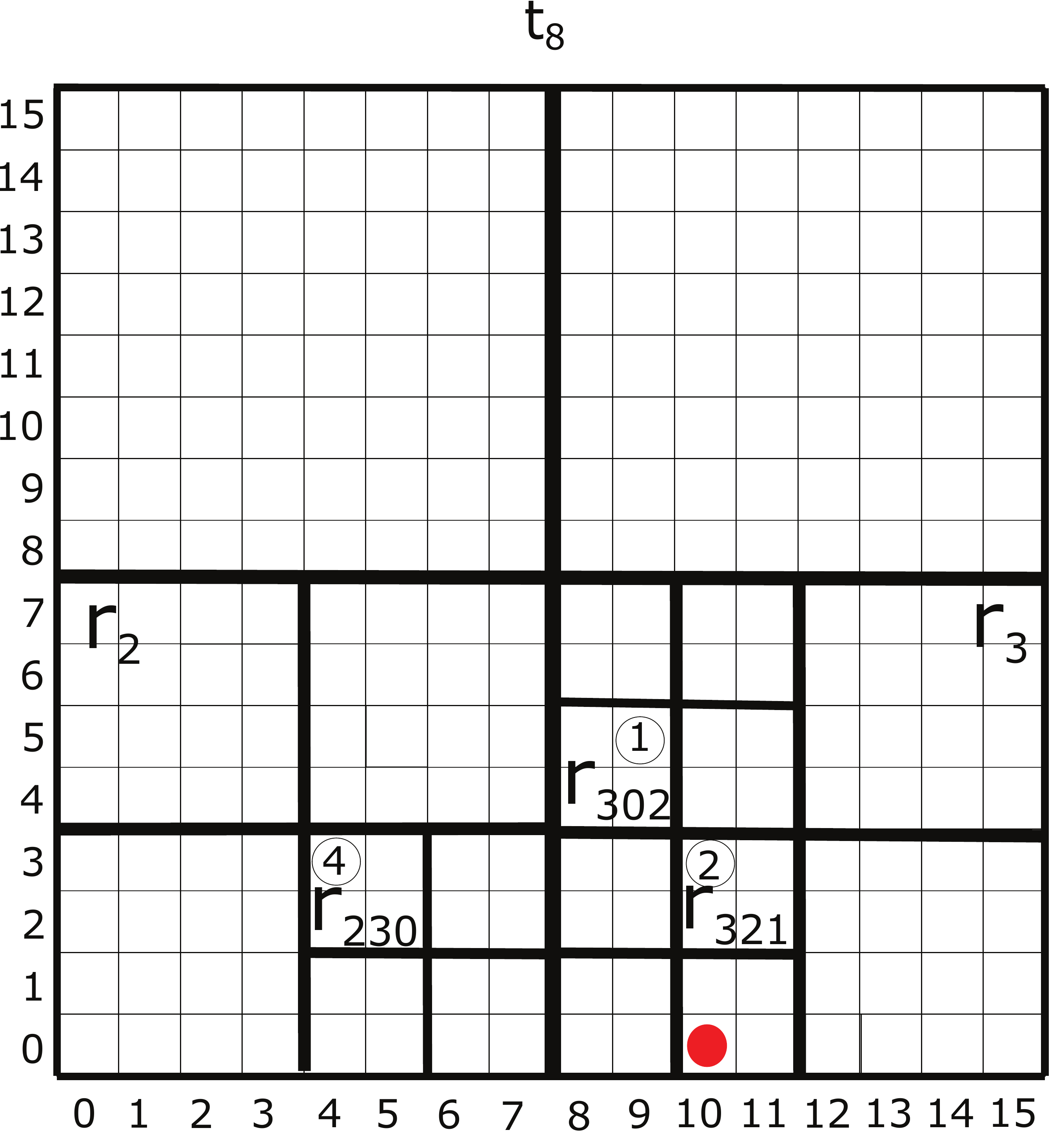}
	\end{center}
	\vspace*{-5mm}
	\caption{Example of regions of the space inserted in the queue.}\label{knn}
\end{figure}

%
%
%
Let us assume that $k=2$, $M_{sp}=1$, $K=1$, $t_c=t_8$, $t_q=t_9$, $p_q=(10,0)$, and the grid ${\cal S}_8$ of Figure \ref{knn}. We number the regions with a sequence of subindices in $\{0,1,2,3\}$ considering the consecutive levels. The children of the root at $r_0$ to $r_3$, the children of $r_0$ are $r_{00}$ to $r_{03}$, and so on. From the root, we insert $r_2$ and $r_3$ in the queue ($r_0$ and $r_1$ are empty). We then extract $r_3$, the closest one to $p_q$ (indeed, containing it), and reinsert $r_{30}$ and $r_{32}$. The next most promising region is $r_{32}$, from whose children we only reinsert $r_{321}$, the nonempty one. From the current set, $\{r_2,r_{30},r_{321}\}$, the most promising one is $r_{321}$, at distance $\overline{p_q r_{321}} = 2$. So we extract it and insert object $O_2$ as our first candidate, at distance $\overline{p_q O_2} =3$. Thus, at time $t_9$, $O_2$ could be at maximum distance $\overline{p_q O_2}=4$. Since the next most promising area, $r_2$, is at distance $\overline{p_q r_2}=3$, it is still worth considering, so we reinsert its nonempty child $r_{23}$.
This is still more promising than $r_{30}$, so we extract it and reinsert $r_{230}$, its nonempty child. Now our candidate regions are $\{r_{30},r_{230}\}$, and the most promising is $r_{30}$. We extract it and reinsert $r_{302}$, its nonempty child, which becomes the first region to extract. Since $\overline{p_q r_{302}} = \sqrt{17} \approx 4.1$, a point inside $r_{302}$ might reach distance $\approx 3.1$ to $p_q$ at $t_9$, and thus it might still be interesting. We then extract it and obtain $O_1$, which is at distance $\overline{p_q O_1} = \sqrt{26} \approx 5.1$, and can then get as close as distance $\approx 4.1$ in $t_9$. Thus, there is no point in considering $O_1$, because it cannot get closer to $p_q$ than $O_2$. Similarly, there is no point in considering the remaining region $r_{230}$ (nor any other region that would have remained to process), because it is at distance $\overline{p_q r_{230}} = \sqrt{29} \approx 5.4$. In this case, we only have one candidate.


From the priority queue of candidates, the algorithm extracts the first one (with position $p_c$, at time $t_c$, with minimum $d_{min}$) and calculates the next position of the object using \textit{moveJ}. The object is then reinserted in the queue with its new position $p'_c$ and time instant $t'_c$, and with $d_{min}$ recomputed accordingly. As we iterate, the candidates advance in their log ${\cal L}_{h,h+d}$, with the most promising objects advancing sooner towards $t_q$. At some point, we will start extracting objects with time $t_c = t_q$. For those objects, $d_{min}$ is their exact distance to $p_q$ at time $t_q$. We store them in a third priority queue, of maximum size $K$ and sorted by decreasing $d_{min}$. When we have $K$ answers in this third queue, the insertion of a new result may displace the currently $K$-th result (the first in the priority queue). Further, we can use the value $d_{min}(O)$ of the $K$-th result $O$ to preempt the processing of the queue of candidates: if for the candidate $O_j$ with the minimum $d_{min}$ value it holds that $d_{min}(O_j) \ge d_{min}(O)$, then there is no chance that $O_j$ (nor any other candidate object remaining in the second priority queue) alters the current set of results. We can then stop the processing and return the results in the third queue as the definitive answer. 
%
%
%
%


\begin{algorithm}[t!]
	\footnotesize
	\caption{{\bf Knn}($K$,  $p_q$, $t_{q}$,)}\label{alg.knn}
	
	${\cal S}_h \gets findNearestPreviousSnapshot(t_q);$\\  
	$Q_C \gets ObtainCandidates({\cal S}_h,K, t_q,p_q)$;\\
	$Q_F \gets \emptyset$; \tcp*[h]{Priority queue of pairs (obj,dist) sorted by decreasing dist, max size $K$} \\
    $d_{max} \gets +\infty$; \tcp*[h]{Infinite until obtaining $K$ results}\\
	\lIf{${\cal S}_h.time = t_{q}$}{
		\Return first $K$ objects of $Q_C$
	}
		\While {$Q_C$ is not empty}
		{
			$\langle O_j,t_c, p_c,d_{min}\rangle \gets Q_C.extractMin$; \\
			\lIf(\tcp*[h]{Best candidate cannot beat $K$-th result}){$|Q_F| \ge K$ and $d_{min} \ge d_{max}$ }{\textbf{break}}
			\If(\tcp*[h]{We have a result}){$t_c=t_q$}{
            $Q_F.add(\langle O_j,d_{min}\rangle)$; \tcp*[h]{$d_{min}$ is $\overline{p_qp_c}$}\\
            \lIf{$|Q_F| > K$}{$Q_F.extractMax$; \tcp*[h]{maximum size is $K$, discard one}}
            \lIf{$|Q_F| \ge K$}{$\langle O,d_{max}\rangle \gets Q_F.inspectMax$; \tcp*[h]{update $d_{max}$, $O$ is dummy}}
			}
			\Else(\tcp*[h]{Advance one position in the log; we have a pointer to $CT_{O_j}$}){
				$\langle t_c,p_c\rangle \gets moveJ(p_c,t_c,t_q,CT_{O_j}(t_{c+1},t_q)[0])$;\\
                $d_{min} \gets \overline{p_q p_c} - M_{sp}\cdot(t_q-t_c)$;\\
				$Q_C.add(\langle O_j,t_c,p_c,d_{min}\rangle)$; \\
			}		
	}
	\Return $Q_F$

\end{algorithm}

Algorithm \ref{alg.knn} shows the pseudocode of the {\em knn} query. To illustrate it,  we use the example from Figures \ref{components} and \ref{graCT} and obtain the nearest object ($K = 1$) with respect to location $p_q= (10,0)$ at $t_q= t_9$. After obtaining the previous closest snapshot to $t_9$ (${\cal S}_8$), Algorithm \ref{alg.knn} issues the call $ObtainCandidates({\cal S}_8,1,t_9)$ to obtain the candidates from the snapshot. 
As we have already seen, the traversal of regions yields only one candidate, $O_2$, with minimum distance $d_{min} = \overline{p_q O_2} - M_{sp}\cdot(t_q-t_c) = 2$. Therefore, the queue of candidates contains only $Q_C = \{ \langle O_2,t_8,(10,3),2\rangle\}$. However, objects $O_3$, $O_5$, $O_6$, and $O_7$ appear in the log after ${\cal S}_8$. From those, only $O_5$ appears before or at $t_9$, and thus it can be added to $Q_F$. Indeed, $O_5$ appears at $t_9$, in position $(7,2)$. Its minimum (and exact) distance to $p_q$ in $t_q$ is $d_{min} = \overline{p_q (7,2)} = \sqrt{13} \approx 3.6$. Since $d_{max}=4$, this object cannot yet be discarded, so the final queue of candidates is $Q_C = \{ \langle O_2,t_8,(10,3),2\rangle, \langle O_5,t_9,(7,2),3.6\rangle \}$.

Once Algorithm \ref{alg.knn} has all the candidates, it starts processing them in lines 6-16. It first extracts $\langle O_2,t_8,(10,3),2\rangle$. Since $t_8$ is not the target time, the algorithm considers $CT_{O_2}(t_9,t_9)=\langle 5\rangle$ and applies {\em moveJ} on the first element, $5$. This converts $t_8$ into $t_9$ and $(10,3)$ into $(9,3)$. It then recomputes $d_{min} = \sqrt{10} \approx 3.2$ and reinserts $O_2$ into $Q_C$ with these new values. The queue of candidates now contains $Q_C = \{ \langle O_2,t_9,(9,3),3.2\rangle, \langle O_5,t_9,(7,2),3.6\rangle \}$. The next best candidate is again $O_2$, but now its time is $t_9$, the same as the query, and this means that we have our first result, $Q_F = \{ \langle O_2,3.2 \rangle\}$. Since now $|Q_F| \ge K$,
we can compute $d_{max} = 3.2$. In the next step, when we extract $O_5$ from $Q_C$, its $d_{min}$ value is $3.6$, larger than $d_{max}$, and thus disregard it. We have exhausted $Q_C$ and the answer is in $Q_F$.

\section{Experimental Evaluation}\label{sect:exp}

In this section we experimentally evaluate the space and time performance of GraCT and compare it with related work. It has been very hard to obtain implementations of previous indexes. Further, compression-only systems developed for trajectories like Trajic \cite{Trajic} and SQUISH \cite{Muckell2014} did not compress well on our data, reducing the space of a plain text representation, but not the space of a binary representation of the integer coordinates. We have, nevertheless, compared GraCT with two baselines:

\begin{description}
\item[\texttt{scdcCT:}] This structure is similar to GraCT with a different representation of the log structure. The sequence of spiral codes that differentially encode the movements is compressed with a statistical zero-order compressor, namely byte-oriented $(s,c)$-Dense Codes \cite{RodrguezBrisaboa07} (SCDC) optimized to $s=183$ and $c=73$. SCDC combines good compression ratios with fast decoding, thereby supporting fast scanning of the logs. We use scdcCT as a representative of the various systems that use statistical compression, instead of the grammar-based compression used by GraCT. We aim to demonstrate that grammar compression is more effective, and that the ability of GraCT to skip over whole nonterminal symbols makes it faster than statistical encoders that, no matter how fast, must process each movement individually. Note that sampling absolute values at regular intervals to speed up the processing of logs in scdcCT is equivalent in space to shortening the sampling period $d$ of the snapshots.

\item[\texttt{MVR-tree:}]
This is the only classic spatio-temporal structure for which we could obtain code. The complete structure is the MV3R-tree \cite{PapadiasT01}, which combines an MVR-tree with an auxiliary 3DR-tree, but only the MVR-tree part was available. The 3DR-tree is used to speed up the most basic queries, like obtaining the position of an object at a given time instant or retrieving its trajectory. We aim to show that a classic structure uses orders of magnitude more space than GraCT, without being much faster (actually, the MVR-tree is slower for some queries). We used the MVR-tree implementation provided by the spatialindex library\footnote{\texttt{http://libspatialindex.github.io}}. This is a disk-based structure, but we make it run in main memory for a direct comparison with GraCT. We also compare both structures on disk, as explained later. 
\end{description}

We implemented GraCT and scdcCT in C++, using components from the SDSL library\footnote{\texttt{https://github.com/simongog/sdsl-lite}} \cite{gbmp2014sea}.  The experiments were run on an Intel\textsuperscript{\textregistered} Core\textsuperscript{TM} i7-3820 CPU @ 3.60GHz (4 cores) with 10MB of cache and 64 GB of RAM, running Ubuntu 12.04.5 LTS with kernel 3.2.0-115 (64 bits), gcc version 4.6.4 with \texttt{-O9} optimization. The code for GraCT and scdcCT is publicly available at {\tt https://gitlab.lbd.org.es/adriangbrandon/ct.git}.

We used an implementation for supporting rank operations that adds 6.25\% space on top of the bit sequence to accelerate queries. The sample value $1/\epsilon$ of the permutations was set to 5, which increases the size of the data structure by 20\% but yields fast $\pi^{-1}$ queries. Our grammar compressor was the Re-Pair implementation provided by G. Navarro\footnote{\texttt{http://www.dcc.uchile.cl/gnavarro/software/repair.tgz}}, using the balanced version. The extra information associated with the Re-Pair nonterminals was encoded using DACs with an unlimited number of levels and without a predefined chunk size. Arrays ${\cal P}$ and ${\cal D}$, in contrast, are represented using DACs configured to use chunks of one byte and with a maximum of two levels.

\subsection{Datasets}\label{datatset}

We used real-world and pseudo-real data from three sources.

\begin{itemize}
\item \texttt{Ships}, a real dataset storing the movements of 3,654 boats sailing in UTM Zone 10 over one month in 2014, obtained from MarineCadastre\footnote{\texttt{http://marinecadastre.gov/ais}}.
\item \texttt{Planes}, a real dataset composed by trajectories of aircrafts from 30 different airlines and flights between 30 European airports. This data was obtained from the OpenSky Network\footnote{\texttt{https://opensky-network.org}}; we discarded the altitude component and retained latitude and longitude.
\item \texttt{Taxis}, a larger synthetic dataset simulating trajectories of taxis in New York City during 2013. We download the data from \textit{NYC Taxis: A Day in the life}\footnote{\texttt{http://chriswhong.github.io/nyctaxi}}. Since this includes only the start and end location of each taxi trip, we computed the trajectory as the shortest route between these two points. 
\end{itemize}

GraCT stores positions in a discrete grid, therefore each position emitted by an object is discretized into a matrix. We chose cell sizes that seem appropriate in each application, in the sense that more detail is probably useless: $50 \times 50$ meters in \texttt{Ships}, $5000 \times 5000$ meters in \texttt{Planes} and $10 \times 10$ meters in \texttt{Taxis}. In addition, our structure deals with object positions at regular timestamps, but in the real datasets, the frequency of signals varied. For this reason, the signals were preprocessed and discretized, in order to obtain the position of each object at regular timestamps: 1 minute for \texttt{Ships} and 15 seconds for \texttt{Planes} and \texttt{Taxis}.

After this normalization step, some location signals were observed to be incorrect, because they implied object movements at an extremely high speed (these kind of errors are not rare in GPS measurements). We set a maximum speed for each dataset, 800km/h for \texttt{Planes} and 234km/h for the other datasets. If the signal of an object at a timestamp implies exceeding the maximum speed, that signal is deleted. This amounts to deleting 0.01\% of the signals from \texttt{Ships} and 0.83\% from \texttt{Planes}. In addition, the frequency of signals is variable. For example, in \texttt{Ships} the frequency of signals from most boats was regular when they were sailing, but it stopped for long periods, typically when they were at a port.
This produces disappearances and appearances, which amount to 0.21\% of the entries in \texttt{Ships} and 0.96\% in \texttt{Planes}.
Finally, since our setup requires the position of each object at regular timestamps, we interpolated the two closest signals if they were less than 15 time instants apart, since an object cannot move far during a short period of time; otherwise we consider it a disappearance/appearance. 

Each dataset is stored in a plain text file comprising four columns: \textit{object identifier}, \textit{time instant}, \textit{coordinate x} and \textit{coordinate y}. Each value in the columns is stored as a string. In order to obtain a better compression comparison, the data was alternatively stored in binary form, with the minimum number of bytes. For example, in \texttt{Ships}, two bytes were used to represent object identifiers (max value 3,654), two for the time instant column (max value 44,642), two for the x-axis (max value 12,719) and three for the y-axis (max value 368,186). Table \ref{table:datasets} shows the resulting sizes.

To illustrate how compressible the data is, the plain data files were compressed with \textit{p7zip}\footnote{\texttt{http://p7zip.sourceforge.net}} with default settings. As we see, \textit{p7zip} compresses the data to 9.5\%--11.8\% of its binary representation.

\begin{table}[t]
\setlength{\tabcolsep}{2pt}
 \begin{center}
 \begin{tabular}{l @{~~} | @{~~~} r @{~~~} r @{~~~} r }
 	
  \hline
     & \texttt{Ships} & \texttt{Planes} & \texttt{Taxis}\\
  \hline
   {Total objects}& 3,654 & 2,263  & 1,283\\
   {Total points} & 44,304,802
& 35,315,697 & 2,524,018,668\\
   {Max $x$} & 12,719 & 3,313& 1,090,125
\\
   {Max $y$} & 368,186 & 1,148 & 1,160,194
\\
    {Max $time$} & 44,642 & 172,547 & 2,102,879\\
    \hline
    {Size Plain} & 947.35 MB & 671.02 MB & 59,263.57 MB
\\
    {Size Bin} & 380.27 MB & 303.12 MB & 26,478.01 MB
\\
Size \textit{p7zip} & 36.14 MB & 35.82 MB & 3,075.93 MB
\\
  \hline

 \end{tabular}
\end{center}

 \caption{Datasets and their dimensions.}
 \label{table:datasets}
\end{table}


\subsection{Compression}
The first experiment compares the compression ratio of GraCT with scdcCT on the two real datasets, \texttt{Ships} and \texttt{Planes}. GraCT and scdcCT data structures were built on both datasets using snapshot intervals of $d=120$, 240, 360 and 720 time instants. The construction time for the complete structure was approximately one minute.

Table \ref{table:experiment1} shows the results of compression. The first row gives the size of both structures for each snapshot period $d$, whereas the next two rows separate the space due to the snapshots (i.e., the $k^2$-trees) and to the compressed logs. It can be seen that the space of the snapshots is much lower than that of the logs, even with the highest sampling rate. Still, the compression of the logs themselves improves for larger $d$. In this aspect, Re-Pair improves faster than SCDC: the quotient of the log space with $d=720$ versus $d=120$ is 0.68 on \texttt{Ships} and 0.47 on \texttt{Planes}, whereas with scdcCT they are 0.78 and 0.60, respectively.

  \begin{table}[t]
\setlength{\tabcolsep}{3pt}
 \begin{center}
 \begin{tabular}{l @{~~~} l | r r r r | r r r r}
 	
  \hline
   & {Index} & \multicolumn{4}{c|}{{GraCT}} & \multicolumn{4}{c}{{scdcCT}} \\
  \hline
   & {Period} & {120} & {240} & {360} & {720} & {120} & {240} & {360} & {720}\\
  \hline
  \multirow{5}{1cm}{\texttt{Ships}} & {Size} & 34.72 & 26.92 & 24.27 & 21.58 & 65.93 & 56.02 & 52.65 & 49.28\\
   & {Snapshot} & 4.05 & 2.13 & 1.43 & 0.74 &  4.05 & 2.13 & 1.43 & 0.74 \\
  & {Log} & 30.66 & 24.79 & 22.84 & 20.85 & 61.88 & 53.89 & 51.22 & 48.55\\
  & {Ratio (plain)} & 3.66\% & 2.84\% & 2.56\% & 2.28\% & 6.96\% & 5.91\% & 5.56\% & 5.20\%\\
  &{Ratio (bin)} & 9.13\% & 7.08\% & 6.38\% & 5.68\% & 17.34\% & 14.73\% & 13.85\% & 12.96\%\\
  \hline
  \multirow{5}{1cm}{\texttt{Planes}} & {Size} & 49.24 & 32.23 & 26.55 & 20.83 & 82.61 & 59.80  & 52.21 & 44.56 \\
   & {Snapshot} & 3.85 & 1.95 & 1.31 & 0.68 & 3.85 & 1.95 & 1.31 & 0.68 \\
  & {Log} & 41.06 & 28.08 & 23.75 & 19.38 & 74.43 & 55.66 & 49.41 & 44.56\\
  & {Ratio (plain)} & 7.34\% & 4.80\% & 3.96\% & 3.10\% & 12.31\% & 8.91\% & 7.78\% & 6.64\%\\
  &{Ratio (bin)} & 16.25\% & 10.63\% & 8.76\% & 6.87\% & 27.25\% & 19.73\% & 17.23\% & 14.70\%\\
  \hline

 \end{tabular}
\end{center}

 \caption{Structure sizes (in MB) and compression ratios.}
 \label{table:experiment1}
\end{table}

%
%
%

The last two rows give the compression ratios obtained with respect to the plain and binary representations, respectively.
The compression achieved by scdcCT with $d=720$, 13\%--15\% of the binary representations, shows that statistical compression of the differences is competitive with state-of-the art compression like the one offered by {\em p7zip}. The grammar-compression used by GraCT, on the other hand, obtains less than half the space of scdcCT, and less than 60\% of the space used by {\em p7zip}.

We remark that, within this space, GraCT not only represents the trajectory data, but it offers efficient access and queries on it. Devoided of the extra information it stores on the nonterminals to speed up queries, the space used by the grammar compression of the logs would be halved or so. This shows that grammar compression is much more efficient than statistical compression to exploit the redundancy of real-world trajectory data.

\subsection{Query performance}\label{subsec:query-types}

\begin{figure}[p]
	\centering     
\subfigure[\textit{Object}]
{\label{fig:time-ot}\includegraphics[width=0.43\textwidth]{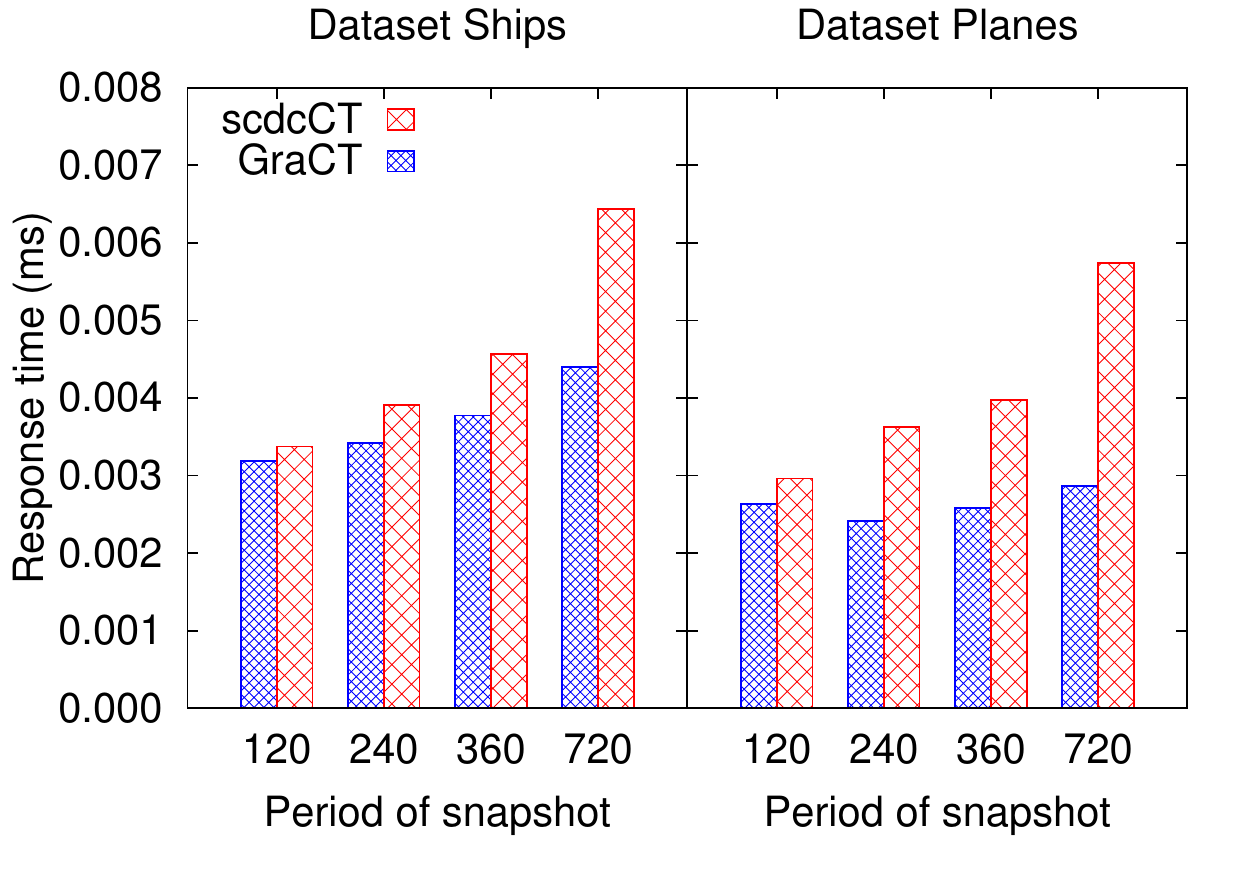}}
\subfigure[\textit{Trajectory}]
{\label{fig:time-trajectory}\includegraphics[width=0.43\textwidth]{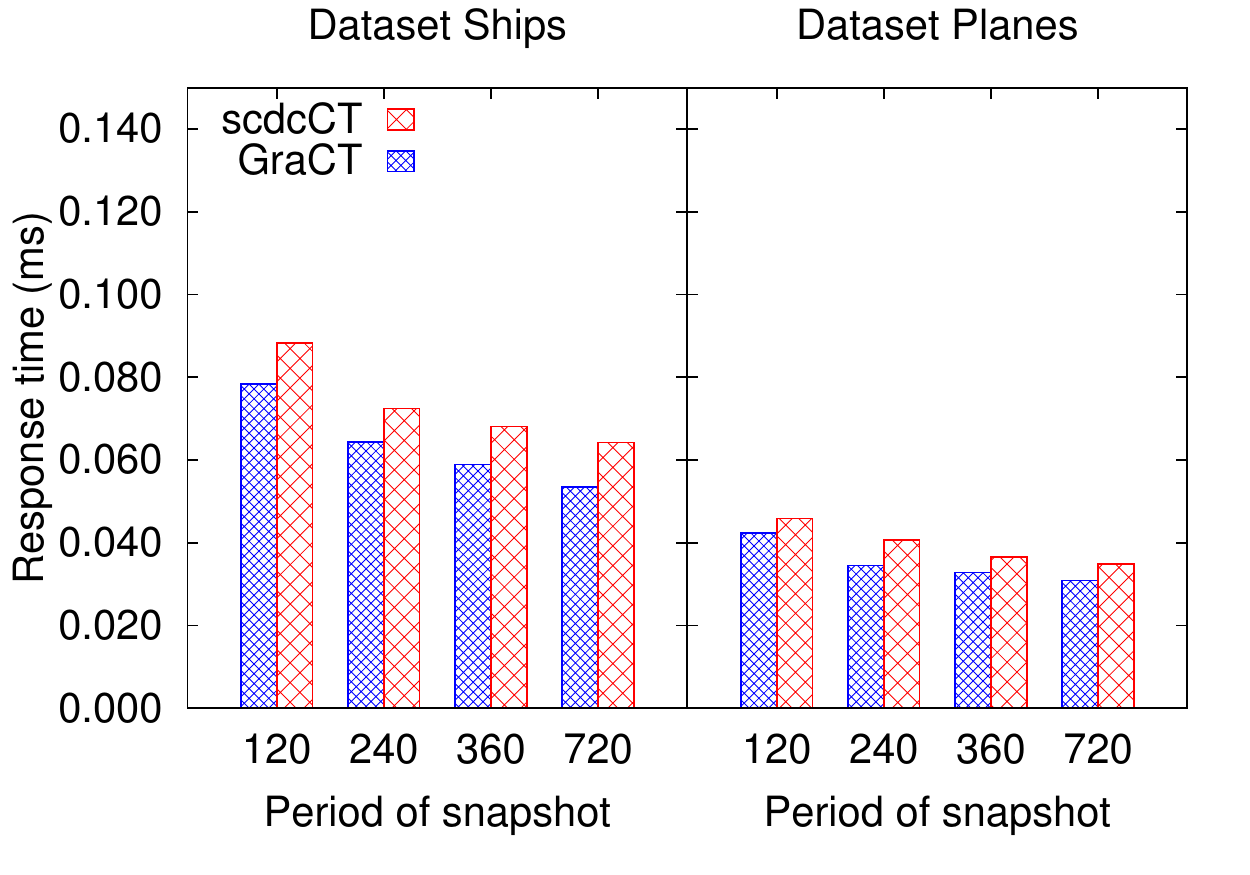}}
\subfigure[\textit{Time-slice S}]
{\label{fig:time-sliceS}\includegraphics[width=0.43\textwidth]{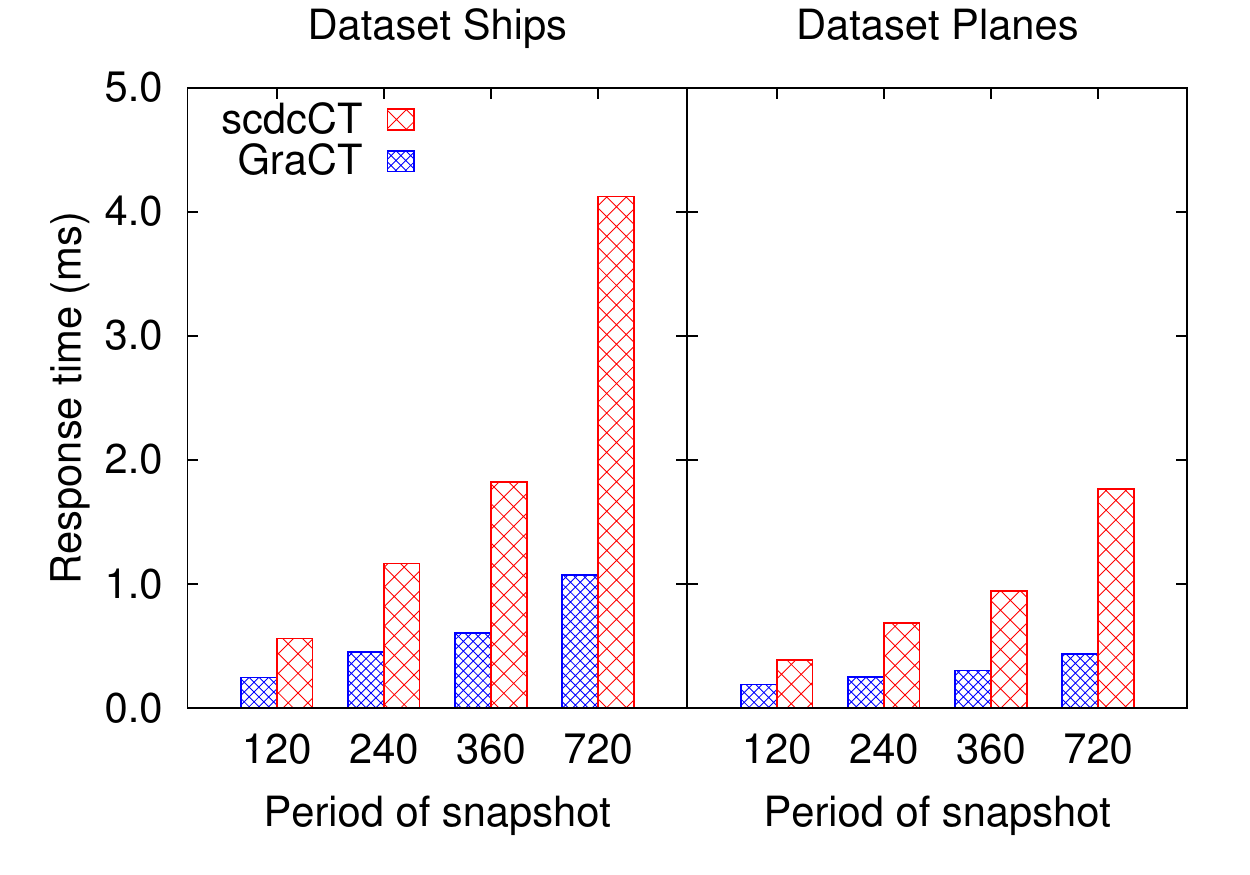}}
\subfigure[\textit{Time-slice L}]
{\label{fig:time-sliceL}\includegraphics[width=0.43\textwidth]{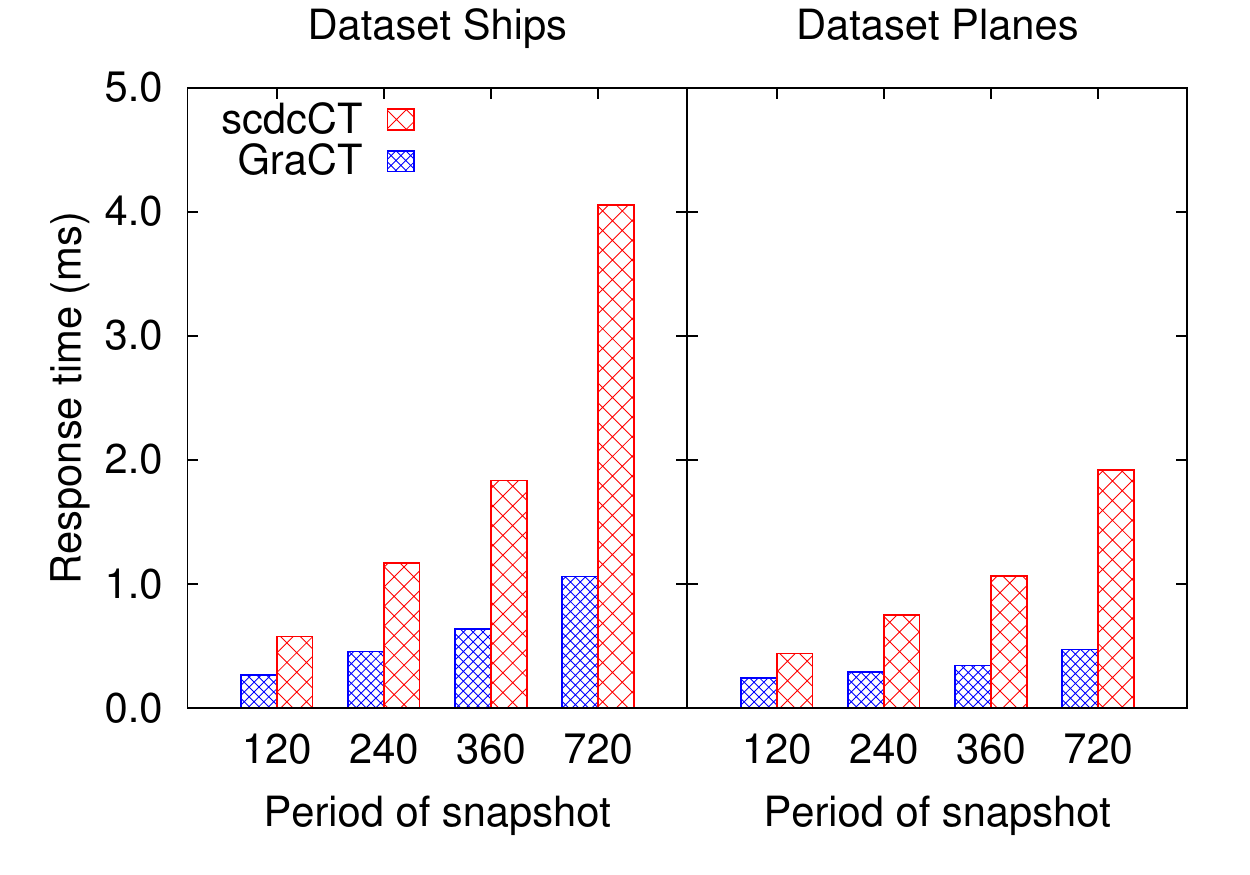}}
\subfigure[\textit{Time-interval S}]
{\label{fig:time-intervalS}\includegraphics[width=0.43\textwidth]{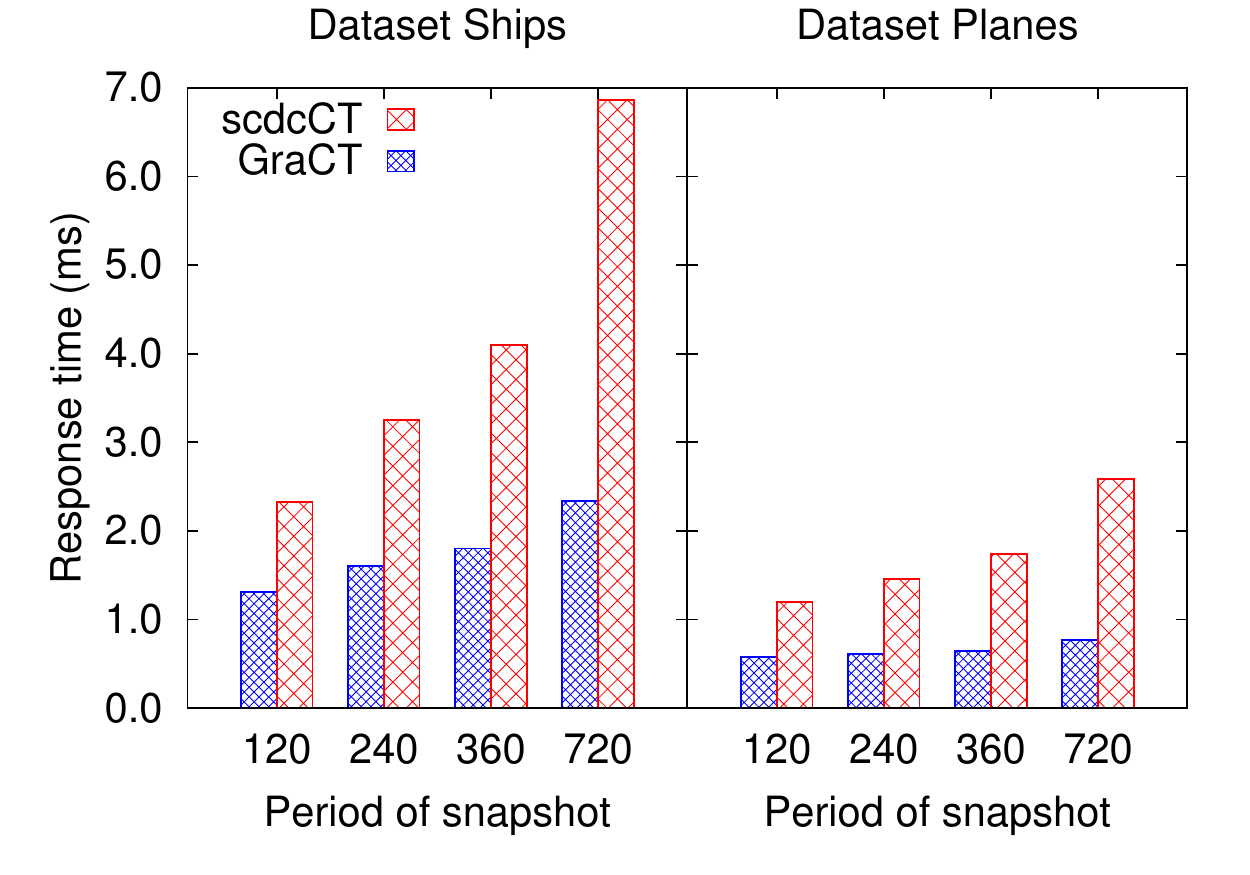}}
\subfigure[\textit{Time-interval L}]
{\label{fig:time-intervalL}\includegraphics[width=0.43\textwidth]{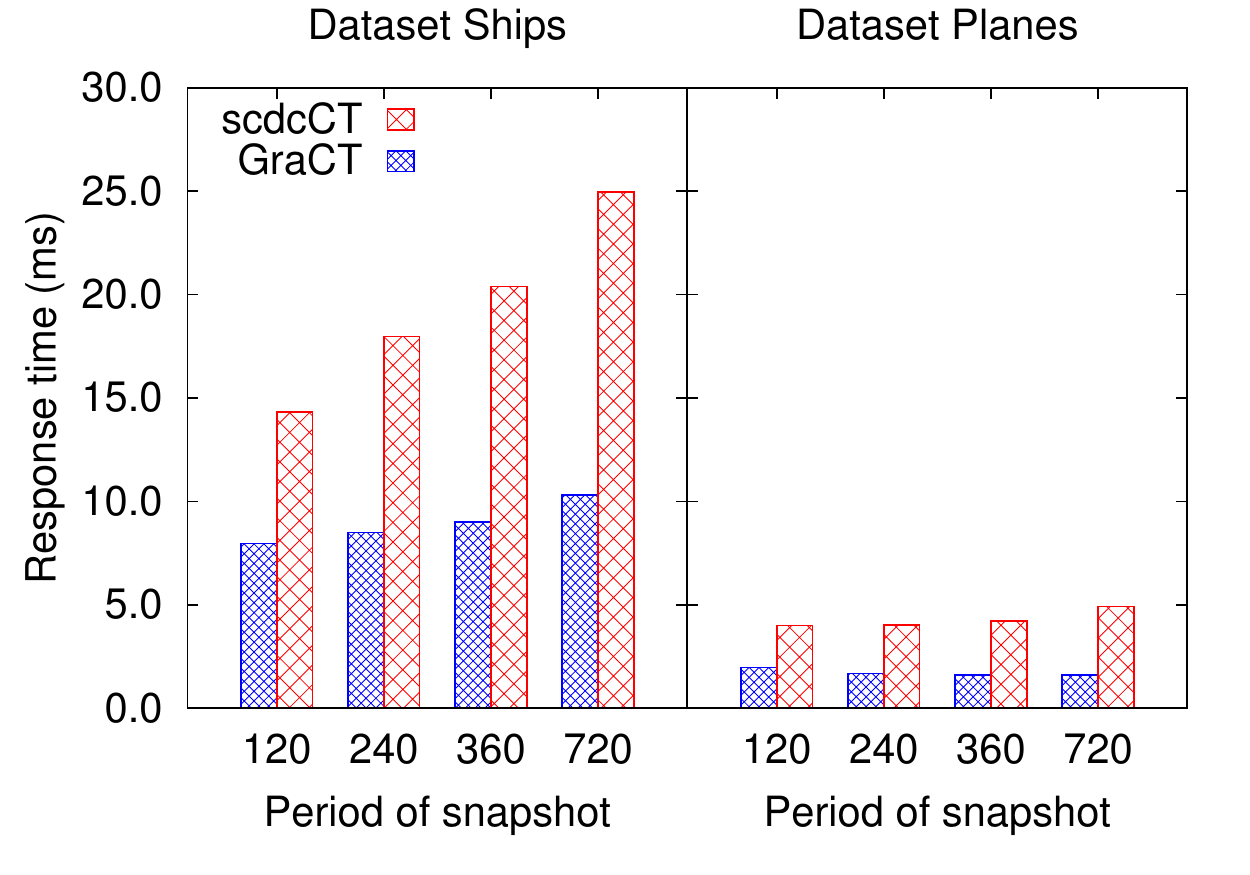}}
\subfigure[\textit{Knn}]
{\label{fig:time-knn}\includegraphics[width=0.43\textwidth]{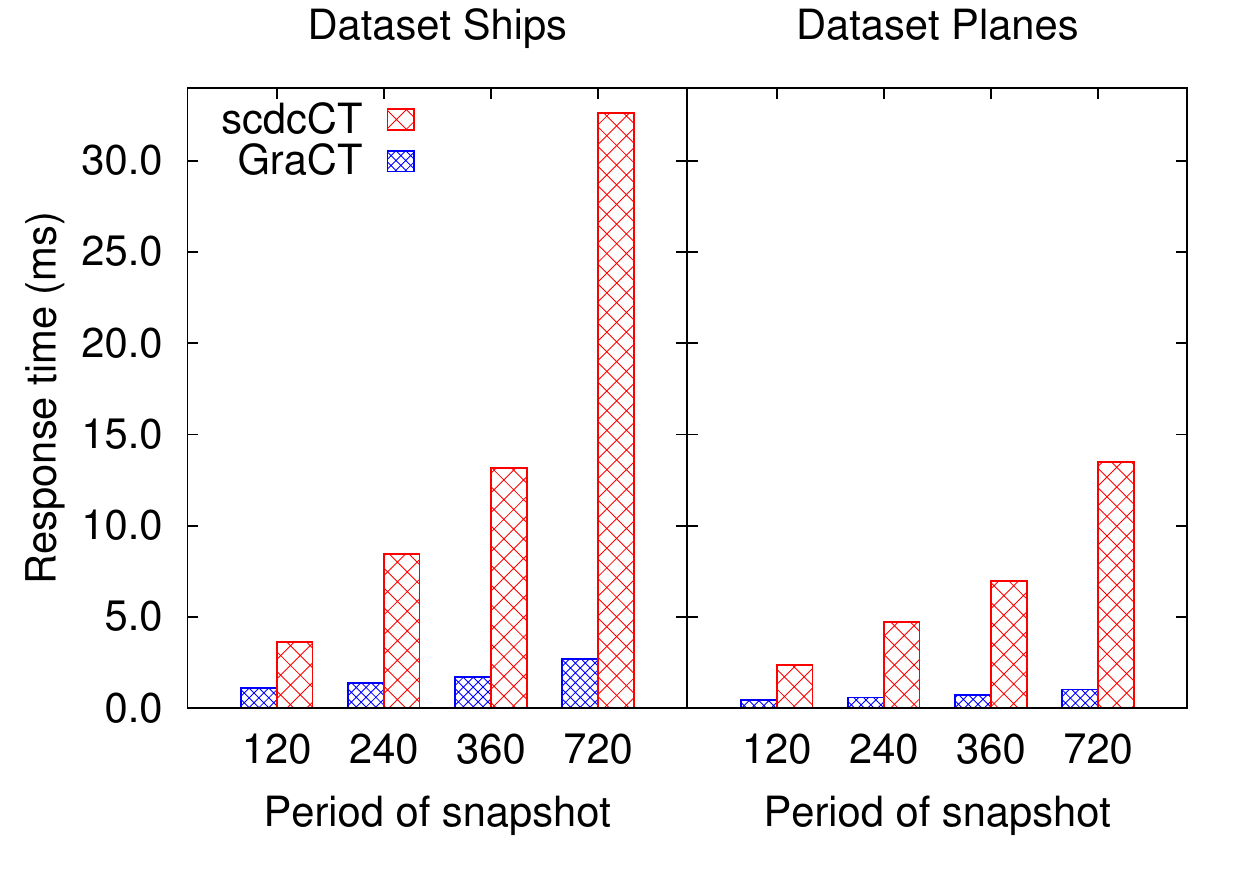}}

\vspace*{-1mm}
\caption{Response time for different queries, in milliseconds per query.}
\label{fig:experiment2}
\end{figure}

This experiment studies the response times of GraCT and scdcCT, with different snapshot periods, on the real datasets. Figure \ref{fig:experiment2} shows the average time (in milliseconds) per query for the following seven types of queries:
\begin{itemize}
 \item {\em object}: this query obtains the position of a given object at a given time instant $t_q$. We averaged over 20,000 queries for random objects and time instants.
 \item {\em trajectory}: this query returns the trajectory followed by an object during an interval $[t_b, t_e]$, where $t_e-t_b$ is fixed at 2,000 time instants. We averaged over 10,000 queries for random objects and time instants $t_b$.
\item {\em time-slice S}: this query obtains the identifiers and positions of the objects lying within a small region ($40 \times 40$\ cells) at a given time instant. We averaged over 1,000 queries for random region positions and time instants.
\item {\em time-slice L}: this query obtains the identifiers and positions of the objects lying within a larger region ($320 \times 320$\ cells). We averaged over 1,000 queries for random region positions and time instants.
\item {\em time-interval S}: this query obtains the objects present in a small region ($40 \times 40$\ cells) at any time instant between $t_b$ and $t_e$, where the interval size is $t_e-t_b = 100$ instants. We averaged over 1,000 queries with random region positions and instants $t_b$.
\item {\em time-interval L}: this query obtains the objects present in a larger region ($320 \times 320$\ cells) over a longer interval of size $t_e-t_b=500$ instants. We averaged over 1,000 queries with random region positions and instants $t_b$.
\item {\em knn}: this query obtains the $K$ nearest neighbors to a given position at a given time instant, where $K$ is a random value between 1 and 50. We averaged over 1,000 queries with random objects and time instants.
\end{itemize}

\begin{figure}[!t]
	\centering     
	\subfigure[{Period 120}]
	{\label{fig:snaplog-120}\includegraphics[width=0.49\textwidth]{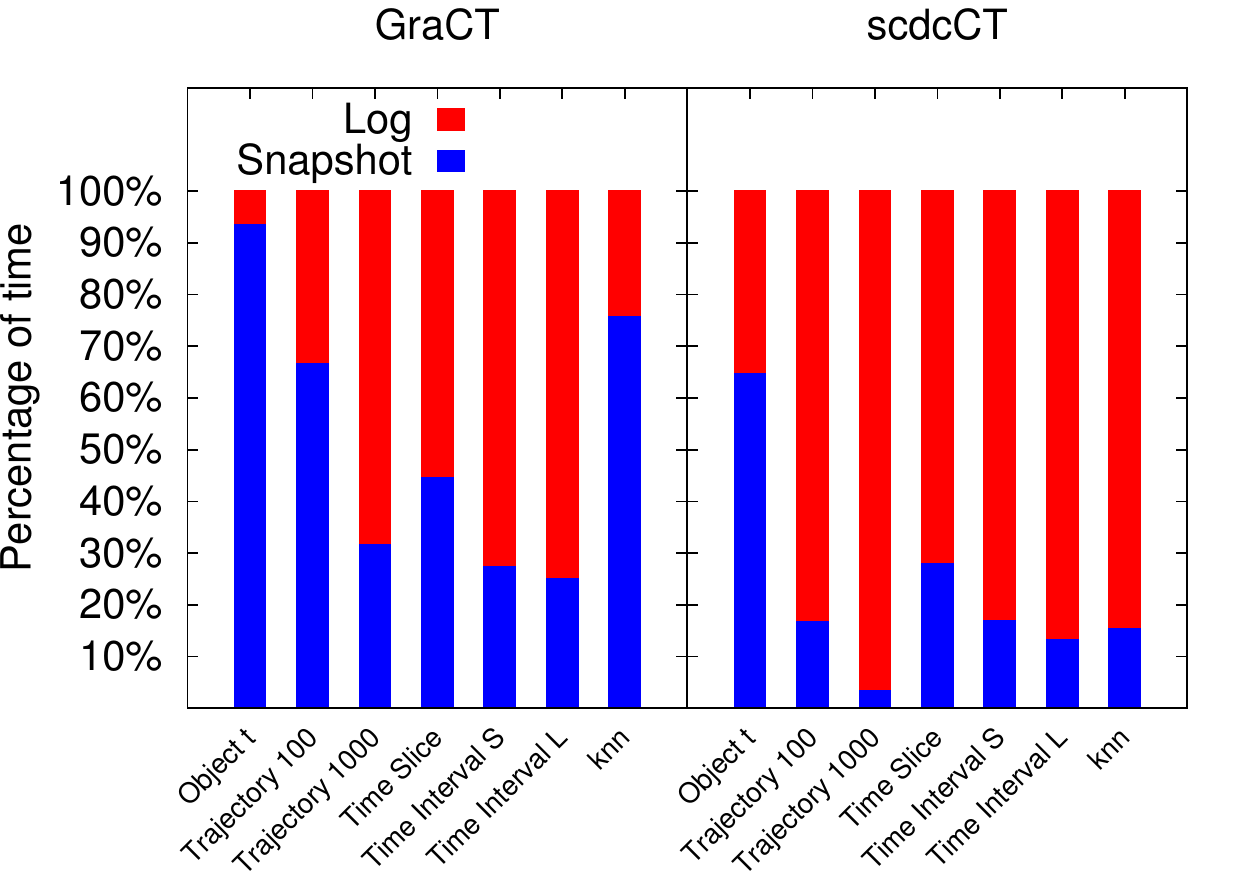}}
	\subfigure[{Period 720}]
	{\label{fig:snaplog-720}\includegraphics[width=0.49\textwidth]{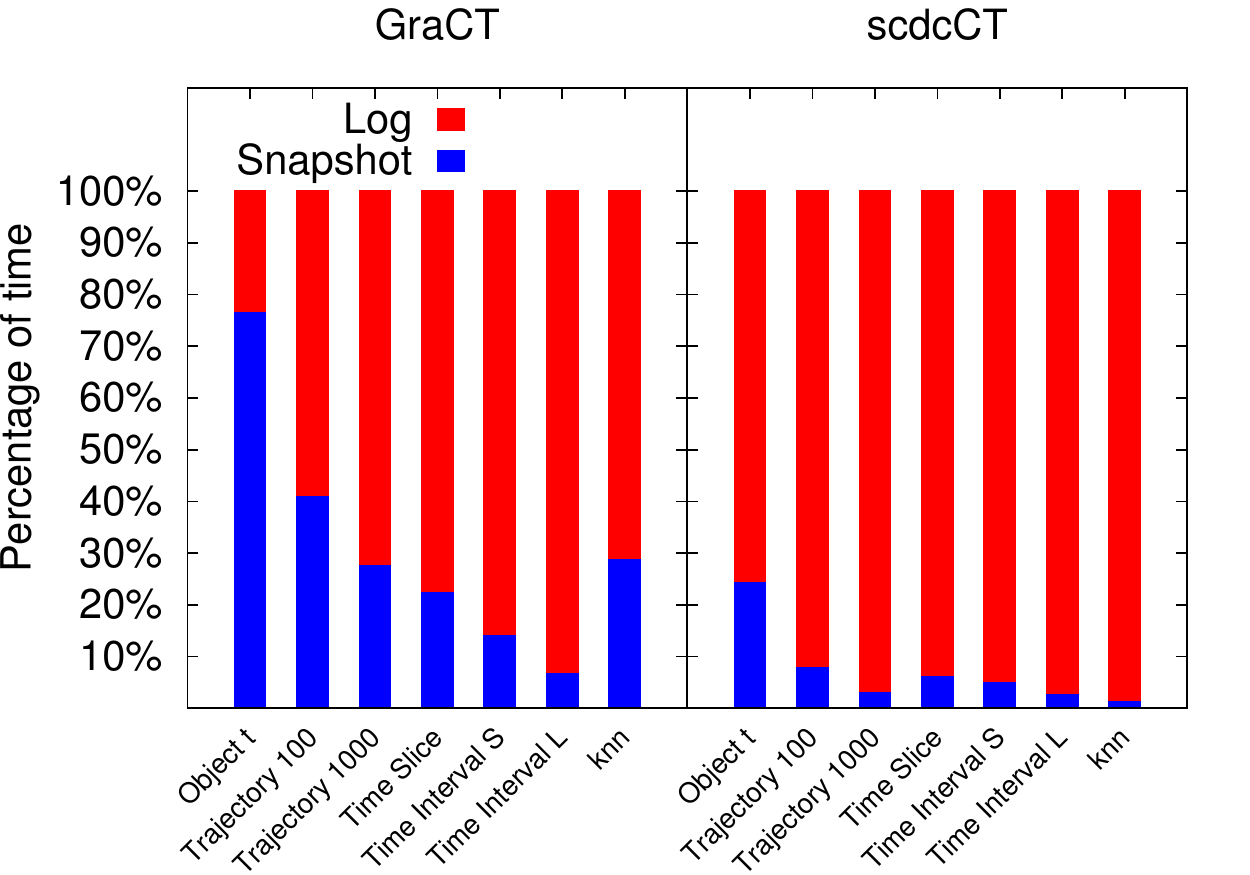}}
	\caption{Time spent on the snapshot versus the logs for different queries.}
	\label{fig:experiment3}
\end{figure}

\subsubsection{Object}

As seen in Figure \ref{fig:time-ot}, GraCT outperforms scdcCT in this query, answering within 2--4 microseconds. The logs in GraCT have far fewer entries than those in scdcCT, and in many cases the extra information stored on nonterminals allows processing of each GraCT log entry in constant time, just as a simple scdcCT log entry. Further, as previously demonstrated, Re-Pair compression improves faster as the logs increase, which makes GraCT times less sensitive than scdcCT to the increased distance between snapshots. The improvements of GraCT over scdcCT range from 5.9\% with a snapshot period of $d= 120$ to 46.3\% with $d=720$ on \texttt{Ships}, and from 12.3\% to 100.4\% on \texttt{Planes}.

Figure \ref{fig:experiment3} shows the relative time spent on the $k^2$-tree of the snapshot versus the time spent on the logs for the different queries. Since the times spent on the snapshots are identical, it follows that GraCT processes the logs way faster than scdcCT.
This demonstrates that, provided with the proper enhanced data, the use of grammar compression also yields good query times, outperforming the statistical encoding of trajectories, because it processes fewer entries at about the same speed per entry. 

\subsubsection{Trajectory}

In trajectory queries, as shown in Figure \ref{fig:time-trajectory}, GraCT also outperforms scdcCT, but the difference in response times is small: 12\%--20\% on \texttt{Ships} and 8\%--13\% on \texttt{Planes}. The majority of the time in this query is spent in decompressing the trajectory (of 2,000 entries), which is done at about the same rate in statistical (scdcCT) and grammar-based (GraCT) compression: 25--45 nanoseconds per decompressed datum on \texttt{Ships} and 15--20 on \texttt{Planes}. Since the trajectories to recover are significantly longer than the snapshot periods, snapshots mostly disrupt the processing of the logs, which explains why times improve with fewer snapshots.

Figure \ref{fig:experiment3} shows the case of shorter trajectories, of lengths 100 (\texttt{Trajectory 100}) and 1,000 (\texttt{Trajectory 1000}), on \texttt{Ships}. As expected, more time is spent on the logs to retrieve longer trajectories or with longer sampling periods. 

\subsubsection{Time-slice S and time-slice L}

In time-slice queries, the difference in performance between GraCT and scdcCT becomes much more significant. As shown in Figures \ref{fig:time-sliceS} and \ref{fig:time-sliceL}, GraCT performs 2--4 times faster than scdcCT in both datasets, running each query in less than a millisecond.

This query can be regarded as running an {\em object} query on each of the candidate objects, but as shown in Figure \ref{fig:experiment3}, the algorithm spends more time working on the log than for {\em object} queries, because the $k^2$-tree amortizes its traversal cost when retrieving many points. Thus both indexes spend more time in the traversal of the log, where GraCT outperforms scdcCT, and less on the $k^2$-tree, where they perform similarly. As a result, GraCT can be up to 4 times faster than scdcCT on {\em time-slice} queries, whereas in {\em object} queries it was up to twice as fast.

\subsubsection{Time-interval S and time-interval L}

As observed in Figures \ref{fig:time-intervalS} and \ref{fig:time-intervalL}, time-interval queries are very similar to time-slice queries with respect to the comparison between GraCT and scdcCT, although this time GraCT is only up to 3 times faster than scdcCT. On longer time intervals, the difference is smaller because both structures may have to decompress the movements along the interval $[t_b,t_e]$. The difference is larger than on a {\em trajectory} query, however, because GraCT can avoid decompressing nonterminals if their MBRs are disjoint or contained in the query range. 

\begin{table}[t]
\small
\begin{center}
\begin{tabular}{l|rr|rr|rr|rr}
Period & \multicolumn{2}{c|}{120} & \multicolumn{2}{c|}{240} &\multicolumn{2}{c|}{360} & \multicolumn{2}{c}{720} \\
\hline
       & MBR & no MBR & MBR & no MBR & MBR & no MBR & MBR & no MBR \\
\hline
{\em Slice S}    & 0.25 &  0.25 & 0.46 &  0.47 & 0.61 &  0.65 &  1.08 &  1.13 \\
{\em Slice L}    & 0.27 &  0.27 & 0.46 &  0.47 & 0.64 &  0.67 &  1.06 &  1.11 \\
{\em Interval S} & 1.31 &  7.40 & 1.60 & 12.66 & 1.80 & 17.39 &  2.34 & 29.68 \\
{\em Interval L} & 7.96 & 39.23 & 8.50 & 50.84 & 9.00 & 59.43 & 10.32 & 76.67 \\
\end{tabular}
\end{center}
\caption{The impact of using MBRs to preempt time-slice and time-interval queries, in milliseconds.}
\label{tab:mbr}
\end{table}

Table \ref{tab:mbr} evaluates the improvement we obtain due to the use of MBRs of nonterminals on \texttt{Ships}. As it can be seen, the use of MBRs has a relatively small impact on time-slice queries (below 5\%), but a very significant one on time-interval queries: increasing with the sampling period, the speedup factor goes from 5.6 to 12.7 on {\em time-interval S} and from 4.9 to 7.4 on {\em time-interval L} queries. Without using MBRs, indeed, GraCT would be slower than scdcCT on time-interval queries.

\subsubsection{Knn}

Nearest neighbor queries also require processing the logs of many candidate objects. However, the process done on each log entry is significantly heavier than on previous queries: objects are extracted from and reinserted in priority queues, their $d_{min}$ distances are recalculated, etc. In this scenario, the fewer number of entries processed by the grammar-compressed logs of GraCT, which can skip whole nonterminals, is even more valuable. As a result, GraCT is up to an order of magnitude faster than scdcCT for these queries, answering them in less than 3 milliseconds.

Figure \ref{fig:experiment3} shows that the prioritized traversal of the $k^2$-tree consumes a significant part of the query time on GraCT, but on scdcCT it is counterweighed by the more expensive processing of the log.

\subsection{Scalability}\label{sec:scalability}

\begin{figure}[!t]
	\centering     
	\subfigure[{Compression ratio.}]
	{\label{fig:scala-ratio}\includegraphics[width=0.49\textwidth]{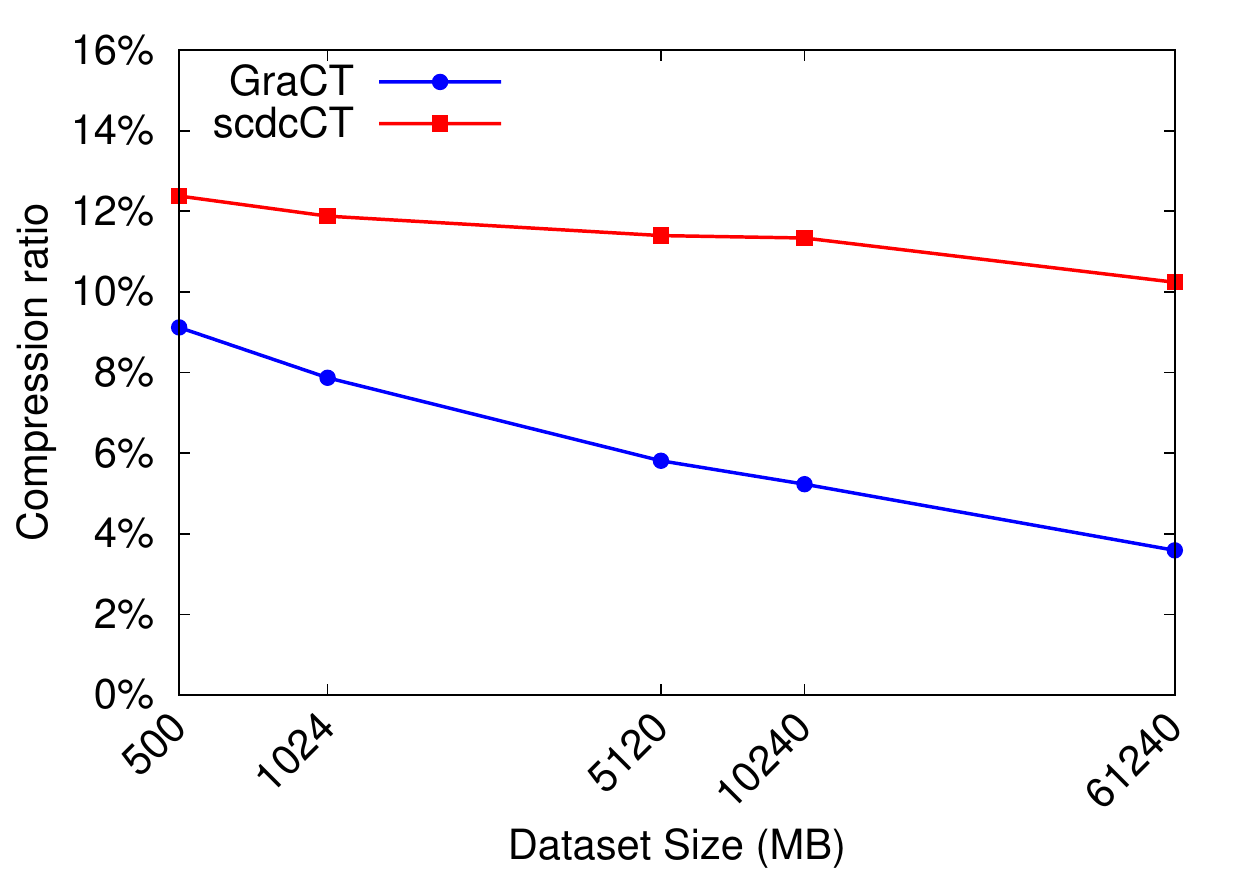}} 
	\subfigure[{Query times for {\em object}.}]
{\label{fig:scala-query}\includegraphics[width=0.49\textwidth]{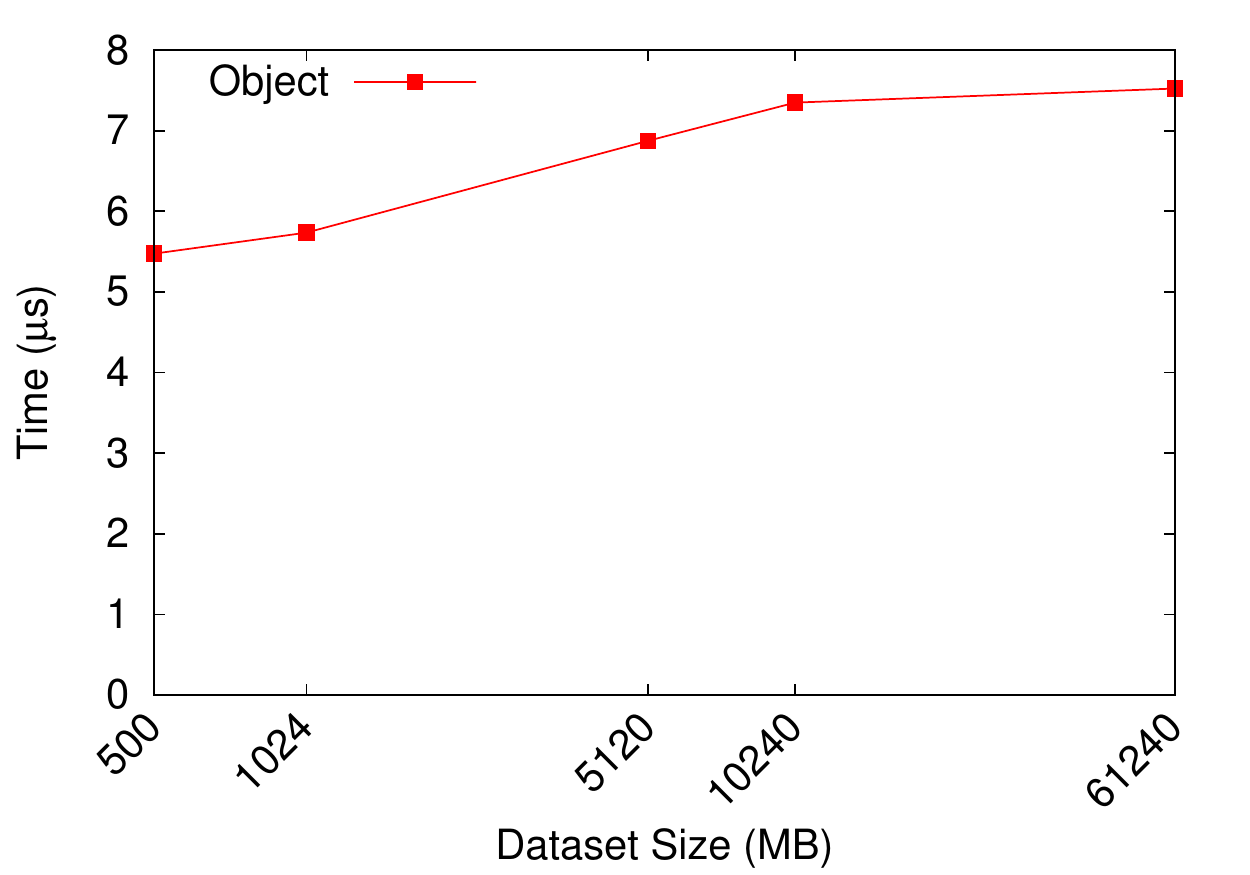}}
	\subfigure[{Query times of {\em trajectory}.}]
	{\label{fig:scala-query2}\includegraphics[width=0.49\textwidth]{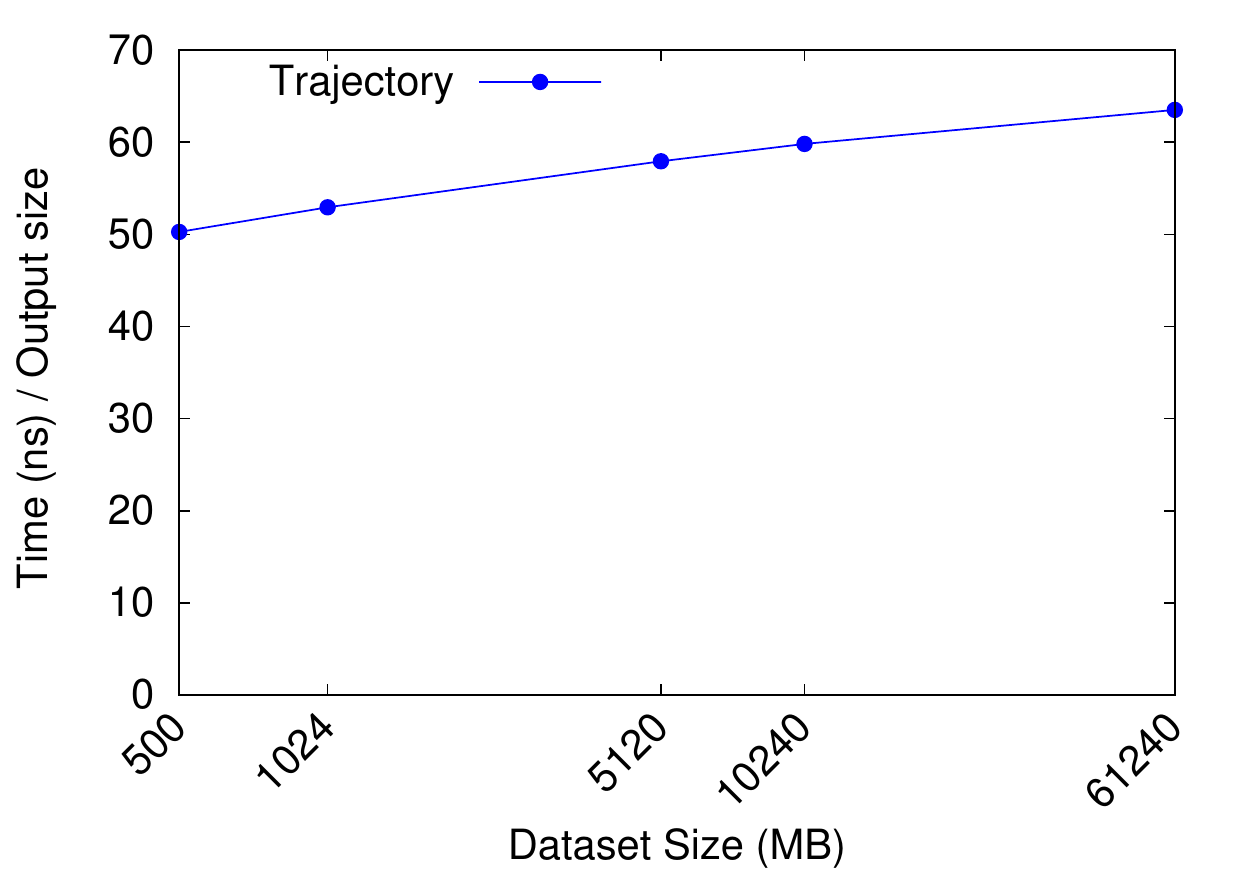}}
	\subfigure[{Other query times, over input size.}]
	{\label{fig:scala-query3}\includegraphics[width=0.49\textwidth]{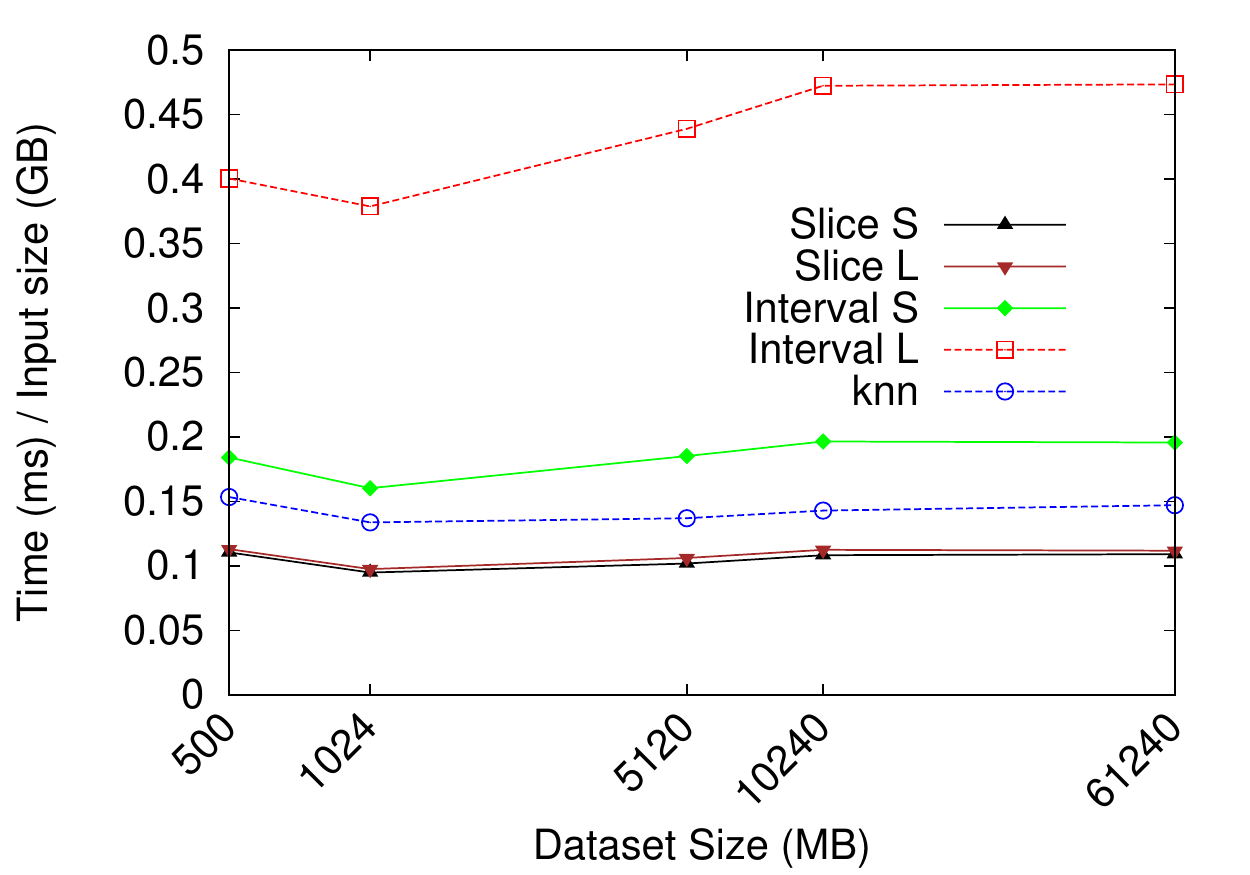}}
	\caption{Evolution of compression ratio and query times as the dataset grows. Dataset sizes are shown in logscale.}
\end{figure}

In order to evaluate the scalability of our structure in compression and query times, we create five prefixes of \texttt{Taxis} with sizes 500 MB, 1,024 MB, 5,120 MB, 10,240 MB, and 61,240 MB. In each longer prefix, more objects are incorporated to the dataset. We built GraCT and scdcCT on each dataset, with the period of snapshots set to 720. Over the GraCT indexes, we ran queries with the same settings of Section \ref{subsec:query-types}.

As shown in Figure \ref{fig:scala-ratio}, the compression ratio of GraCT improves as the dataset grows, from 9\% to 4\%. This is a consequence of exploiting repetitiveness via grammar compression. As the number of objects and trajectories of the same kind increases, finding similar trajectories is more likely. Instead, as it can be seen, statistical compression is much less sensitive to repetitiveness, improving only from 12\% to 10\%. This shows another advantage of grammar compression when handling very large datasets.

With the growth of the data, on the other hand, the depth of the grammar used by GraCT also grows, logarithmically in case the grammar is balanced. This increases the cost of navigating inside a nonterminal, for example. This logarithmic growth is apparent in the {\em object} queries (Figure \ref{fig:scala-query}; since the $x$ axis is logarithmic, a logarithm looks like a straight line), though in {\em trajectory} queries (Figure \ref{fig:scala-query2}) the time is amortized over the 2,000 output positions (those are decompressed in constant time each from the grammar, with only a logarithmic additive penalty for the whole query). The other queries grow linearly because, as more objects are added on a fixed grid, the output size and/or number of candidates to consider grows proportionally to the data size. This is apparent in Figure \ref{fig:scala-query3}, where the query times divided by dataset size are essentially constant.

\subsection{Comparison with a classical uncompressed index}

\begin{figure}[!t]
	\centering     
	\subfigure[\textit{Time-slice S}]
	{\label{fig:time-mvr-sliceS}\includegraphics[width=0.49\textwidth]{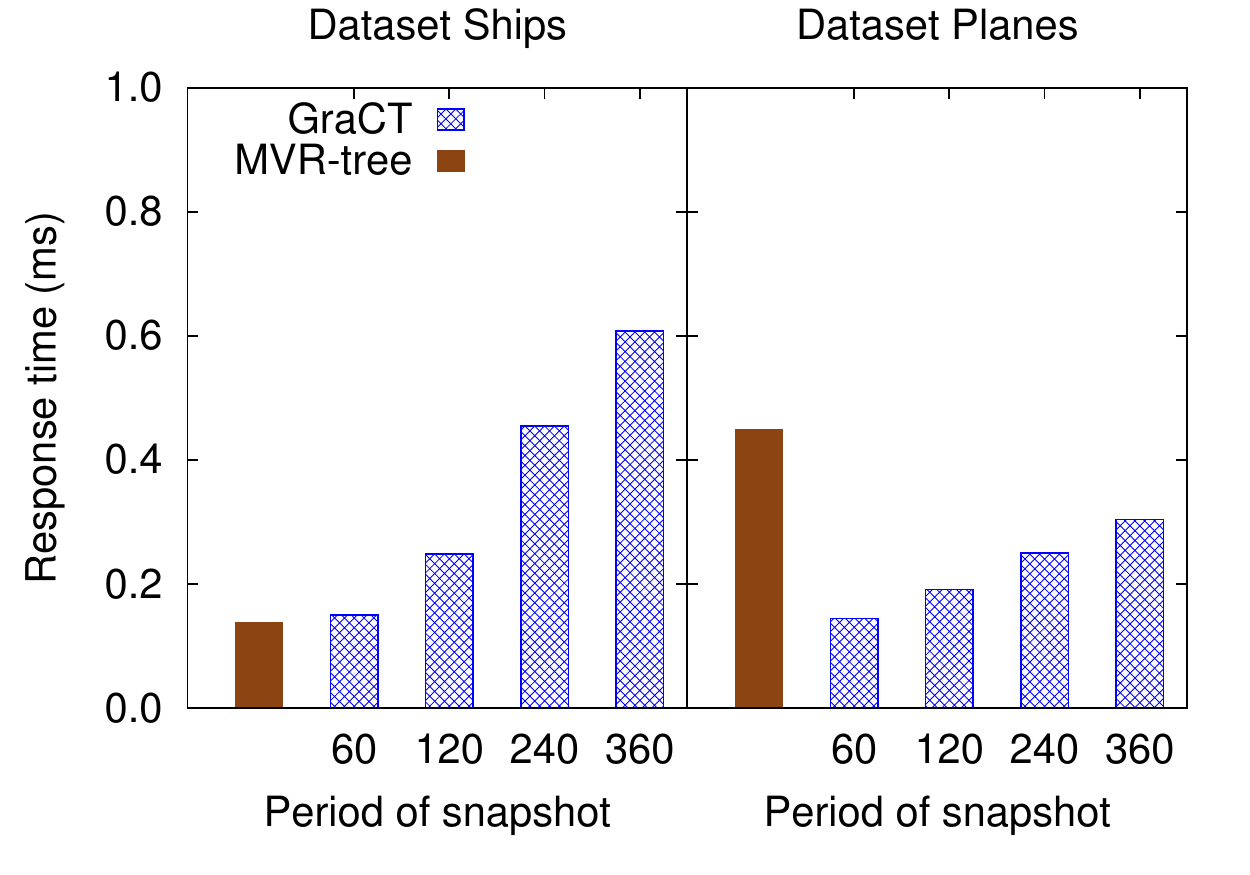}}
	\subfigure[\textit{Time-slice L}]
	{\label{fig:time-mvr-sliceL}\includegraphics[width=0.49\textwidth]{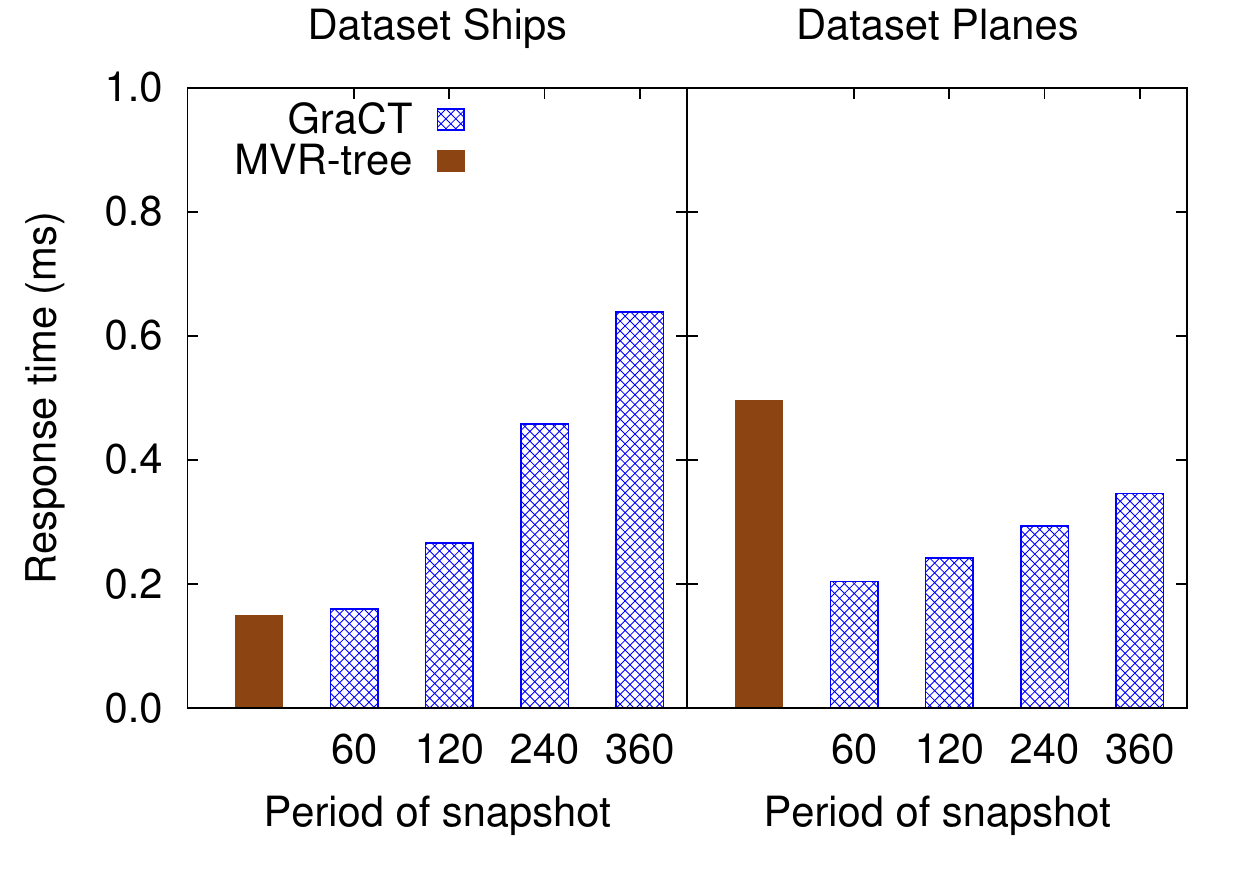}}
	\subfigure[\textit{Time-interval S}]
	{\label{fig:time-mvr-intervalS}\includegraphics[width=0.49\textwidth]{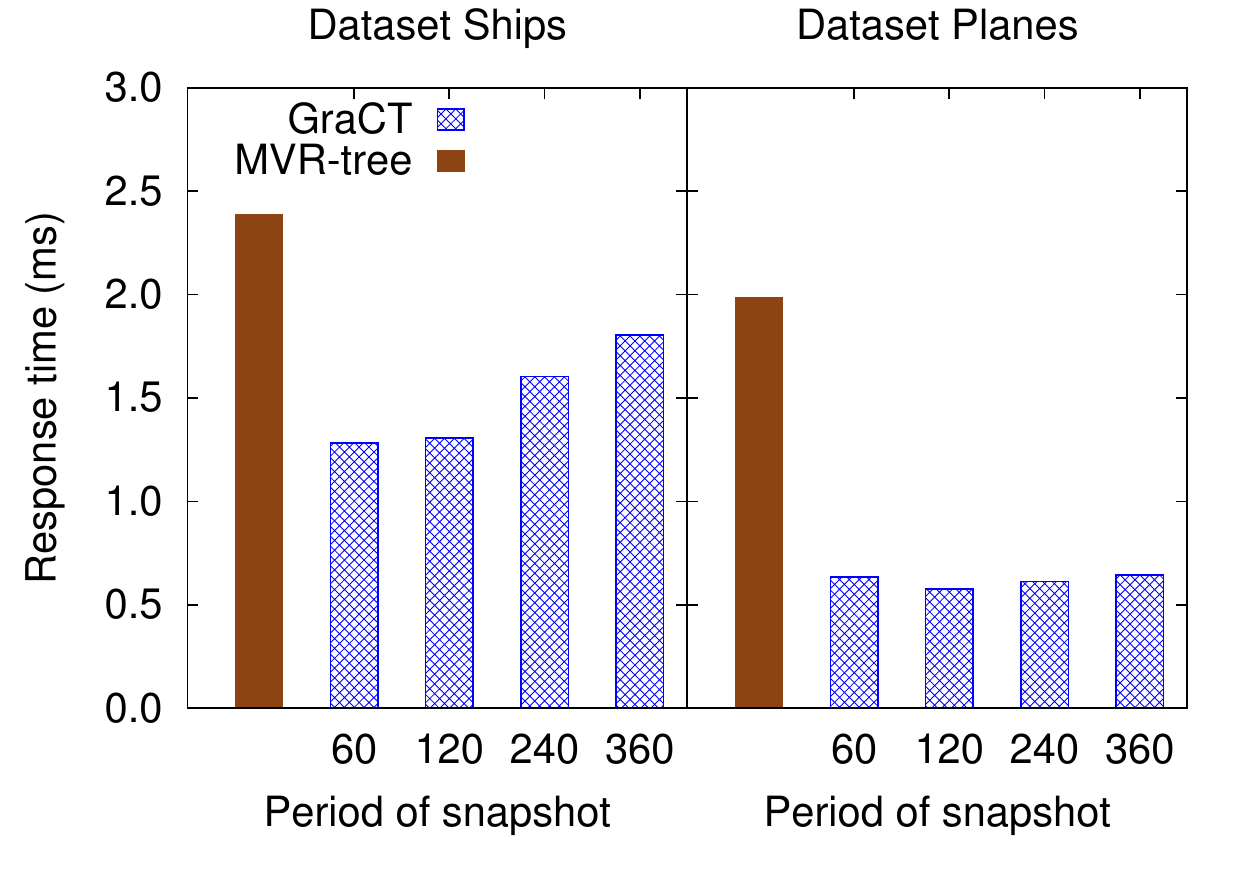}}
	\subfigure[\textit{Time-interval L}]
	{\label{fig:time-mvr-intervalL}\includegraphics[width=0.49\textwidth]{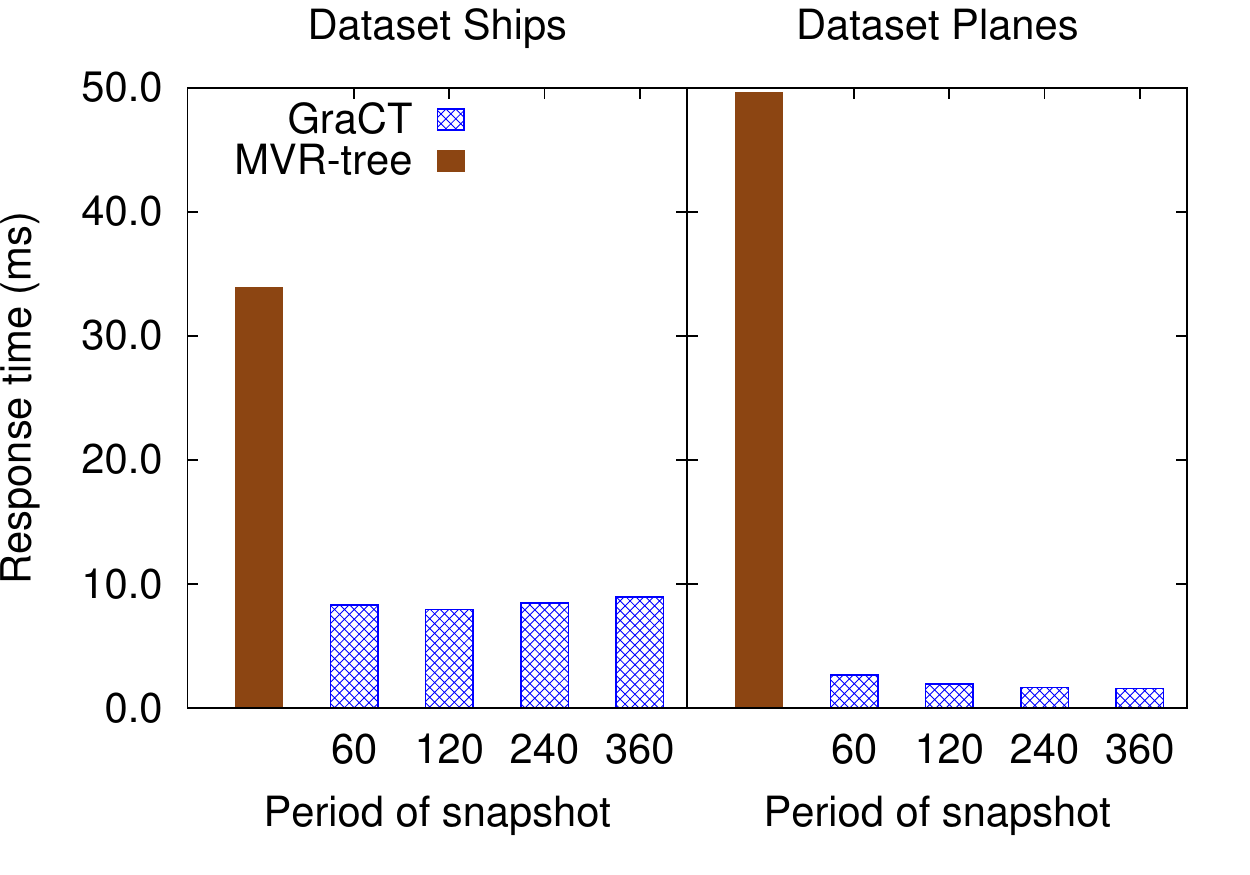}}
	\subfigure[\textit{Knn}]
	{\label{fig:time-mvr-knn}\includegraphics[width=0.49\textwidth]{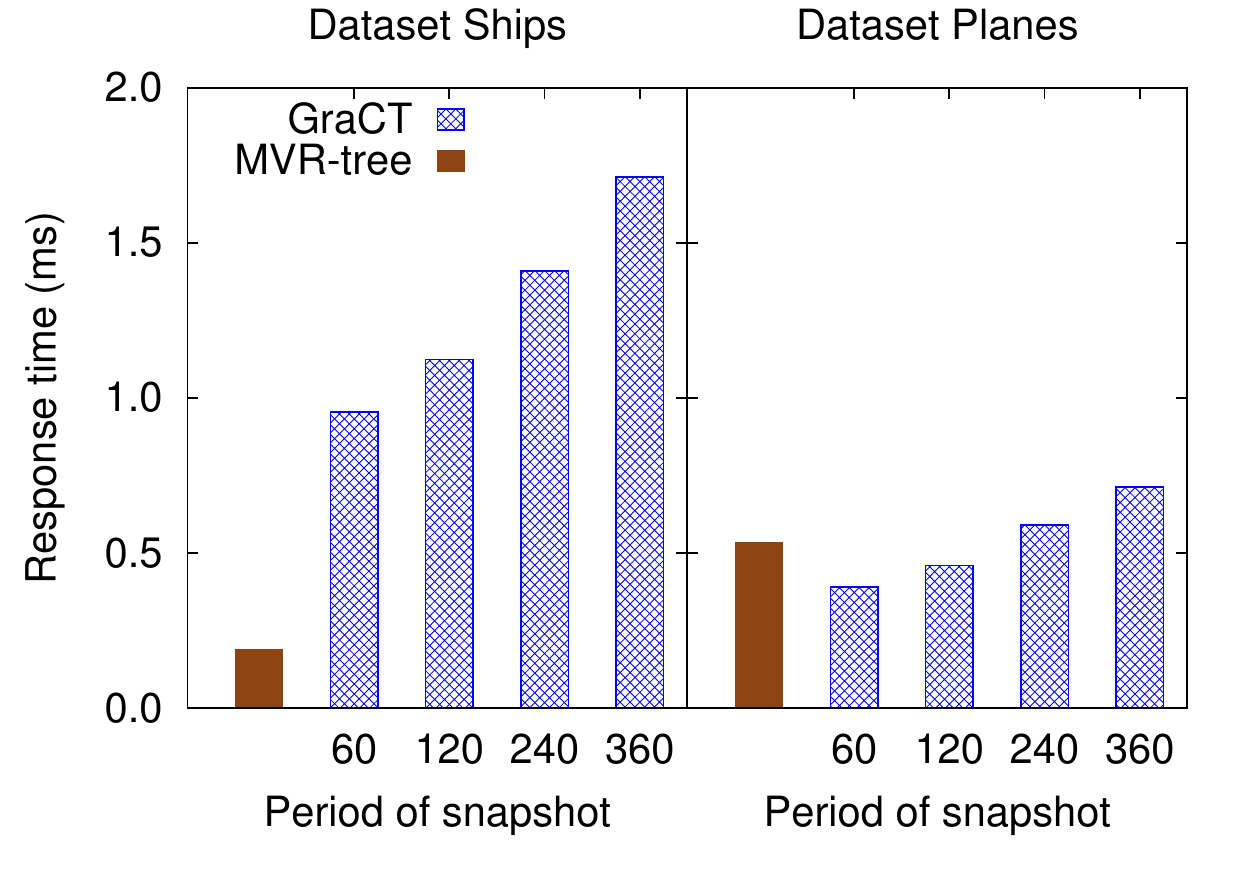}}
	\subfigure[Growing time-interval queries on \texttt{Ships}]
	{\label{fig:time-mvr-interval-growing}\includegraphics[width=0.49\textwidth]{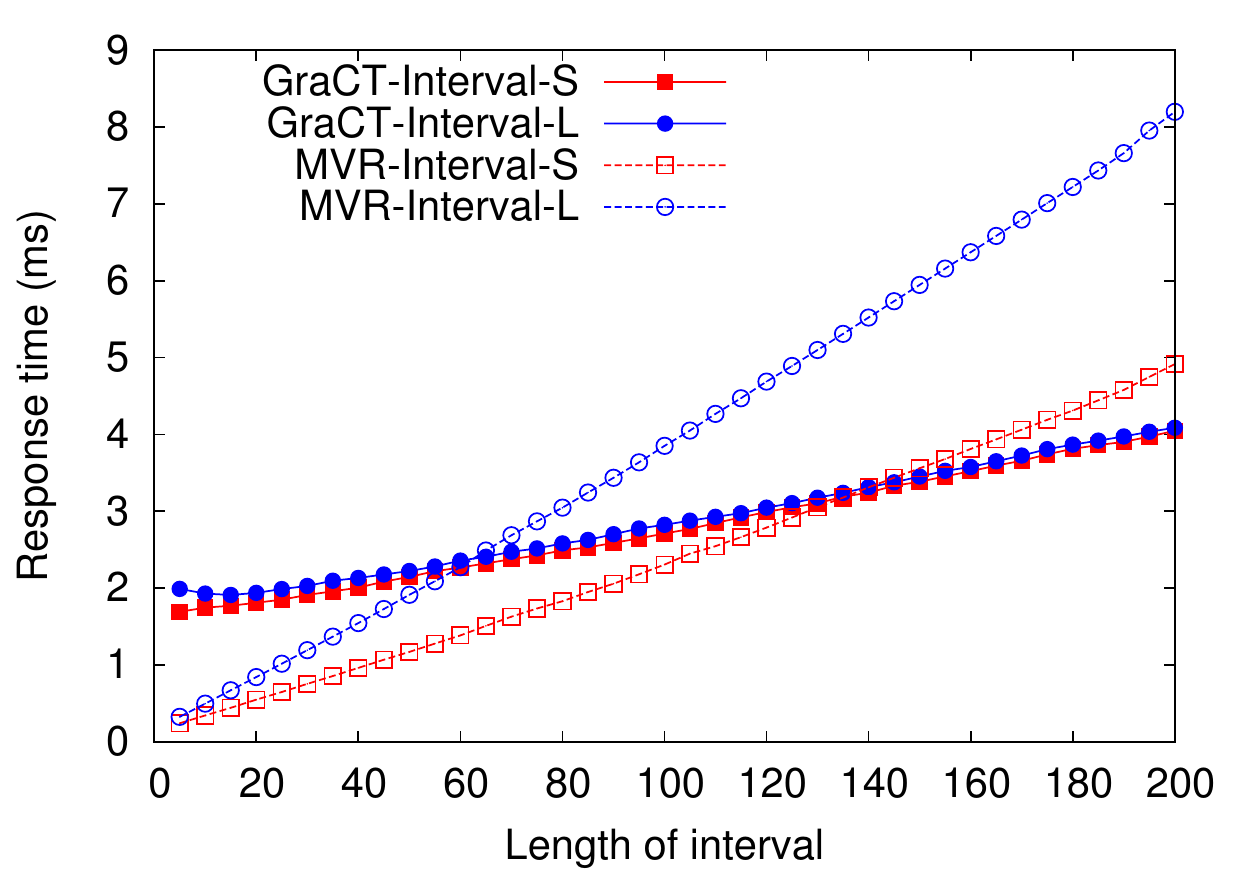}}
	\caption{Query time comparison of GraCT with the MVR-tree, running in main memory.}
	\label{fig:experiment4}
\end{figure}

As anticipated, we compare GraCT with the MVR-tree \cite{PapadiasT01}, a classic spatio-temporal structure based on the R-tree. The MVR-tree comprises multiple versions of the same data, each associated with a different interval of time, and an R-tree representing the positions of the objects during that interval. The intervals of the versions are disjoint. The MVR-tree exploits the repetition of nodes along consecutive R-trees, by sharing common nodes between versions. It is important to note that this spatio-temporal structure is efficient at solving time-slice, time-interval, and knn queries only, which are solved by identifying versions that overlap with the queried time interval. The algorithm traverses the R-tree of each such version, following the nodes involved in the spatial query range. Queries {\em object} and {\em trajectory}, instead, are costly for the MVR-tree. These are speeded up in the full structure, the MV3R-tree, by using an auxiliary 3DR-tree structure \cite{PapadiasT01} (which requires even more space). We therefore omit these queries in our comparison.

The MVR-tree was configured to run in main memory. Both structures were built on the \texttt{Ships} and \texttt{Planes} datasets described in Section \ref{datatset}, and GraCT was configured with the same snapshot distances used in the previous experiments (120, 240, 360 and 720). The size of the MVR-tree in \texttt{Ships} was 12.16 GB and in \texttt{Planes} it was 11.72 GB, while the maximum-space configuration of GraCT used 36.40MB and 51.63MB, respectively, two orders of magnitude less (more precisely, GraCT uses 342 times less space on \texttt{Ships} and 232 times less space on \texttt{Planes}). The MVR-tree is also 30--40 times larger than the raw data represented in binary form. 

We now study if this gigantic difference in space induces a significant impact on query time performance. 
Figure \ref{fig:experiment4} shows the times of both structures averaged over 1,000 random queries of each type: {\em time-slice S}, {\em time-slice L}, {\em time-interval S}, {\em time-interval L}, and {\em knn} queries. Instead of the sampling $d=720$, we include a new one, $d=60$, where GraCT uses 50.70 MB on \texttt{Ships} and 83.14 MB on \texttt{Planes}. This is still 245 and 144 times smaller than the MVR-tree, respectively, and is sufficient for GraCT to be as fast as MVR-tree on most queries.

Time-slice queries are the most efficient ones for the MVR-tree. Since they retrieve  objects within a given region at a single time instant only, the MVR-tree needs to traverse just one R-tree. When the traversal of the tree reaches the leaves, it can retrieve more than one object per leaf, which is faster than traversing one log of GraCT per candidate. Figures \ref{fig:time-mvr-sliceS} and \ref{fig:time-mvr-sliceL} show that, even in this case, the maximum-space configuration of GraCT matches the performance of the MVR-tree on \texttt{Ships}, and outperforms it on \texttt{Planes}. It is likely that the higher locality of reference when accessing a structure that is two orders of magnitude smaller plays in favor of GraCT.

This advantage of the MVR-tree vanishes on time-interval queries. In an  interval of time, the MVR-tree may have to traverse a number of R-trees, which makes it much slower and always loses to GraCT, as seen in Figures \ref{fig:time-mvr-intervalS} and \ref{fig:time-mvr-intervalL}. 
Figure \ref{fig:time-mvr-interval-growing} studies the turning point on {\tt Ships} by varying the length of the queried time interval, using the smallest space configuration of GraCT (snapshot distance 720). As the time span of the queries increases, both structures slow down. The MVR-tree times, however, increase much faster, becoming slower than GraCT when the time interval exceeds 140 time instants on {\em time-interval S} and 60 on {\em interval L}. GraCT is thus less sensitive to the length of the queried interval, outperforming MVR-tree already in short time intervals even on its minimum-space configuration.

Nearest neighbor queries refer again to a single time instant, and therefore the MVR-tree can handle them within a single R-tree, even if it usually has to traverse many more nodes than for a time-slice query. In this query, GraCT also has to traverse a large number of $k^2$-tree nodes and verify many candidates because, unlike the MVR-tree, it does not have a snapshot of the precise query time instant. As it can be seen in Figure \ref{fig:time-mvr-knn}, the MVR-tree clearly outperforms GraCT on \texttt{Ships}, though not on \texttt{Planes}, where GraCT always improves and the MVR-tree always degrades with respect to \texttt{Ships}. 

\subsection{Disk and full precision}

With the minimum-space configuration (snapshot sampling of 720), GraCT uses nearly 600 times less space than the MVR-tree. We can then use GraCT to index in main memory datasets for which the MVR-tree would  by far exceed the available memory, and would then have to operate on disk. As we show next, even a time-slice query (where the MVR-tree most prominently excels) requires around 50 milliseconds to be solved by the MVR-tree on disk, which is 50 times slower than the time required by GraCT in main memory. If we use a snapshot sampling of 60, then GraCT becomes more than 300 times faster, and is still 150--250 times smaller than the MVR-tree.

The compression of GraCT is achieved, however, by means of discretizing the data and therefore losing precision. In most real systems, full precision is anyway unnecessary. For example, using eight decimals for the GPS position of an object discriminates micrometers, which is much finer than what most physical devices can acquire, and only useful in very specialized fields like tectonic plate mapping. 

Still, if full precision is required, GraCT can be used as an in-memory {\em cached index} of the full-precision data stored on disk, exploiting its very small main memory footprint. To answer a query, GraCT is first used to obtain a list of {\em candidate} answers, which satisfy the query up to where the precision of GraCT can discriminate. Then, in a second step, those candidate answers are verified by reading the full-precision data from disk.

As a proof-of-concept of the feasibility of this approach, we designed a simple arrangement for time-slice queries, where the MVR-tree performs best. We use the GraCT data structures of the \texttt{Ships} dataset with 720 time instants between snapshots. In addition, we store the full-precision data in a relational table of PostgreSql DBMS v9.1.24.\footnote{{\tt https://www.postgresql.org}} The table stores tuples $\langle object, time, x, y \rangle$ to represent all the trajectories of all the objects. To efficiently verify the answers to time-slice queries, a hash index is set on the first two columns (together). 
For simplicity, we still assume that the time instants are discretized; a B-tree can be used for the second column in case they are real numbers as well.

As a baseline, we use the MVR-tree data structure on disk. GraCT, including the full-resolution data and the hash indexes, occupies 4.18 GB, whereas the MVR-tree occupies 78.61 GB. That is, GraCT still requires 18 times less disk space than the MVR-tree.

To emulate a scenario where the full-resolution data does not fit in main memory, we configured our machine to use 2 GB for the kernel, so that neither the MVR-tree nor the PostgreSql table can be completely cached by the operating system or the DBMS. 
 
The experiment begins in cold state, with all the data structures residing on disk and the caches of the operating system and the DBMS completely empty. 
 The first row of Table \ref{TS-DBMS} shows the average time of 1,000 executions of the first {\em time-slice} query in cold state. That query requires completely loading the GraCT data structures into main memory. In the case of the MVR-tree, the execution only loads the nodes of the R-tree traversed by the query.
 The second row shows the times resulting from running 1,000 {\em time-slice} queries after running the first one, that is, in warm state with the GraCT index already loaded in memory. We independently tested the second, third, and so on, seeing no significant differences. 
 
Even in cold state, GraCT is 10\% faster than the MVR-tree for {\em time-slice S} queries, but once in warm state, the response time of GraCT drops to 3.57 milliseconds, whereas the MVR-tree requires around 45 milliseconds, which is about 12 times slower. GraCT is much slower than the MVR-tree for {\em time-slice L} queries in cold state, on the other hand, but again, once we enter in warm state, GraCT becomes about 8 times faster. That is, a disk-based system on regime, with full precision using GraCT as an in-memory cached index is an order of magnitude faster than the MVR-tree and uses much less disk space. 


\begin{table}[]
\centering
\begin{tabular}{l|c|c|c|c}
\hline
& \multicolumn{2}{c|}{Slice S} & \multicolumn{2}{c}{Slice L}\\ \hline
                   & GraCT+DBMS & MVR-tree & GraCT+DBMS & MVR-tree \\ \hline
Cold state        & 53.59      & 59.24    & 313.79 & 45.76 \\ \hline
Warm state        & ~3.57       & 45.01 & ~~5.39 & 43.64   \\ \hline
\end{tabular}
\caption{Response times of {\em time-slice} queries on disk, in milliseconds. }\label{TS-DBMS}
\end{table}

Other queries, like {\em time-interval}, require a more sophisticated arrangement to use GraCT as an in-memory cache. For example, we can store the consecutive points of each trajectory in contiguous form, indexed by time with a B-tree, and a hash-indexed table of objects pointing to their corresponding trajectories. This enables an efficient traversal of the trajectories of candidate objects in the time intervals where GraCT estimates that they may intersect the query range. Those arrangements also support the simpler queries {\em object} and {\em trajectory}. Our experiment suggests that it is much more efficient to verify potential answers on disk than to find them using a disk-based structure.

\section{Conclusions}\label{sect:con}

This article introduces GraCT, a compact data structure for the representation of moving objects. In our experiments on real-world trajectory datasets, GraCT compresses the data to 4\%--7\% of its raw representation size. Within this space, GraCT supports not only direct access to the data, but also various queries looking for objects in spatio-temporal windows. The trajectory data is manipulated directly in compressed form, with only small portions requiring decompression in order to answer queries. The space also includes the storage of spatial indexes representing snapshots at regular time intervals, which speeds up queries. The main algorithmic novelty of GraCT is the representation of the logs of movements between snapshots using grammar-based compression. On our real-life datasets, grammar-based compression captures the repetitiveness arising from recurrent trajectories, outperforming statistical compression by a factor of 2 and improving as the datasets become larger. A second novelty is to enrich the grammar by associating information with the nonterminals, such as their relative MBR. This enables GraCT to traverse the logs very quickly without decompressing nonterminals in most cases, thereby answering queries within a few milliseconds. A comparison with the MVR-tree, a classical index, shows that GraCT is two orders of magnitude smaller and still competitive in query times when both run in main memory. When the MVR-tree cannot fit in main memory, GraCT is one or two orders of magnitude faster. This includes an arrangement where GraCT is used as a low-precision main-memory cached index that must verify its result on disk, where the full-precision data is kept.

Future work in this area will explore other compression mechanisms that exploit repetitiveness, looking for ways to process the trajectory data without decompressing it. Some alternatives are the use of block trees \cite{BGGKOPT15} or relative Lempel-Ziv \cite{KPZ10}. We also plan to further use GraCT as a main-memory cached index, extending our arrangement to handle all the queries considered in this article.

\section*{Acknowledgments}

This work has been funded by the European Union Horizon 2020 Marie Sk{\l}odowska-Curie Action Fund [grant agreement EU H2020 MSCA RISE BIRDS: 690941]; Ministerio de  Econom\'{\i}a y Competitividad (PGE and FEDER) [grant number TIN2016-78011-C4-1-R], Centro para el desarrollo Tecnol\'ogico e Industrial Programa CIEN 2014 (co-founded with FEDER) [grant number ITC-20151247], Ministerio de Educaci\'{o}n y Formaci\'{o}n Profesional (FPU program) [grant number FPU16/02914]; Xunta de Galicia (co-founded with FEDER) [grant numbers ED431C 2017/58; ED431G/01]; the Chilean National Science and Technology Development Fund (Fondecyt Grant 1-170048); and the Millennium Institute for Foundational Research on Data (IMFD), Chile.

\section*{References}
\bibliographystyle{elsarticle-harv}
\bibliography{bibliografia}

\end{document}